\documentclass{article} 
\usepackage{amsmath,amsfonts,amssymb}
\usepackage{graphicx,wrapfig}
\usepackage[hyphens]{url}
\usepackage{hyperref}
\usepackage{multirow}
\usepackage[font=small]{caption}
\usepackage{subcaption}
\usepackage{float}
\usepackage{longtable}

\usepackage{pdflscape}
\usepackage[margin=0.75in]{geometry}
\usepackage{bm}
\usepackage{afterpage}
\usepackage{multicol}
\usepackage{dblfloatfix}
\usepackage[toc,page]{appendix}
\usepackage{breqn}
\usepackage{mathtools}
\usepackage[numbers,sort&compress]{natbib}
\usepackage{authblk}
\usepackage{multicol}
\usepackage{breqn}
\usepackage{booktabs}
\usepackage[export]{adjustbox}
\usepackage{comment}
\usepackage{hyperref}

\usepackage[siunitx]{circuitikz}
\usepackage{tikz}

\usepackage{arydshln}

\usepackage{xcolor}

\usepackage{titlesec}
\usepackage{enumitem}

\titleformat{\paragraph}[runin]
{\normalfont\normalsize\bfseries}{\theparagraph}{1em}{}[:]
\titlespacing*{\paragraph}
{0pt}{3.25ex plus 1ex minus .2ex}{1.5ex plus .2ex}

\title{Multiscale modelling of reversed Potts shunt as a potential palliative treatment for suprasystemic idiopathic pulmonary artery hypertension in children}

\author[1,$\dagger$]{Sanjay Pant}
\author[2,3]{Aleksander Sizarov}
\author[1]{Angela Knepper}
\author[4]{Ga\"{e}tan Gossard}
\author[4]{Alberto Noferi}
\author[5]{Younes Boudjemline}
\author[4]{Irene Vignon-Clementel}

\affil[1]{Faculty of Science and Engineering, Swansea University, United Kingdom}
\affil[2]{Department of Pediatrics, Maastricht University Medical Centre, Maastricht, the Netherlands}
\affil[3]{Pediatric Cardiology, Necker University Hospital for Sick Children, Paris, France}
\affil[4]{Inria, Saclay Ile-de-France, France}
\affil[5]{Cardiac Catheterization Laboratories, Sidra Heart Center, Sidra Medicine, Doha, Qatar \vspace{2mm}}

\affil[$\dagger$]{Correspondence to: Sanjay.Pant@swansea.ac.uk}

\date{\vspace{-5ex}} 

\providecommand{\keywords}[1]{\textbf{\textit{Keywords:}} #1}

\begin{document}
\maketitle

\begin{abstract}
\noindent 
Reversed Potts shunt (PS) was suggested as palliation for patients with suprasystemic pulmonary arterial hypertension (PAH) and right ventricular (RV) failure. PS, however, can result in poorly understood mortality. Here, a patient-specific geometrical multiscale model of PAH physiology and PS is developed for a paediatric PAH patient with stent-based PS. In the model, 7.6mm-diameter PS produces near-equalisation of the aortic and PA pressures and $Q_p/Q_s$ (oxygenated vs deoxygenated blood flow) ratio of 0.72 associated with a 16\% decrease of left ventricular (LV) output and 18\% increase of RV output. The flow from LV to aortic arch branches increases by 16\%, while LV contribution to the lower body flow decreases by 29\%. Total flow in the descending aorta (DAo) increases by 18\% due to RV contribution through the PS with flow into the distal PA branches decreasing. PS induces 18\% increase of RV work due to its larger stroke volume pumped against lower afterload. Nonetheless, larger RV work does not lead to increased RV end-diastolic volume. Three-dimensional flow assessment demonstrates the PS jet impinging with a high velocity and wall shear stress on the opposite DAo wall with the most of the shunt flow being diverted to the DAo. Increasing the PS diameter from 5mm up to 10mm results in nearly linear decrease in post-operative $Q_p/Q_s$ ratio. In conclusion, this model reasonably represents patient-specific haemodynamics pre- and post-creation of the PS, providing insights into physiology of this complex condition, and presents a predictive tool that could be useful for clinical decision-making regarding suitability for reversed PS in PAH patients with drug-resistant suprasystemic PAH.
\\[5pt]
\noindent \keywords{pulmonary artery hypertension, Potts shunt, lumped parameter model, multiscale model, computational haemodynamics}
\end{abstract}

\section{Introduction}
Pulmonary arterial hypertension (PAH) is a rare disease in paediatric patients that is associated with significant morbidity and mortality. In the majority of paediatric patients, PAH is idiopathic or associated with congenital heart disease \cite{ivy2013pediatric}. 
Idiopathic pulmonary artery hypertension (iPAH) is a rare, chronic disorder of the pulmonary vasculature characterised by cellular changes in the vascular walls, which cause progressive constriction, obstruction or obliteration of the small pulmonary vessels in the lungs, thereby increasing the resistance to pulmonary blood flow.  In response, the right ventricle (RV) progressively adapts to pump blood through the high-pressure pulmonary vasculature by developing hypertrophy and dilatation, eventually resulting in cardiac failure and death.

iPAH in children, probably due to the highly adaptive RV myocardium, is characterised by the ability to sustain very high pulmonary arterial pressures above systemic levels for a long time \cite{barst2011pulmonary}. Furthermore, due to often non-specific symptoms, iPAH in children is typically diagnosed relatively late \cite{hoeper2013definitions,ivy2013pediatric}. Although the implementation of the so-called triple therapy strategy in paediatric iPAH has lead to significant improvement in prognosis of these patients \cite{shu2021efficacy}, there is a subgroup showing refractoriness of vascular resistance to medical treatment. In the settings of donor organ shortage, for children with progressive suprasystemic PAH presenting with an inadequate response to drug therapy and progressive RV failure, there is a need for alternative approaches to avoid further RV deterioration.

Based on the superior long-term survival and lower RV failure incidence in PAH patients with Eisenmenger syndrome due to untreated congenital heart disease compared to iPAH  \cite{baruteau2014palliative,grady2016potts,sizarov2016vascular}, surgical creation of a post-ventricular right-to-left shunt was suggested as a promising palliative treatment for severely ill children with drug-refractory suprasystemic iPAH \cite{blanc2004potts,baruteau2014palliative,grady2016potts}. The so-called Potts shunt (PS), an anastomosis between the left pulmonary artery (LPA) and the descending aorta (DAo) \cite{potts1946anastomosis}, in cases where pulmonary arterial pressures are above systemic ones, allows blood to flow from the LPA to the DAo, thereby nearly equalising pressures and partially decompressing the RV. Furthermore, the blood flow through the PS allows to  limit desaturation and the risk of paradoxical embolism to the lower body, thus sparing the brain \cite{hansmann2017pulmonary,boudjemline2013patent}. The vicinity of the DAo to LPA in humans creates an attractive possibility to place a covered stent between the lumens of these two vessels and create an anastomosis percutaneously \cite{sizarov2016vascular}. Recently, feasibility of percutaneous stent-based PS creation in adults and children with iPAH has been reported demonstrating an approach potentially allowing substantial reduction of the treatment invasiveness in these patients \cite{boudjemline2017safety, esch2013transcatheter}.

Whilst small surgical and interventional series of PS procedures have demonstrated sustained improvement in functional capacities and prolonged survival in the majority of patients, the clinical response to PS creation is mixed with substantial mortality \cite{baruteau2014palliative,grady2016potts,boudjemline2017safety}, the mechanisms of which are poorly understood. Although cardiovascular magnetic resonance imaging can provide useful insights \cite{schafer2019close}, the full-spectrum of haemodynamic changes due to the PS creation is not accessible. In contrast, computational models that can accurately and efficiently predict the patient-specific post-procedural haemodynamic changes based upon pre-operative characteristics can prove to be a useful tool in understanding this complex condition. Furthermore, such models could present a substantial aid in clinical decision-making to assess the suitability of PS creation in individual patients with particular haemodynamics. With the exception of a recent non patient-specific  study using the well-known CircAdapt lumped parameter model (LPM) \cite{delhaas2018potts}---which simulated a reference patient, its adaptations with increased pulmonary arterial pressures, and subsequent creation of PS with varying shunt diameters---till now there is no comprehensive computational investigation of this subject. In this study, the development and output of two models---a geometric multiscale model (GMM) model and an LPM---is reported. For the first time, 3D flow features are comprehensively assessed along with changes in global haemodynamics in response to stent-based PS creation of varying diameters and lengths. The fully-tuned GMM is representative of the patient-specific pre- and post-procedural haemodynamics and provides insights into blood flow features through and around the PS. Furthermore, the results from the computationally expensive GMM are compared to the less demanding LPM exploring the validity of the latter in providing solutions for patient-specific parameter estimation and assessment of global haemodynamics in such a complex condition.

\section{Materials and Methods}

\subsection{Patient characteristics and measurements}
\label{sec:patient_measurements}

Clinical measurements were obtained for a 13-year old patient with morphologically normal heart and suprasystemic iPAH complicated by the RV failure despite the triple vasodilator therapy, who received the PS using a covered stent implanted through percutaneous approach at the Necker University Hospital for Sick Children, Paris, France, as published previously \cite{boudjemline2017safety}. In this particular patient, despite the substantial technical difficulties during the procedure, there was no acute circulatory deterioration after PS creation with near equalization of the systemic and pulmonary artery (PA) pressures. Pre-operative measurements included an electrocardiogram, pressure tracings from heart catheterisation, Doppler flow velocity tracings from echocardiography, and CT-angiography imaging. Body surface area (BSA) of the patient was 1.13 m\textsuperscript{2}.

Pre-operatively measured pressures were available for the right atrium (RA), RV, DAo, main PA (mPA), and the pulmonary capillary wedge pressure, which was used as an indirect estimate of the left atrial (LA) pressure. Post-operative hemodynamic measurements were limited to pressure tracings for the DAo and mPA. Depending upon the completeness of the measurements, the pressure traces are averaged over 3-5 cardiac cycles. The measured pre-operative pressures used in this study are summarised in Table \ref{tab:measpress}.

The average pre-operative cardiac output (CO) for the patient was 3.4 L/min, based upon three sets of thermodilution measurements under sedation with no anaesthesia and a further three measurements within the same catheterisation but under nitric oxide inhalation and 100\% oxygen. Durations of pre-operative atrial and ventricular systole were obtained from Doppler echocardiography and electrocardiography, from which a cardiac cycle time-period of 0.9s was extracted. 
Pre-operative end-diastolic ventricular volumes (EDV), corresponding end-systolic atrial volumes (ESV), and the myocardial wall volumes for each cardiac cavity, $V_w$, were determined by 3D reconstruction of the contrast medium-stained cavities and their surrounding walls as visible on the cardiac CT imaging. 
The measured pre-operative CO and cardiac cycle time-period provided stroke volumes (SV) for the LV and the RV, which yielded corresponding ejection fractions (EF) when combined with the CT-reconstructed EDVs. These measurements are summarised in Table \ref{tab:pre_heart_data}.

\begin{table}[tb]
\centering
\def\arraystretch{1.2}
\resizebox{0.55\textwidth}{!}
{  
\begin{tabular}{l|cccc|cc}
\hline
 & \multicolumn{4}{c}{Pre-Operative}  &  \multicolumn{2}{c}{Post-Operative} \\
\hline
 & $P_{\rm{DAo}}$ & $P_{\rm{mPA}}$ & $P_{\rm{LA}}$ &$P_{\rm{RA}}$ &  $P_{\rm{DAo}}$ & $P_{\rm{mPA}}$\\
 \hline
 Systolic [mmHg]& 94 & 112 &-&-&  97 & 102\\
 Diastolic [mmHg]& 53 & 67 &-&-&  51 & 54\\
 Mean [mmHg]& 69 & 85 & 4 & 6 &  71 & 76\\
 \hline
\end{tabular}
}
\caption{Measured pressures in the pre- and post-operative states.}
\label{tab:measpress}
\end{table}

\begin{table}[tb]
\centering
\def\arraystretch{1.2}
\resizebox{0.5\textwidth}{!}{  
\begin{tabular}{l|cccc}
\hline
  Pre-operative & $\text{LA}$ & $\text{RA}$ & $\text{LV}$ & $\text{RV}$ \\
  \hline
    Cardiac output(CO) [L/min] & - & - & - & 3.4 \\
     Activation duration ($t_{\rm{max}}$) [s] & 0.16 & 0.14 & 0.4 & 0.38\\
  End-diastolic volume (EDV) [ml] & - & - & 66.9 & 91.8 \\
  End-systolic volume (ESV) [ml] & 22.7 & 51.9 & - & - \\
  Ejection fraction (EF) [-] & - & - & 0.76 & 0.55 \\
  Myocardial volume ($V_w$) [ml] & 6.4 & 10.3 & 51.7 & 69.0\\
   \hline
\end{tabular}
}
\caption{Pre-operative measurements for heart chamber output, end-diastolic volumes, ejection fractions, myocardial volumes, and systole durations.}
\label{tab:pre_heart_data}
\end{table}

\subsection{Models}
Two closed-loop computational models have been developed: a stand-alone LPM; and a GMM model consisting of a reconstructed patient-specific three-dimensional (3D) flow domain, coupled to an LPM. The fast-to-compute LPM provides quick solutions (in the order of seconds) for patient-specific parameter estimation and global haemodynamics assessment, and the GMM, while the computationally expensive (run-time in the order of days), provides detailed local flow information in the vessels and the PS.

 
\subsubsection{Geometric Multiscale Model (GMM)} \label{multiscalemodelsection}

\begin{figure}[tb]
\centering
\includegraphics[width=1.0\textwidth]{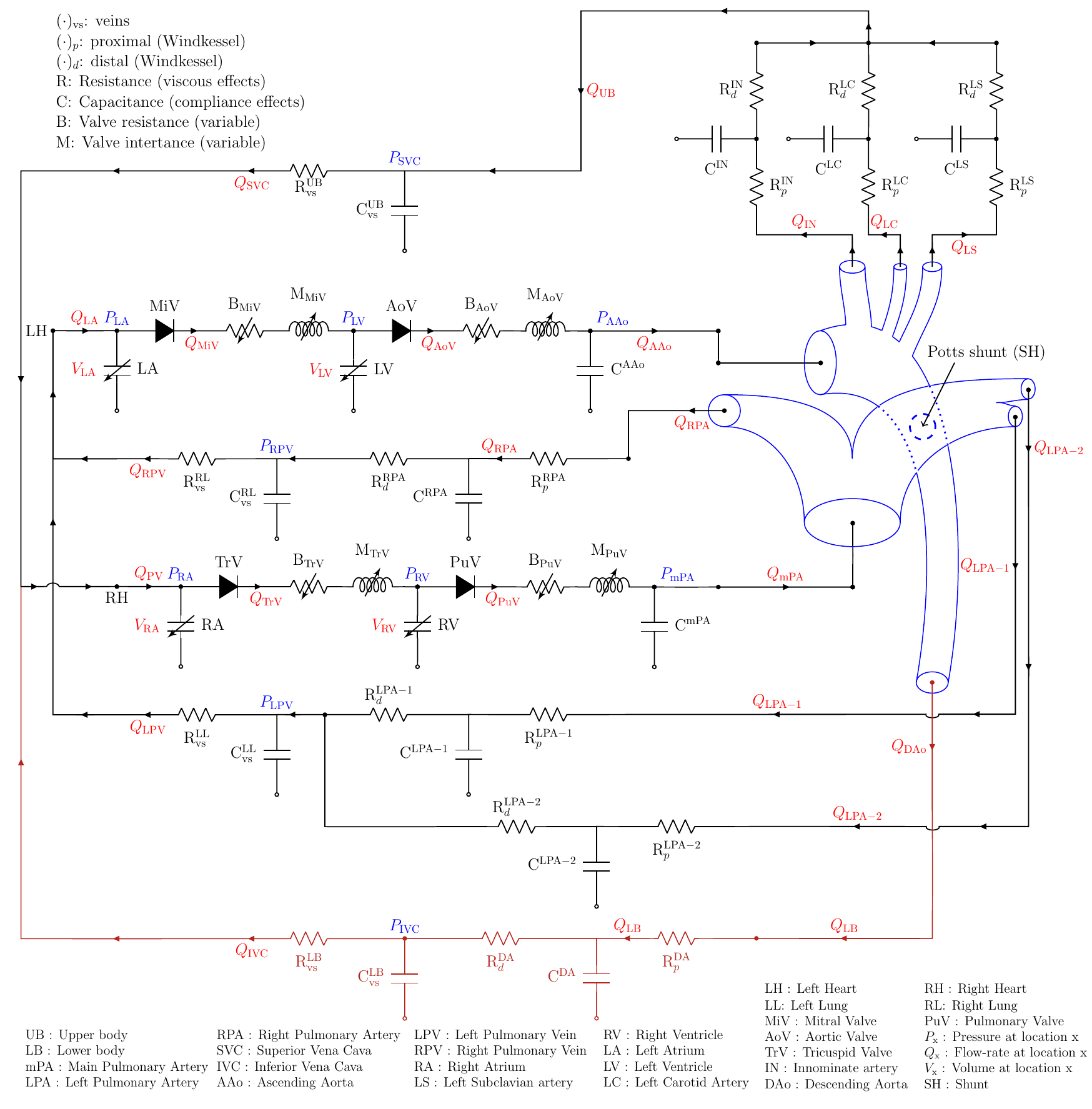}
\caption{GMM for the iPAH and PS physiology: the 3D geometry (see Figure \ref{fig:AOPABasicImage}) is represented in blue and the remaining circulation is represented as a lumped parameter model (LPM) with en electrical analogy to blood-flow. $Q_{(\cdot)}$ represents volumetric flow-rate; $P_{(\cdot)}$ represents pressure, $V_{(\cdot)}$ represents volume.}
\label{fig:multiscalemodel}
\end{figure}

\begin{figure}[tb]
\centering
    \begin{subfigure}[t]{0.45\textwidth}
    \centering
        {\includegraphics[width=\textwidth]{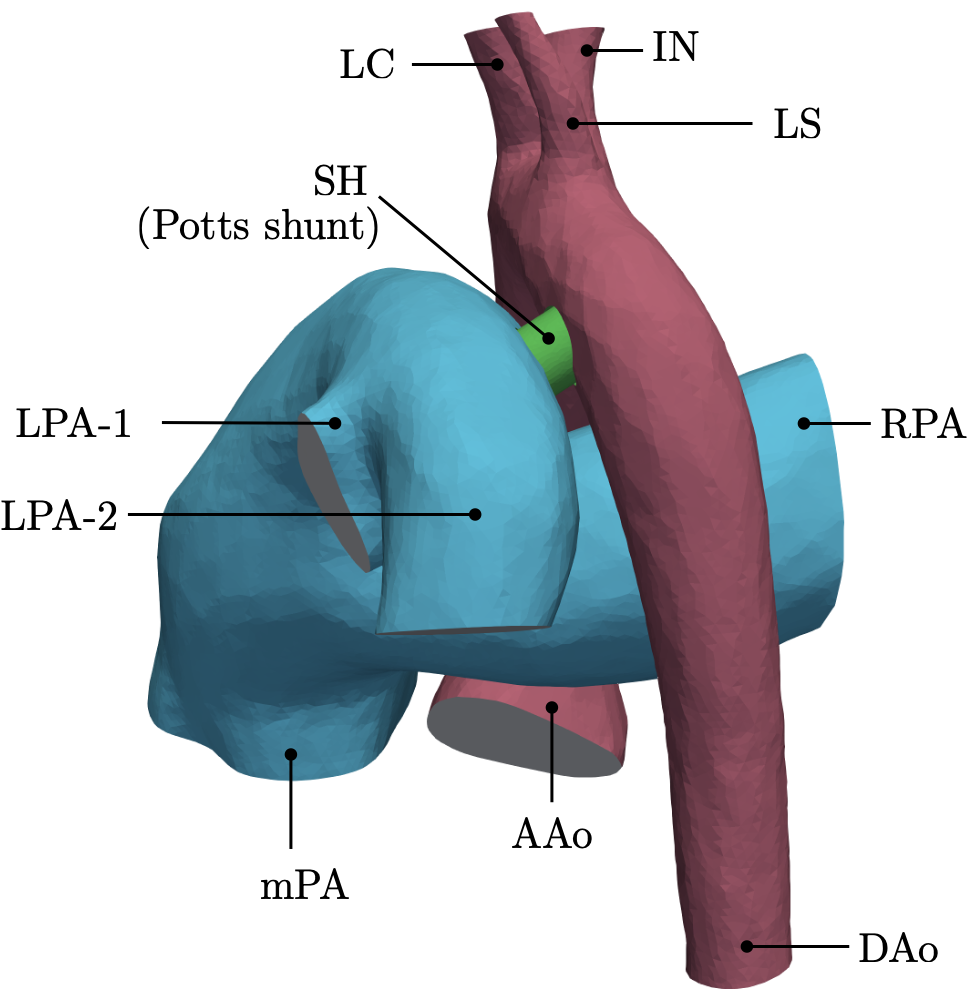}}
        \caption{3D model showing pulmonary vessels (cyan), systemic vessels (magenta), and the Potts shunt (green).}
        \label{fig:AOPABasicImage}
    \end{subfigure}
    \hfill
    \begin{subfigure}[t]{0.45\textwidth}
    \centering
        {\includegraphics[width=0.8\textwidth]{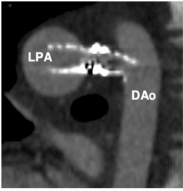}}
        \caption{CT angiogram showing the long and protruding PS between LPA and DAo.}
        \label{fig:Doppler3Dimage}
    \end{subfigure}
\caption{3D reconstruction of the patient anatomy and CT angiogram. Labels are defined in Figure \ref{fig:multiscalemodel}.}
\label{fig:3dmodel_and_angio}
\end{figure} 

A schematic of the GMM  is shown in Figure \ref{fig:multiscalemodel}.  The 3D domain is shown in blue and comprises sections of the large systemic and pulmonary vasculature relevant to the PAH physiology.
The 3D anatomical geometry is prepared for computational modelling using Amira software (Thermo Fisher Scientific, Darmstadt, Germany) and Mimics Innovation Suite \citep{Mimics}.
The systemic vessels represented in the 3D model include the ascending aorta (AAo), aortic arch with its branches---the innominate artery (IN), the left carotid artery (LC), and left subclavian artery (LS)---and the thoracic portion of the descending aorta (DAo).  The pulmonary vasculature is represented by the main pulmonary artery (mPA), the right pulmonary artery (RPA), and the left pulmonary artery (LPA) with its two branches (denoted as LPA-1 and LPA-2).  
Post-operative models are generated by modelling the PS created by a covered stent as a cylinder of constant diameter between the LPA to the DAo adjacent walls (Figure \ref{fig:AOPABasicImage}).

Initially, a stent-based PS with 7.6 mm diameter PS is generated, corresponding to the clinical case under consideration  (see Section \ref{sec:patient_measurements}). To determine the effect of the stent diameter on flow patterns and global haemodynamics, additional post-operative geometries are created using cylinders with diameters of 5, 6, 7, 8, 9 and 10 mm. For simplicity, the simulations of the shunts with varying diameters were performed with the length of modelled stents constrained by the distance between the LPA and DAo, thus, not protruding into the vascular lumens, a situation corresponding to either surgically created side-to-side anastomosis or a spool-shaped covered stent with  complete apposition of its flaring ends to the vessel walls \cite{chigogidze2006intervascular}. The currently available stents used to create the PS in a clinical setting, however, invariably produce protrusion of the stent ends into the vessel lumens, as was the also the case in the patient considered in this study (Figure \ref{fig:Doppler3Dimage}). Therefore, additional simulations of the PS creation were run with maintaining diameter at 7 mm, while varying the lengths of the modelled stent: four further post-operative geometries were created to investigate the effect of stent protrusion into the LPA and DAo, with the stent lengths of 10, 15, 20 and 25 mm. All models were meshed using TetGen \cite{si2015tetgen,sahni2008adaptive} available within the SimVascular Suite \citep{updegrove2017simvascular,lan2018re}, with additional local and regional mesh refinement ensuring preservation of key geometrical features.

The 3D flow domain is coupled to an LPM, which represents the global circulatory system with a hydraulic-electric analogy \citep{shi2011review}.  
The LPM partitions the circulatory system into multiple blocks or compartments, each of which is described by a series of elements representing the linear viscous losses ($R$), non-linear viscous losses ($K$), blood inertia ($L$) and/or vessel compliance ($C$), depending upon the salient characteristics of the local vasculature.  The resulting instantaneous pressure-flow relationships are shown in Table \ref{pressflowequ}. The four heart chambers are each described by a single fibre model (see Section \ref{heartmodel}) and the atrioventricular and semilunar valves are modelled dynamically (see Section \ref{valvemodel}), with the contraction of the heart and valve function represented by time-varying capacitors (compliances) and diodes, respectively.

In the GMM, Figure \ref{fig:multiscalemodel}, downstream of each 3D outlet (see for example the highlighted path in orange downstream of the DAo), the arterial structure is represented by a three-element RCR Windkessel model, while the subsequent venous structure is represented by an additional two-element RC Windkessel model.  Specifically, the three-element RCR Windkessel model comprises a proximal resistance of the large vessels, $R_p$, a distal resistance of the microvasculature, $R_d$, and the total arterial compliance, $C$, representing the elastance of the large vessels.  The two-element RC Windkessel model comprises a venous resistance,  $R_{\mathrm{vs}}$, and the total venous compliance, $C_{\mathrm{vs}}$.

\begin{table}[tb]
\centering
\def\arraystretch{1.0} 
\resizebox{0.55\textwidth}{!}{  
\begin{tabular}{ccc}
\hline
 Component & & Pressure-Flow Relationship\\
\hline
\includegraphics[scale=1.0]{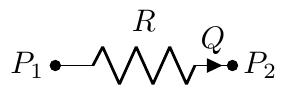}
& & 
\raisebox{0.5 em}{$P_1$ -- $P_2$ = $RQ$} \\[5pt]

\includegraphics[scale=1.0]{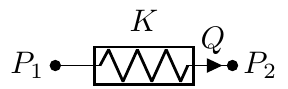}
 & &
 \raisebox{0.5 em}{$P_1$ -- $P_2$ = $KQ|Q|$} \\[5pt]

\includegraphics[scale=1.0]{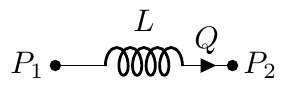}
& & 
\raisebox{0.5 em}{$P_1$ -- $P_2$ = $L\dfrac{dQ}{dt}$} \\[5pt]

\includegraphics[scale=1.0]{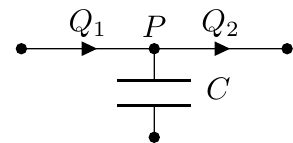}
& &
\raisebox{2.0 em}{$Q_1$ -- $Q_2$ = $C\dfrac{dP}{dt}$} \\[5pt]
\hline
\includegraphics[scale=1.0]{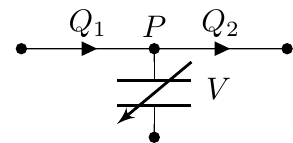}
& &
\raisebox{2.0 em}{$Q_1$ -- $Q_2$ = $\dfrac{dV}{dt}$} \\[5pt]

\includegraphics[scale=1.0]{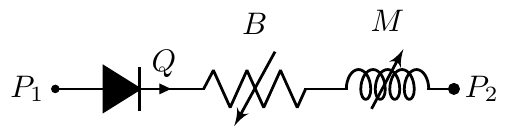}
& & 
\raisebox{0.5 em}{$P_1$ -- $P_2$ = $BQ|Q| + M \dfrac{dQ}{dt}$} \\[5pt]
\hline
\end{tabular}
}
\caption{Pressure ($P$) and flow rate ($Q$) relationships applying the hydraulic-electrical analogy. The last two rows show the heart chambers and heart valves, respectively, where $V$ represents volume of the chamber.}
\label{pressflowequ}
\end{table}

\subsubsection{Deriving the stand-alone LPM from the GMM}
\label{sec:stand-alone_LPM}

For creating the stand-alone LPM, the 3D flow-domain of the GMM is substituted by lumped parameter components while minimising the error between the GMM solution and the stand-alone LPM \cite{pant2013multiscale, pant2014methodological}. Based on the GMM solution, it was found that the while the pressure drops in the PAs and the DAo section downstream the PS could be ignored,  the pressure drop in the AAo, supraoartic branches, and DAo section upstream the Potts shunt required an appropriate lumped representation. Each of these segments is described by a linear resistance $R_{\mathrm{3D}}$, a non-linear quadratic resistance $K_{\mathrm{3D}}$, and an inductance $L_{\mathrm{3D}}$. The values of these parameters are found through a regression analysis on the GMM solution \cite{pant2013multiscale, pant2014methodological}. The resulting stand-alone LPM, excluding the components that yielded near-zero values in the regression analysis, is shown in Figure \ref{fig:lpm_schematic}. The post-operative stand-alone LPM includes a shunt law model, shown in magenta colour in Figure \ref{fig:lpm_schematic}, which is described in further detail in Section \ref{shuntmodel}. 

\subsubsection{Single Fibre Model for Heart Chamber} \label{heartmodel}
Each heart chamber is described by a single fibre wrapped around the cavity \citep{arts1991relation, bovendeerd2006dependence, pant2016data, pant2017inverse}.  This model describes the following relationship between the cavity pressure $P_{cav}$, cavity volume $V_{cav}$, the wall volume $V_w$, and the stress in the fibre, $\sigma_f$:
\begin{equation}
\frac{\sigma_f}{P_{cav}} = \left(1+\frac{3V_{cav}}{V_w}\right).
\label{singlefibre_start}
\end{equation}
The dependence on volume alone, as opposed to shape of the cavity, makes the single fibre model particularly suitable for 0D modelling, as it has been shown that under rotational symmetry, the shape of the chamber has little effect upon the cavity pressure and fibre stress \citep{arts1991relation}.  The fibre stress $\sigma_f$ comprises an active component $\sigma_a$ and a passive component, $\sigma_p$, whereby:
\vspace{-1ex}
\begin{equation}
\sigma_f = \sigma_a + \sigma_p.
\end{equation}
The active component of stress is described by
\begin{equation}
\label{eqn:active_stress}
\sigma_a = c\;T_{a0}\;f(l)\;g(t_a)\;h(v_s),
\end{equation}
where $c$ represents contractility, $T_{a0}$ represents the maximum active sarcomere stress, $f(l)$ represents the force-length relationship in a sarcomere of length $l$, $g(t_a)$ represents the time-dependent variation of active stress with $t_a$ representing time elapsed since activation, and $h(v_s)$ represents the active viscous stress dependent on the sarcomere shortening velocity.
For a sarcomere length $l$, $f(l)$ is described as
\begin{equation}
f(l) =
\begin{cases}
0, & \text{if } l<l_{a0}\\
(l-l_{a0})/(l_{am}-l_{a0}), & \text{if } l_{a0}<l \leq l_{am}\\
1, & \text{if } l_{am}\leq l \leq l_{ae}\\
(l_{af}-l)/(l_{af}-l_{ae}), & \text{if } l>l_{ae}
\end{cases},
\end{equation}
where the length parameters $l_{a0}$, $l_{am}$, $l_{af}$ and $l_{ae}$ are sarcomere material constants.  With $V_0$ and $l_0$ representing the cavity volume and sarcomere lengths at zero transmural pressure, the sarcromere stretch $\lambda$ for a generic sarcomere length $l$ and cavity volume $V_{cav}$ is given by
\begin{equation}
\label{eqn_stretch}
\lambda  = \frac{l}{l_0} = \left(\frac{1+(3V_{cav}/V_w)}{1+(3V_0/V_w)}\right)^{1/3}.
\end{equation}
The time-dependent activation $g(t_a)$ is described as
\begin{equation}
\label{eqn:gta}
g(t_a) =
\begin{cases}
\left[\frac{1}{2}\left(1-\cos\left(2\pi{t_a}/{t_{max}}\right)\right)\right]^{E_a}, & \text{if } t_a<t_{\rm{max}}\\
0, & \text{otherwise }
\end{cases},
\end{equation}
where $t_a$ is the time since the beginning of cavity activation, $t_{\rm{max}}$ is the total systole duration of the specific cardiac chamber, and $E_a$ controls the shape of the activation curve.  The active viscous stress $h(v_s)$ is described as:
\begin{equation}
h(v_s) = \frac{1-(v_s/v_0)}{1+c_v(v_s/v_0)},
\end{equation}
where $v_0$ is the initial sarcomere shortening velocity and $c_v$ governs the shape of the velocity-stress relation, and the shortening velocity $v_s$ is
\begin{equation}
v_s= -\frac{dl}{dt}= -\frac{1}{V_w}\left(1+\frac{3V_{cav}}{V_w}\right)^{-1}\frac{dV_{cav}}{dt}.
\end{equation}
Lastly, the passive stress component, $\sigma_p$, is modelled by:
\begin{equation}
\sigma_p =
\begin{cases}
0, & \text{if } \lambda<1\\
T_{p0}(\exp\;[c_p(\lambda-1)\;]-1), & \text{if } \lambda \geq 1
\end{cases}
\label{singlefibre_end}
\end{equation}
where $T_{p0}$ and $c_p$ are sarcomere material constants.  Thus, Equations \eqref{singlefibre_start}--\eqref{singlefibre_end} define the relationship between pressure, volume and flow rate for each cardiac chamber.

\subsubsection{Heart Valve Model} \label{valvemodel}
The adopted model \cite{mynard2012simple, pant2016data, pant2017inverse} determines valve dynamics based upon the instantaneous pressure difference across the valve.  Neglecting viscous losses, the relationship between the pressure drop $\Delta P$ across the valve and the volumetric flow-rate $Q$ through the valve is described by the Bernoulli relation as:
\begin{equation}
\Delta P=B \; Q\; |Q| \; + \; M \; \frac{dQ}{dt},\\
\end{equation}
where $B$ and $M$ represent the valve non-linear resistance and inertance of the valve, respectively, and depend on the geometrical parameters of the valve and the fluid density $\rho$ as
\begin{equation}
B=\frac{\rho}{2A^2_{\text{eff}}} \quad \text{and} \quad M=\frac{\rho \; l_{\text{eff}}}{A_{\text{eff}}},
\end{equation}
where $A_{\text{eff}}$ and $l_{\text{eff}}$ are the effective area and effective length of the valve, respectively.  The time varying dynamics of $A_{\text{eff}}(t)$ is described as:
\begin{equation}
A_{\text{eff}}(t) = (A^{\text{max}}_{\text{eff}}-A^{\text{min}}_{\text{eff}}) \; \xi (t) + A^{\text{min}}_{\text{eff}}
\end{equation}
where $\xi (t) \in [0,1]$ is a variable describing the valve state, and $A^{\text{min}}_{\text{eff}}$ and $A^{\text{max}}_{\text{eff}}$ are the minimum and maximum effective areas, respectively.  Based on \cite{mynard2012simple}, the valve state $\xi (t)$ depends on the pressure difference across the valve and is described by:
\begin{equation}
\dot{\xi} =
\begin{cases}
(1-\xi) \; K_{vo} \; \Delta P, & \text{if } \Delta P \geq0\\
\xi \; K_{vc} \; \Delta P, & \text{otherwise}
\end{cases},
\end{equation}
where $K_{vo}$ and $K_{vc}$ are  parameters that govern the valve opening and closing rates, respectively.  The valve starts to open when $\Delta P>0$ and is fully open when $\xi (t)$ is equal to 1. Similarly, the valve starts to close when $\Delta P<0$ and is fully closed when $\xi (t)$ is equal to 0.

\subsubsection{LPM for shunt} \label{shuntmodel}
Here, the lumped model for shunts proposed in literature \citep{migliavacca2000computational,migliavacca2001modeling} is adopted for modelling the PS. The pressure drop across the PS is described as:
\begin{equation}
P_{\rm{LPA}} - P_{\rm{DAo}} = R_{\mathrm{3D}}^{\mathrm{SH}} \; Q_{\rm{SH}} \; + \; K_{\mathrm{3D}}^{\mathrm{SH}} \; Q_{\rm{SH}} \; |Q_{\rm{SH}}| \; + \; L_{\mathrm{3D}}^{\mathrm{SH}} \; \frac{dQ_{\rm{SH}}}{dt},
\label{shunt_law}
\end{equation}
where $P_{\rm{LPA}}$ is the pressure in the LPA, $P_{\rm{DAo}}$ is the pressure in the DAo, $Q_{\rm{SH}}$ is the instantaneous volume flow-rate in the shunt and $R_{\mathrm{3D}}^{\mathrm{SH}}$, $K_{\mathrm{3D}}^{\mathrm{SH}}$ and $L_{\mathrm{3D}}^{\mathrm{SH}}$ are functions of the shunt/vessel geometry and the fluid properties.  Equation \eqref{shunt_law} is often  expressed in terms of the shunt diameter, $D_{\rm{SH}}$, such that:
\begin{equation}
P_{\rm{LPA}} - P_{\rm{DAo}} = \frac{k_1}{D^4_{\rm{SH}}} Q_{\rm{SH}} \; + \frac{k_2}{D^4_{\rm{SH}}} Q_{\rm{SH} }\; |Q_{\rm{SH}}| +\frac{k_3}{D^2_{\rm{SH}}} \frac{dQ_{\rm{SH}}}{dt},
\label{shunt_law_D}
\end{equation}
where $k_1$, $k_2$ and $k_3$ are proportionality constants.  In lieu of experimental data, $k_1$ and $k_2$ are derived computationally using regression analysis on the GMM solutions for all the post-operative geometries with shunt diameters varying from 5 mm to 10 mm. This analysis also showed that the last term in Equation \eqref{shunt_law_D} accounts for only a 0.2\% difference in pressure drop across the PS.  Thus,this term is neglected and the pressure drop across the shunt is described by:
\begin{equation}
P_{\rm{LPA}} - P_{\rm{DAo}} = \frac{k_1}{D^4_{\rm{SH}}} Q_{\rm{SH}} \; + \frac{k_2}{D^4_{\rm{SH}}} Q_{\rm{SH} }\; |Q_{\rm{SH}}|.
\label{shunt_law_reduced}
\end{equation}

\subsection{Solution Methodology}

\begin{figure}[tb]
\centering
\includegraphics[width=1.0\textwidth]{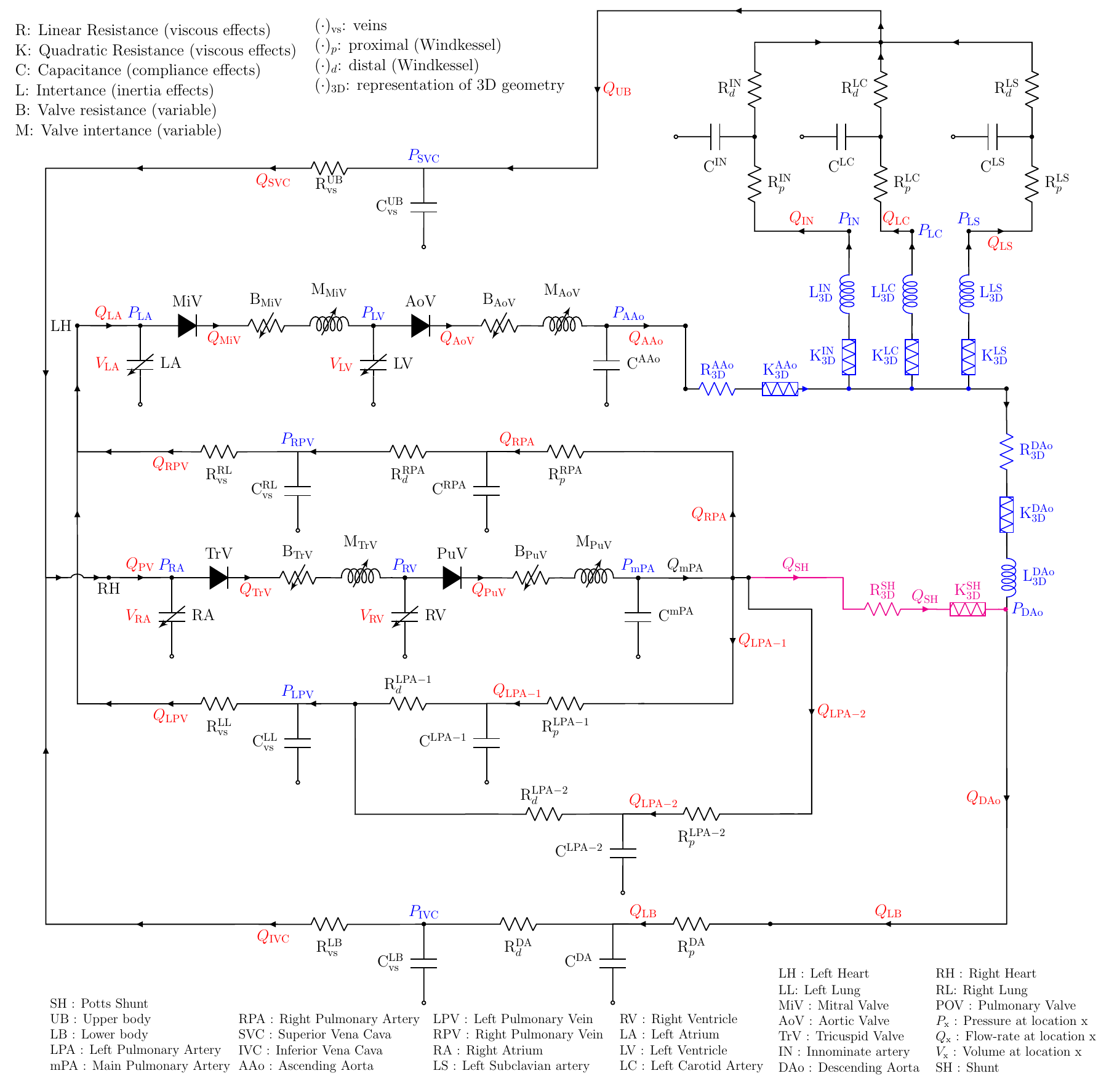}
\caption{Stand-alone LPM for the iPAH and PS physiology. The lumped components that represent the 3D regions (see Figure \ref{fig:multiscalemodel}) are shown in blue. In the post-operative state, the lumped representation of the Potts shunt is shown with components in magenta. $Q_{(\cdot)}$ represents volumetric flow-rate; $P_{(\cdot)}$ represents pressure, $V_{(\cdot)}$ represents volume.}
\label{fig:lpm_schematic}
\end{figure}

For the 3D part of the GMM, blood is assumed to be an incompressible Newtonian fluid with density 1.06 g$\cdot$cm$^{-3}$ and dynamic viscosity 0.04 g$\cdot$cm$^{-1}\cdot$s$^{-1}$, governed by the Navier-Stokes equations and a rigid wall assumption.  A no-slip condition is imposed on the walls (i.e. surfaces that are neither inlets nor outlets) and a backflow stabilisation coefficient of 0.5 is employed to control backflow at the outlets \citep{moghadam2011comparison}.
The inlets and outlets of the 3D model are coupled to the LPM by interface conditions such that time-dependent pressures yielded by the LPM are uniformly applied to the boundaries of the 3D domain and the volumetric flow rates at the 3D inlets and outlets are imposed as forcing terms to the LPM \citep{moghadam2013modular}. The GMM was developed within the SimVascular Suite \citep{updegrove2017simvascular,lan2018re}.
A constant time-step of $2.0\times10^{-4}$ seconds is specified, which was selected following a comparison of results from three different GMM simulations with fixed time increments of 1.0, 2.0 and 5.0 $\times$ 10$^{-4}$ seconds.  
The simulations were run for multiple cardiac cycles until the relative changes in all the variables were less than 0.5\% of those in the previous cardiac cycle.

The stand-alone LPM network, Figure \ref{fig:lpm_schematic}, leads to a set of ordinary differential equations which are solved using a fourth-order Runge-Kutta scheme with constant time-step of 1.0 $\times$ 10$^{-4}$ seconds.  Within 25 cardiac cycles, the relative changes in the state variables were less than 0.1\% when compared to those in the previous cardiac cycle.  These converged (stabilised) values for the pre-operative state variables are then used to define the initial conditions for the post-operative state in order to expedite convergence.

\begin{table}[tb]
\centering
\def\arraystretch{1.6}
\resizebox{0.6\textwidth}{!}{  
\begin{tabular}{cc|cc|cc|cc}
\hline
  \multicolumn{8}{c}{Parameters describing the sarcomere active and passive stresses}\\
  \hline
    $T_{a0}\; \text{[kPa]}$ & 71.0 & $T_{p0}\; \text{[kPa]}$ & 0.9 & $c_{v} \;[-]$ & 0.0 & $c_{p} \; [-]$ & 12.0\\
    $v_{0} \; [\mu \text{m/s}]$ & 10.0 & $l_{0} \; [\mu \text{m}]$ & 2.0 & $l_{a0} \;  [\mu \text{m}]$ & 1.5 & $l_{ar} \; [\mu \text{m}]$ & 2.0\\
    $E_a^{\text{LA}} \;[-]$ & 1.0 & $E_a^{\text{RA}} \;[-]$ & 1.0 & $E_a^{\text{LV}} \;[-]$ & 1.05 & $E_a^{\text{RV}} \;[-]$ & 0.48\\
  \hline
\end{tabular}
}
\caption{Patient-specific pre-operative parameters for the active and passive stress laws.}
\label{tab:mechdata}
\end{table}

\begin{table}[tb]
\centering
\def\arraystretch{1.4}
\resizebox{0.39\textwidth}{!}{  
\begin{tabular}{ccccc}
\hline
  \multicolumn{5}{c}{Valve Model Parameters}\\
  \hline
  & ${\text{MiV}}$ & ${\text{AoV}}$ & ${\text{TrV}}$ & ${\text{PuV}}$\\
  \hline
  $K_{vo} \; \text{[cm s/g]}$ & 0.03 & 0.012 & 0.03 & 0.02\\
  $K_{vc} \; \text{[cm s/g]}$ & 0.04 & 0.012 & 0.04 & 0.02\\
  $A_{\text{eff}} \; \text{[cm\textsuperscript{2}]}$ & 2.1 & 2.2 & 2.1 & 2.3\\
  \hline
\end{tabular}
}
\caption{Patient-specific pre-operative parameters for the valve model: mitral valve (MiV), aortic valve (AoV), tricuspid valve (TrV) and pulmonary valve (PuV).}
\label{tab:valvedata}
\end{table}

\subsection{Patient-specific parameter specification and estimation}
\label{sec:tuning}

For a reference set of parameters, the lumped parameters representing the 3D sections are first estimated through a regression analysis on the GMM solutions, see section \ref{sec:stand-alone_LPM} and the references therein. With this, the stand-alone LPM includes correct 3D representation, but lacks patient-specific parameters. Subsequently, patient-specific parameter estimation is performed on the stand-alone LPM, Figure \ref{fig:lpm_schematic}, as it is significantly cheaper in terms of computation time relative to the GMM. 


For parameter specification and estimation, the measurements acquired in the patient,  Tables \ref{tab:measpress} and \ref{tab:pre_heart_data}, are utilised. In the absence of direct pressure measurements in the ascending aorta, the pre-operative aortic pressure $P_{\rm{AAo}}$ is assumed comparable to the pressure in thoracic portion of the descending aorta $P_{\rm{DAo}}$.  This surrogate measurement of $P_{\rm{AAo}}$ is used for estimation of model parameters as it better constrains the estimates compared to $P_{\rm{DAo}}$, and allows for the computation of systemic vascular resistance for further parameter estimation (see section \ref{sec_tuning_global}). 
In what follows, where manual tuning is used the parameters are adjusted to reproduce the following target measurements: surrogate pre-operative aortic and PA pressures (systolic, diastolic, and mean), the LV and RV EDVs, LV and RV stroke volumes, and the CO.

\subsubsection{Heart and valve models}
From the pre-operative patient measurements (see Section \ref{sec:patient_measurements} and Table \ref{tab:pre_heart_data}), many heart model parameters can be directly specified. In particular, the cardiac cycle time-period of 0.9s, the myocardium wall volumes ($V_w$), and the activation durations ($t_{\rm{max}}$) for each heart chamber are directly specified. For the geometrical parameters of the heart model, the unknown parameters are the chamber volumes at zero transmural pressure, $V_0$, which are manually tuned, and found to be 20.6 ml, 39.1 ml, 28.9 ml, and 43.8 ml for the LA, RA, LV, and RV, respectively. Table \ref{tab:mechdata} displays the parameters pertaining to the sarcomere active and passive stress laws in the single fibre model, which are based upon the experimental values provided in \citep{bovendeerd2006dependence} and then manually tuned. 

Table \ref{tab:valvedata} shows the opening $K_{vo}$ and closing $K_{vc}$ rates of the valves, provided in \citep{mynard2012simple}, together with the effective valve areas, $A_{\text{eff}}$, which are initially approximated from literature  \citep{mynard2012simple} and then manually tuned to ensure that the flow rates for the atrioventricular (AV) valves are consistent with the pre-operative echocardiographic velocity traces.

\subsubsection{Global circulation parameters}
\label{sec_tuning_global}
To specify the remaining circulation parameters, first the total vascular resistances and compliances are computed, which are then appropriately distributed to individual branches and components. 
Pulmonary vascular resistance (PVR) is derived from the measured CO, the measured mean $P_{\text{PA}}$, and the measured mean $P_{\text{LA}}$.  Similarly, the total systemic vascular resistance (SVR) is derived from the measured CO, the measured mean $P_{\text{AAo}}$ and the measured mean $P_{\text{RA}}$.
The total resistance for the entire LPM branch subsequent to the $i^{\mathrm{th}}$ outlet of the 3D flow domain consists of the 3-element Windkessel elements $R_p$, $C$, and $R_d$ for arterial branches and $C_{\mathrm{vs}}$ and $R_{\mathrm{vs}}$ for the venous elements (see Figures \ref{fig:multiscalemodel} and \ref{fig:lpm_schematic}; such a branch corresponding to the descending aorta is highlighted in Figure \ref{fig:multiscalemodel} in brown colour) and is computed as
%
\begin{equation}
R_{i, \mathrm{branch}}=\frac{\sum_j A_j}{A_i}R_{tot}
\label{resistancefraction}
\end{equation}
where $R_{tot}$ is the total resistance (either PVR or SVR depending on whether the outlet belongs to pulmonary or systemic circulation), $A_i$ is the area of the $i^{\mathrm{th}}$ 3D outlet and $\sum_j A_j$ represents the sum of all 3D outlets in the corresponding pulmonary/systemic.  
The resistance $R_d$ is assumed to be 10 times larger than $R_p$ \cite{ladisa2011computational} and the venous resistance is assumed to be 10\% of the total branch resistance. Thus, in each branch
 these resistances $R_p$, $R_d$, and $R_{\mathrm{vs}}$ are assumed to be  8.2\%, 81.8\%, and 10\% of the total branch resistance $R_{i, \mathrm{branch}}$, respectively.

With the resistances defined, the compliances within the 3-element Windkessel models 
are related to the time constant $\tau$ of the exponential pressure decay such that $C$ = $\tau$/$R_d$ \citep{spilker2010tuning}.   For the PAs, $\tau$ is defined by: 
\begin{equation}
\tau = \frac{(PVR \times SV)}{PP}
\label{timeconstequ}
\end{equation}
where $SV$ is the stroke volume and $PP$ is the pulse pressure \citep{mackenzie2013decreased,chemla2016time}, enabling the compliances for the pulmonary artery Windkessels to be calculated.  Similarly, using Equation \eqref{timeconstequ}, the compliances for the systemic branch arteries are calculated using SVR instead of PVR.
The venous compliances are calculated for each branch from the relationship $C_{\mathrm{vs}}$ = $\tau_{\mathrm{vs}}/R_{\mathrm{vs}}$, with $\tau_{\mathrm{vs}}$ is set equal to 1s for the lungs and equal to 2s for the upper and lower body branches based on literature \citep{baretta2014patient, kilner2009pulmonary, presson1998anatomic, spilker2007morphometry}. Finally, the AAo compliance $C^{\text{AAo}}$ and mPA compliance $C^{\text{mPA}}$ were manually adjusted to align with the target pre-operative measurements. 

For the post-operative stand-alone LPM, the proportionality constants in Equation \eqref{shunt_law_reduced} describing the behaviour of flow across the shunt are derived from regression analysis on the GMM solutions for all non-protruding shunt geometries with diameters of 5, 7, and 9 mm (see Appendix \ref{app:lpm}.1). 
This analysis ensures that 3D behaviour of flow across the shunt is well captured in the stand-alone LPM.
 The  set of all patient-specific lumped parameters describing the global circulation are summarised in Table \ref{tab:lumpedparameters}.


\begin{table}[tb]
\centering
\def\arraystretch{1.6}
\resizebox{1.0\textwidth}{!}
{  
\begin{tabular}{lr|lr|lr|lr|lr|lr}
\hline
 \multicolumn{12}{c}{Pre-Operative Parameters: Windkessel and lumped representation of 3D regions}\\
\hline
  $R_p^{\text{IN}}$           & 701.0           & $R_d^{\text{IN}}$           & 7273  & $C^{\text{IN}}$             & 1.90\text{e-}04 & $R_p^{\text{LC}}$           & 1218.4         & $R_d^{\text{LC}}$           & 12375           & $C^{\text{LC}}$             & 1.12\text{e-}04 \\
  $R_p^{\text{LS}}$           & 1965.9          & $R_d^{\text{LS}}$           & 27674 & $C^{\text{LS}}$             & 5.00\text{e-}05 & $R_p^{\text{DAo}}$           & 35.2           & $R_d^{\text{DAo}}$           & 1669            & $C^{\text{DAo}}$             & 8.28\text{e-}04 \\
  $R_p^{\text{LPA-1}}$        & 1001.7          & $R_d^{\text{LPA-1}}$        & 10017 & $C^{\text{LPA-1}}$          & 1.62\text{e-}04 & $R_p^{\text{LPA-2}}$        & 915.2          & $R_d^{\text{LPA-2}}$        & 9152            & $C^{\text{LPA-2}}$          & 1.62\text{e-}04 \\
  $R_p^{\text{RPA}}$          & 231.3           & $R_d^{\text{RPA}}$          & 2313  & $C^{\text{RPA}}$            & 7.00\text{e-}04 & $C^{\text{AAo}}$             & 4.4\text{e-}04 & $C^{\text{mPA}}$             & 1.5\text{e-}04  & $R_{\mathrm{vs}}^{\text{UB}}$        & 480.4  \\
  $C_{\mathrm{vs}}^{\text{UB}}$        & 4.16\text{e-}03 & $R_{\mathrm{vs}}^{\text{LB}}$        & 205.9 & $C_{\mathrm{vs}}^{\text{LB}}$        & 9.71\text{e-}03 & $R_{\mathrm{vs}}^{\text{LL}}$        & 584.5          & $C_{\mathrm{vs}}^{\text{LL}}$        & 1.71\text{e-}03 & $R_{\mathrm{vs}}^{\text{RL}}$        & 282.8    \\
  $C_{\mathrm{vs}}^{\text{RL}}$        & 3.54\text{e-}03 & $R_{\text{3D}}^{\text{AAo}}$ & 10.1  & $K_{\text{3D}}^{\text{AAo}}$ & 0.167           & $K_{\text{3D}}^{\text{IN}}$ & 0.839          & $L_{\text{3D}}^{\text{IN}}$ & 8.04            & $K_{\text{3D}}^{\text{LC}}$ & 1.32  \\
  $L_{\text{3D}}^{\text{LC}}$ & 17.97           & $K_{\text{3D}}^{\text{LS}}$ & 88.7  & $L_{\text{3D}}^{\text{LC}}$ & 38.58           & $R_{\text{3D}}^{\text{DAo}}$ & 21.2           & $K_{\text{3D}}^{\text{DAo}}$ & 0.98            & $L_{\text{3D}}^{\text{DAo}}$ & 8.41 \\
   \hline
\end{tabular}
}
\caption{Patient-specific pre-operative parameters for Windkessel models and the lumped parameters representing 3D regions: $R$ in units of g/cm\textsuperscript{4}s, $K$ in units of g/cm\textsuperscript{7}, $L$ in units of g/cm\textsuperscript{4} and $C$ in units of cm\textsuperscript{4}s\textsuperscript{2}/g.} 
\label{tab:lumpedparameters}
\end{table}

\subsubsection{Heart parameters governing pulmonary artery hypertension}
\label{sec:two_ways}
In the model, the generation of higher RV pressures due to the PAH relative to a healthy state is mechanistically governed by the two phenomena: first, through a combination of RV wall volume $V_w$ and $V_0$, as varying these parameters changes the amount of myocardial muscle and its stretch, see Equation \eqref{eqn_stretch}, which affects the fibre stresses and consequently the pressures; and second, through the active stress produced by the sarcomeres, see Equation \eqref{eqn:active_stress}. Since our study is concerned with immediate effects of the PS creation, the myocardial wall volume, as it was measured in the patient, was fixed in the model. Thus, to generate higher RV pressures corresponding with the clinical measurements, the active stress produced by the sarcomeres in the RV wall must be higher in comparison to a healthy state. To achieve this, the shape of the time-dependent contraction $g(t_a)$, which is governed by the exponent $E_a$ in Equation  \eqref{eqn:gta}, is altered. The alternative approach of varying the contractility coefficient $c$ (or effectively the maximum stress $T_{a0}$) in Equation (3) is also tested, but it is found that changes in global haemodynamics due to PS creation remain unchanged (see Appendix \ref{app:contractility}).


\begin{table} [tb]
\begin{center}
\def\arraystretch{1.2}
\resizebox{1.0\textwidth}{!}{ 
\begin{tabular}{|c|ccc|ccc|cc|cc|cc|cc|}
\hline
\multicolumn{15}{|c|}{\textbf{I.~Pre-operative measurements and model output (parameter estimation)}}\\
\hline
 & \multicolumn{3}{c}{$P_{\rm{AAo}}$ [mmHg]} & \multicolumn{3}{|c|}{$P_{\rm{mPA}}$ [mmHg]} & \multicolumn{2}{c}{EDV [ml]} & \multicolumn{2}{|c}{SV [ml]} & \multicolumn{2}{|c}{EF} & \multicolumn{2}{|c|}{CO [L/min]}\\
\hline
 & Systolic & Diastolic & Mean & Systolic & Diastolic & Mean & LV & RV & LV & RV & LV & RV & LV & RV\\
 \hline
Measurement & 94.0 & 53.0 & 69.0 & 112.0 & 67.0 & 85.0 & 66.9 & 91.8 & 51.0 & 51.0 & 0.76 & 0.55 & 3.4 & 3.4\\
GMM & 95.4 & 52.5 & 65.4 & 111.1 & 68.3 & 84.6 & 67.3 & 92.0 & 50.7 & 50.7 & 0.75 & 0.55 & 3.38 & 3.38\\
\hline
\multicolumn{15}{|c|}{\textbf{II.~Post-operative measurements and model output (validation)}}\\
\hline
Measurement & 97.0 & 51.0 & 71.0 & 102.0 & 54.0 & 76.0 & - & - & - & - & - & - & - & -\\
GMM & 104.1 &	60.8 &	74.3 &	106.9 & 61.3 & 76.8 & 68.9 & 83.4 & 42.7 & 59.6 & 0.62 & 0.71 & 2.85 & 3.97\\
\hline
\end{tabular}
}
\caption{Pre- and post-operative $P_{\rm{AAo}}$, $P_{\rm{mPA}}$, EDV, SV, EF and CO from the GMM against clinical measurements.}
\vspace{0.5cm}
\label{tab:prepost_press_data_ms}
\end{center}
\end{table}

\section{Results}
\label{sec:results}

All results are presented from the last cardiac cycle in the GMM simulation after the solution has converged. In the main manuscript, the results for the GMM are presented, while the stand-alone LPM results are shown in Appendix \ref{app:lpm}.

\subsection{Comparison of simulation output with pre- and post-operative measurements}
\label{res_comp}
\begin{figure}[htbp]
\centering
\includegraphics[width=0.5\linewidth]{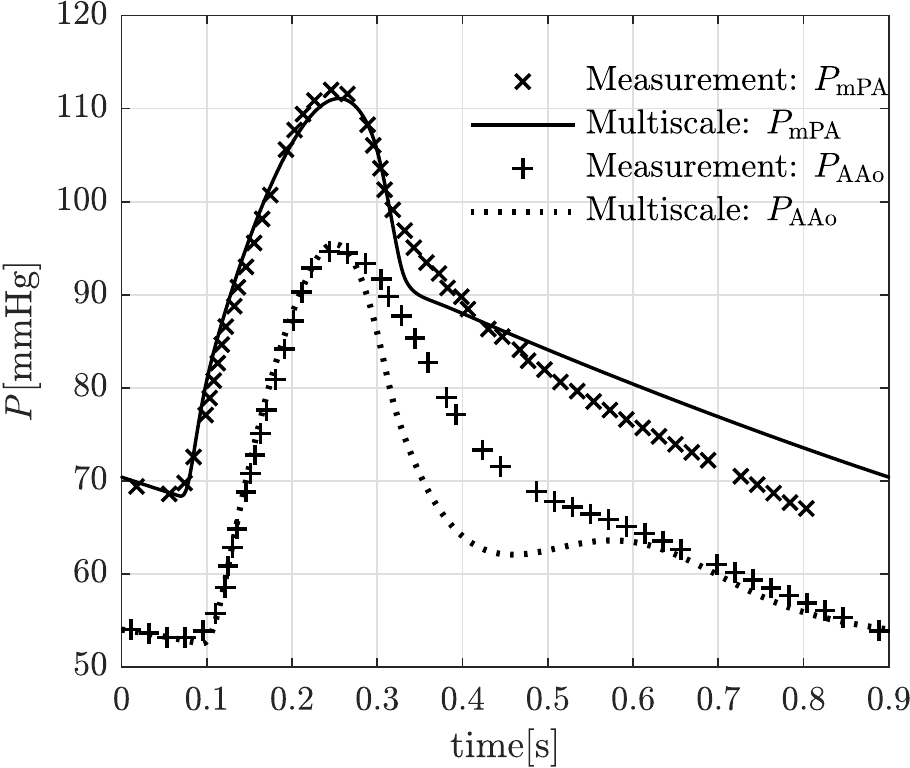}
\caption{Comparison of pre-operative pressure waveforms generated by the GMM against the measurements over one cardiac cycle.}
\label{fig:PaoAndPpa_ms}
\end{figure}

The comparison of simulation output with pre- and post-operative measurements is performed for the 7.6mm diameter PS corresponding to the procedure outcome in the patient, albeit without inclusion of stent's protruding ends. Tables \ref{tab:prepost_press_data_ms}-I shows the GMM  output against the measurements for the quantities that are used for patient-specific parameter estimation. 
For the pre-operative state, the GMM solution is generally within 2.0\% of the clinical measurements and only differs by a maximum of 5\% for the mean aortic pressure.  This agreement between the pre-operative computational solution and the clinical measurements is corroborated by the pressure waveforms presented in Figure \ref{fig:PaoAndPpa_ms}. 

\begin{figure}[b]
     \centering
    \begin{subfigure}[t]{0.32\textwidth}
        \raisebox{-\height}{\includegraphics[width=\textwidth]{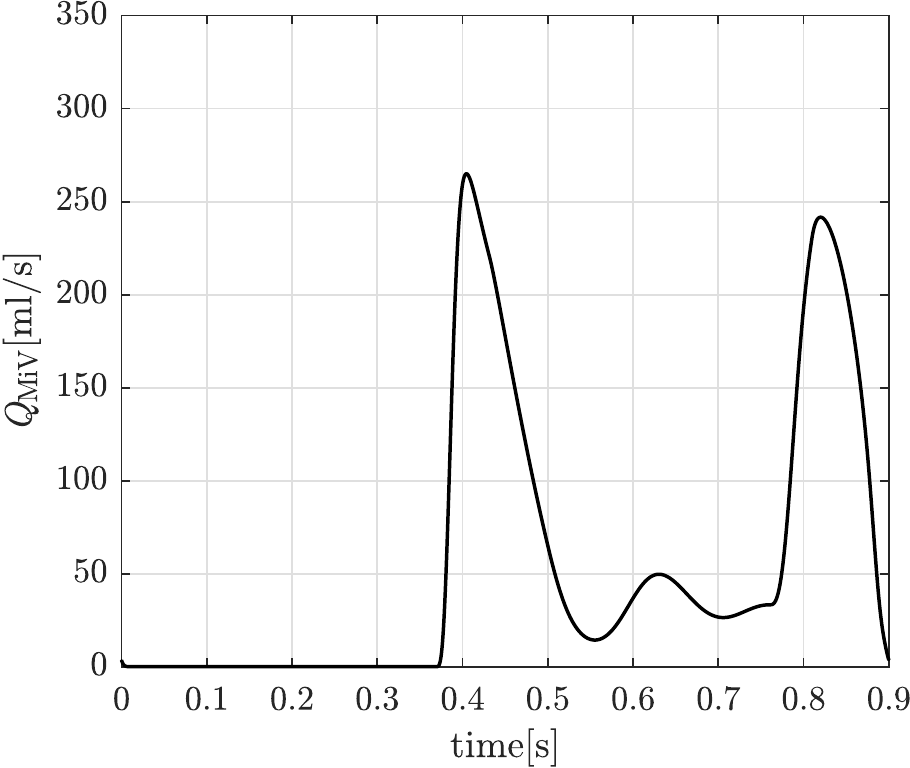}}
        \caption{Flow rate through MiV}
    \end{subfigure}
    \hfill
    \begin{subfigure}[t]{0.32\textwidth}
        \raisebox{-\height}{\includegraphics[width=\textwidth]{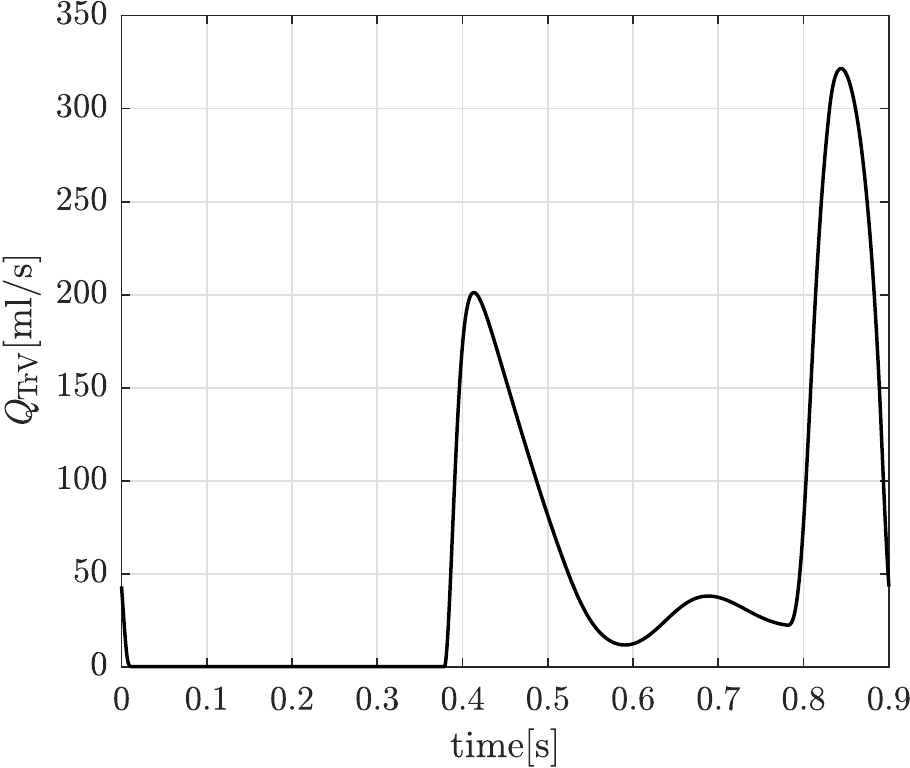}}
        \caption{Flow rate through TrV}
    \end{subfigure}
    \begin{subfigure}[t]{0.33\textwidth}
        \raisebox{-1\height}{\includegraphics[width=1.0\textwidth]{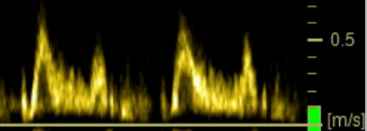}}
        \caption{Measured MiV}
        \raisebox{-0.95\height}{\includegraphics[width=1.0\textwidth]{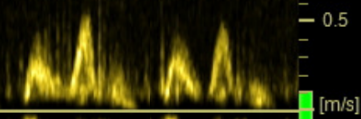}}
        \caption{Measured TrV}
    \end{subfigure}
\caption{Comparison of GMM solution for pre-operative volumetric flow rates through the atrioventricular valves with Doppler measured velocity tracings.}
\label{fig:MVVandTCV}
\end{figure}

For qualitative validation, the simulation output for the pre-operative state is compared against the Doppler velocity tracings for flow across the AV valves, the only clinical data that was not directly used for model parameter estimation (Figure \ref{fig:MVVandTCV}). As can be observed, the model accurately captures the flow velocity profiles across the AV valves. In particular, the flow over mitral valve (MiV) reflects normal LV diastolic function (E$>$A) in contrast to the flow profile across the tricuspid valve (TrV) demonstrating impaired filling and diastolic dysfunction of the pressure-loaded RV (E$<$A).

Further validation of the model and its predictive capabilities are assessed by comparing the post-operative GMM simulation output against the respective clinical measurements (Table \ref{tab:prepost_press_data_ms}-II). The GMM output is in reasonable agreement with the measured pressures, reproducing the mean mPA pressure $P_{\rm{mPA}}$ and mean AAo pressure $P_{\rm{AAo}}$ within 1\% (absolute error = 0.8 mmHg) and 5\% (absolute error = 3.3 mmHg) of the measurements, respectively.  The predicted systolic pressures in the GMM exceed the measurements by 5\% (absolute error = 4.9 mmHg) for the mPA and 7\% (absolute error = 7.1 mmHg) for the aorta. The diastolic pressures exceed the measurements by 13\% (absolute error = 7.3 mmHg) and 19\% (absolute error = 9.8 mmHg) for the mPA and the AAo, respectively. 

\subsection{Global haemodynamic changes due to the PS creation}
\label{res_haemodynamic_changes}
\begin{figure}
\centering
\includegraphics[width=0.87\linewidth]{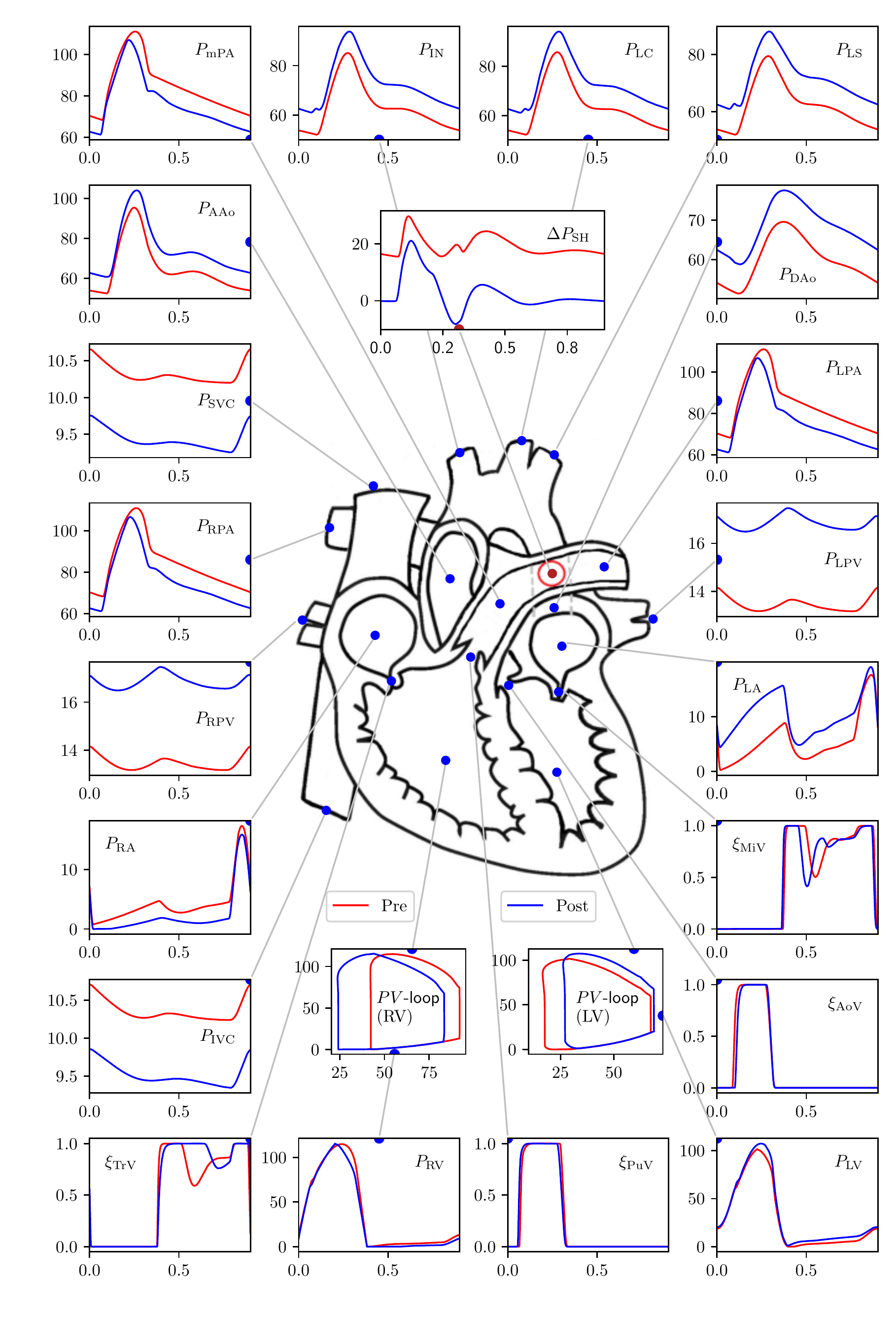}
\caption{GMM results for the 7.6 mm diameter PS showing pre- to post-operative changes in pressure at key locations in the arterial network. PV loops are additionally included. In the PV loop plots, the x-axis represents volume [ml] and y-axis represents pressure [mmHg]. In all other plots the x-axis represents time [s] and the y-axis represents pressure [mmHg]. The valve parameters $\xi$ are dimensionless. For a key to symbol nomenclature, please see Figure \ref{fig:multiscalemodel}. $\Delta P_{\mathrm{SH}}$ represents the pressure gradient across the PS.}
\label{fig:pre_post_pressure_ms}
\end{figure}

\begin{figure}
\centering
\includegraphics[width=0.87\linewidth]{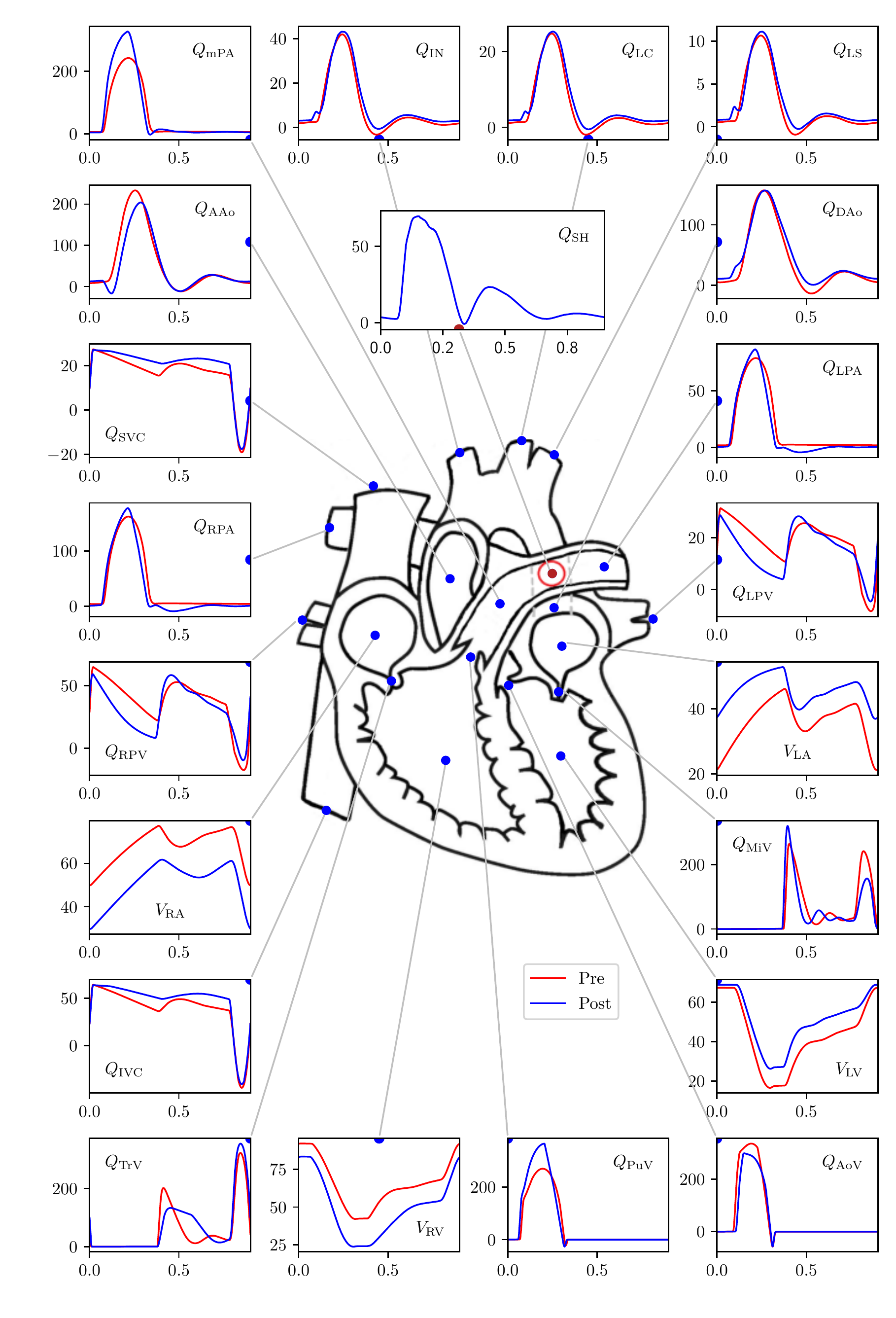}
\caption{GMM results for the 7.6 mm diameter PS showing pre- to post-operative changes in flow-rate and volumes at key locations in the arterial network. In all the plots the x-axis represents time, and y-axis for volumes, $V_{(\cdot)}$, is in [ml], while for the flow-rates, $Q_{(\cdot)}$, is in [ml/s]. For a key to symbol nomenclature, please see Figure \ref{fig:multiscalemodel}.}
\label{fig:pre_post_flow_ms}
\end{figure}

\begin{figure}
\hspace{-1cm}
\includegraphics[width=1.1\textwidth]{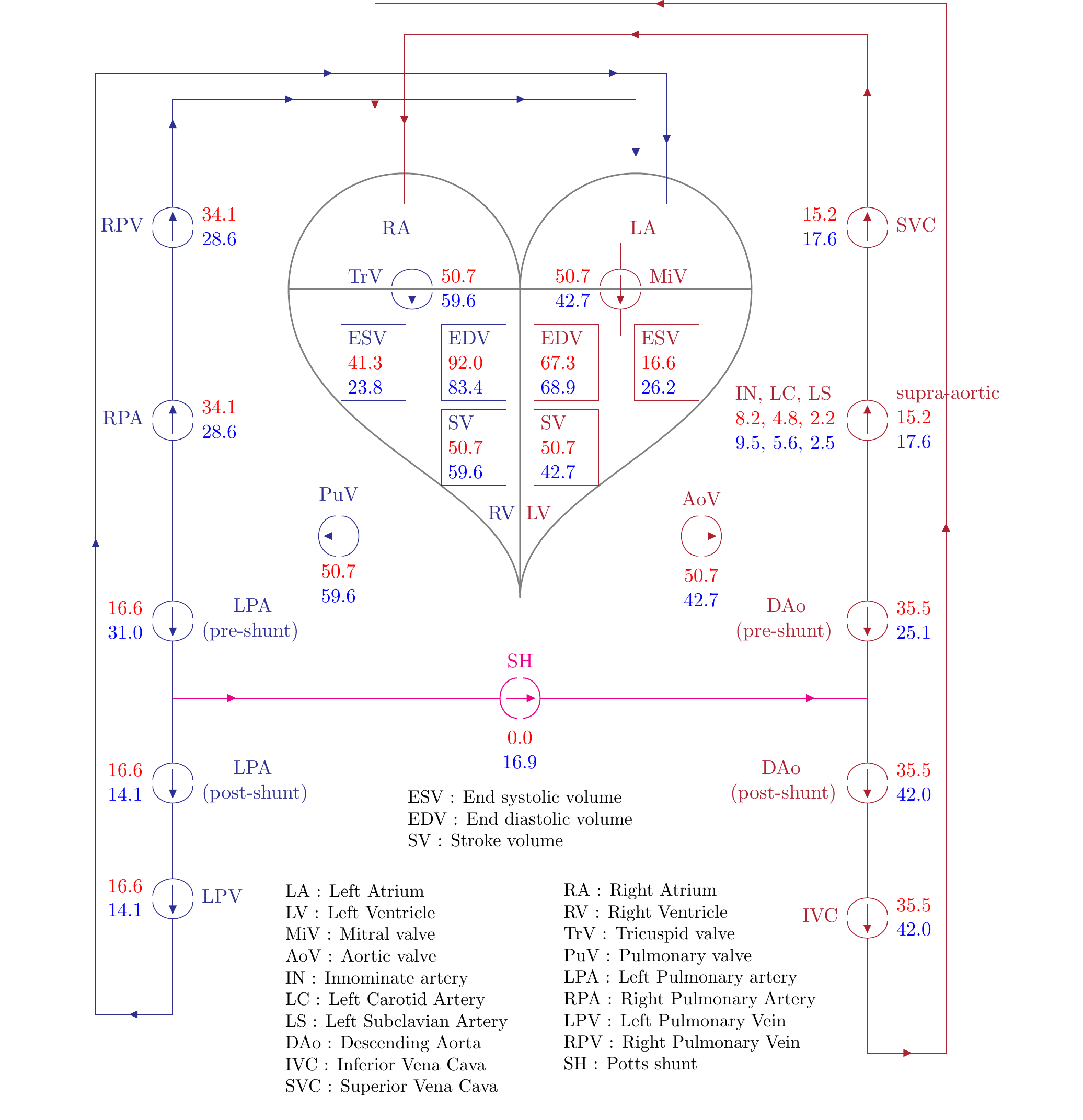}
\caption{GMM results for the 7.6 mm diameter PS showing pre- to post-operative changes in volume of blood flowing in one cardiac cycle through the circulatory system. All numerical values are for volumes in ml, and values in red represent pre-operative state while those in blue represent post-operative state.}
\label{fig:volumes_ms}
\end{figure}

The key advantage of modelling is that detailed haemodynamic changes due to PS creation can be assessed at all locations within circulation. 
Figures  \ref{fig:pre_post_pressure_ms}, \ref{fig:pre_post_flow_ms}, and \ref{fig:volumes_ms} depict such changes occurring in one cardiac cycle in pressures, flow-rates and absolute volumes passing through various compartments, respectively,  for the scenario with the non-protruding PS of 7.6mm diameter. Major changes observed (Figures \ref{fig:pre_post_pressure_ms}--\ref{fig:volumes_ms} and Table \ref{tab:prepost_press_data_ms}) in haemodynamics due to the PS creation as predicted by the GMM are as follows:

\begin{enumerate}
\item The mean flow-rate through the shunt is $\overline{Q}_{\mathrm{SH}}\!=\!18.8$ ml/s (corresponding to a volume displacement of $\overline{V}_{\mathrm{SH}} = 16.9$ ml in one cardiac cycle) with a mean pressure gradient of $\overline{\Delta P}_{\mathrm{SH}} \!=\! 4.6$ mmHg across the PS.
\item The pressure in the PAs decreases while the pressure in the aorta increases, becoming nearly equal to each other.
\item The LV output decreases by 16\% while the RV output increases by 18\%.
\item The LV ejection fraction decreases (from 0.75 pre-PS to 0.62 post-PS) while that of the RV improves (from 0.55 pre-PS to 0.71 post-PS).
\item The LV EDV remains nearly unchanged (increase by 2\%), while LV ESV increases substantially (by 58 \%), resulting in a net decrease in LV SV (by 16\%), and hence the LV output (by 16\%).
\item Even though the LV output decreases, the flow from the LV into the aortic arch branches, and hence to the upper body and the superior vena cava (SVC), increases by 16\%. In contrast, the LV contribution to the flow into the DAo decreases by 29\%. 
\item The total flow in the DAo downstream the PS, however, increases due to RV contribution to the flow through the PS, corresponding to an increase of the flow in the inferior vena cava (IVC) (by 18\%).
\item The work done by the LV (assessed by the area within the ventricular PV-loop) decreases from  $4.13\times 10^3$ mmHg-ml pre-operatively to $3.66\times 10^3$ mmHg-ml post-operatively.
\item The RV EDV decreases by 9\% and its end-systolic volume decreases by an even larger amount (42\%), resulting in a net increase in RV stroke volume (by 18\%), and hence the RV output (by 18\%).
\item Even though the RV output increases, the net flow into the RPA and the LPA downstream the PS decreases (by 16\% and 15\%, respectively) due to the post-ventricular right-to-left shunt. Correspondingly, flow to the left and right pulmonary veins also decreases (by 15\% and 16\%, respectively), reflecting the blood flow diversion through the proximal LPA into the shunt, the volume of which exceeds the volume corresponding to the increase in RV output.
\item The work done by the RV (the area within the PV-loop) increases from  $4.95\times 10^3$ mmHg-ml pre-operatively to $5.87\times 10^3$ mmHg-ml post-operatively.
\item Defining $Q_p$ as the sum of flow-rates in the left and right pulmonary veins and $Q_s$ as the sum of flow-rates in the inferior and superior venae cavae, the  $Q_p/Q_s$ ratio changes from  1.0 pre-operatively to 0.72 in the post-operative state.
\item The valve function remains largely unaffected for all the valves.
\item The operating volumes (minimum and maximum volumes  during a cardiac cycle) of the LA increase, which is accompanied by a corresponding increase in the LA pressures; while the operating volumes of the RA decrease, with a corresponding decrease in the RA pressures.
\item The pressure changes in the inferior and superior venae cavae are less than 2 mmHg, while those in the left and right pulmonary veins are less than 3mmHg.
\end{enumerate}

\begin{figure}[t]
    \begin{subfigure} {1.0\textwidth}
      \centering
      \includegraphics[width = 0.24\textwidth]{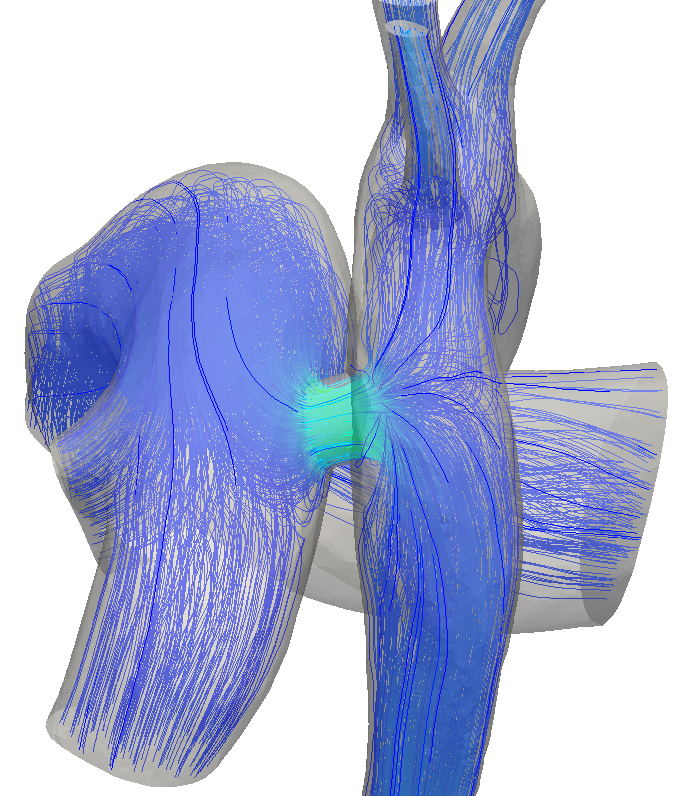}
	\hfill
      \includegraphics[width = 0.24\textwidth]{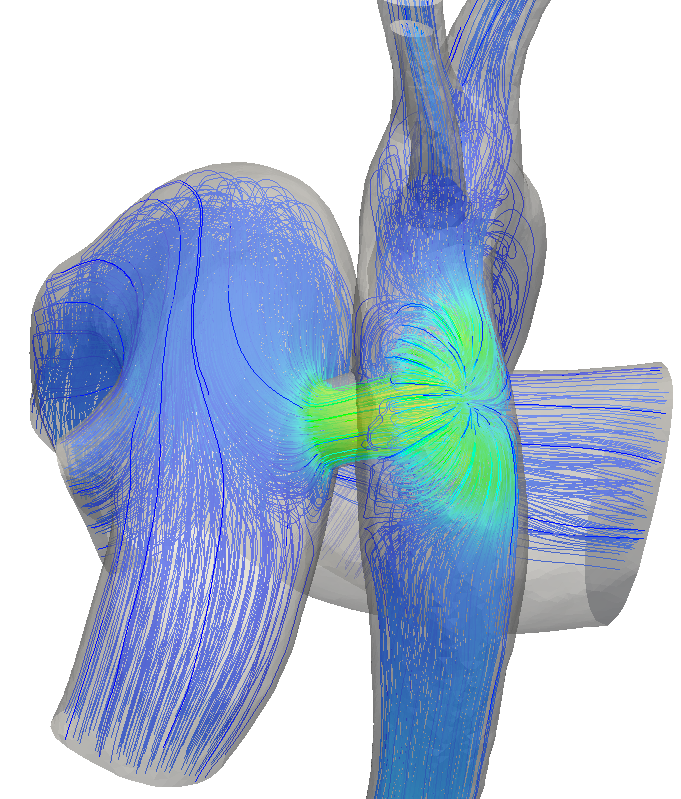}
	\hfill
      \includegraphics[width = 0.24\textwidth]{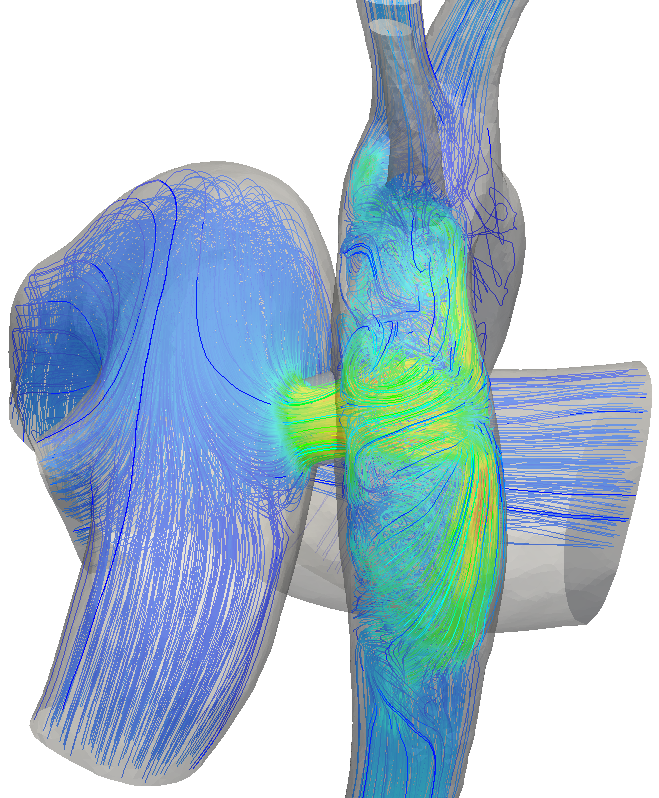}
	\hfill
      \includegraphics[width = 0.24\textwidth]{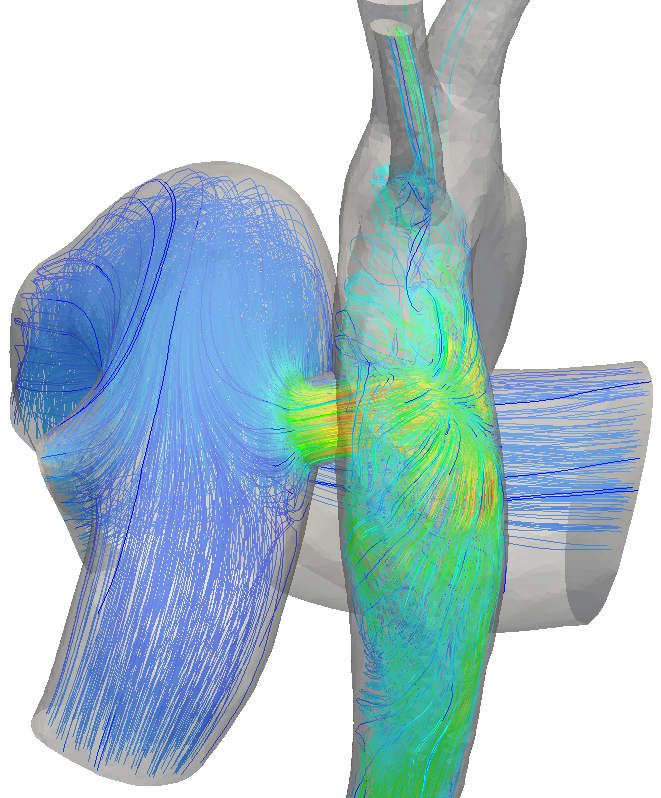}
	\hfill
    \end{subfigure}\\
        \begin{subfigure} {0.24\textwidth}
      \centering
      \caption{time = 0.08 s}
    \end{subfigure}
   \begin{subfigure}{0.24\textwidth}
     \centering
      \caption{time = 0.1 s}
    \end{subfigure}
    \begin{subfigure}{0.24\textwidth}
         \centering
      \caption{time = 0.13 s}
    \end{subfigure}
        \begin{subfigure}{0.24\textwidth}
             \centering
      \caption{time = 0.16 s}
    \end{subfigure}\\
   \begin{subfigure} {1.0\textwidth}
        \centering
      \includegraphics[width = 0.24\textwidth]{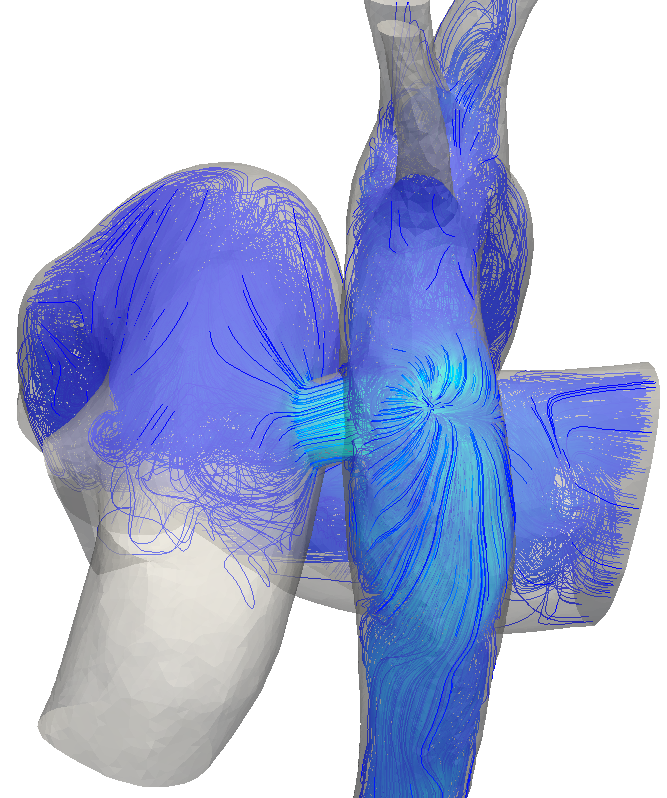}
	\hfill
      \includegraphics[width = 0.24\textwidth]{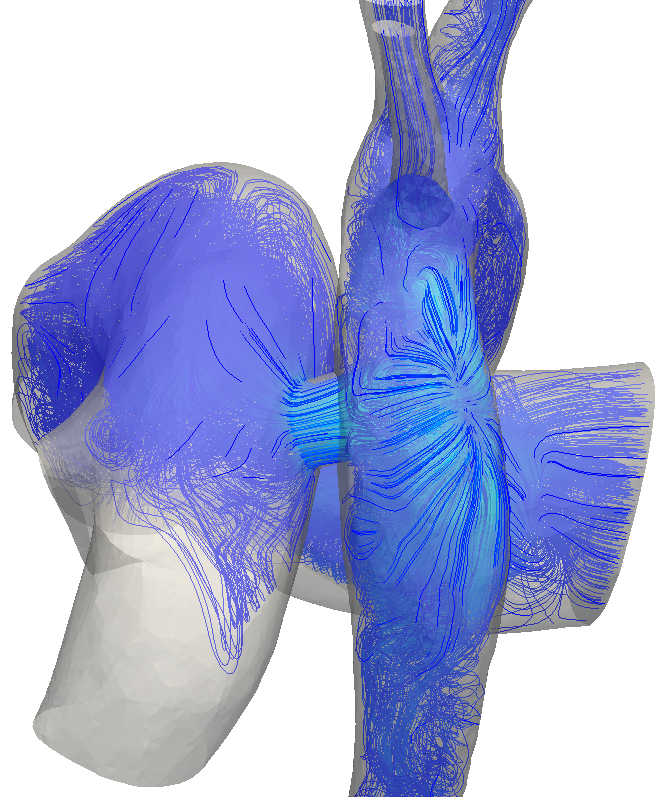}
	\hfill
      \includegraphics[width = 0.24\textwidth]{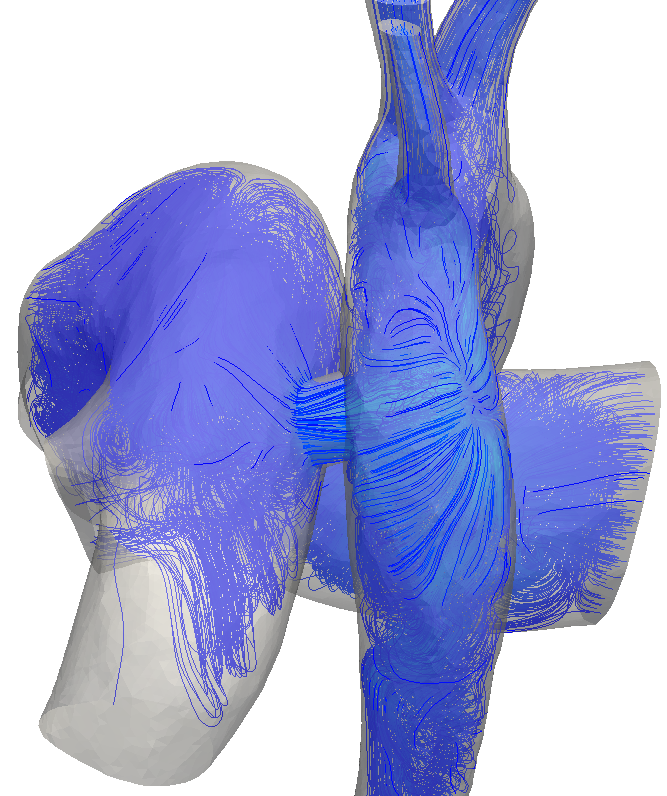}
	\hfill
      \includegraphics[width = 0.24\textwidth]{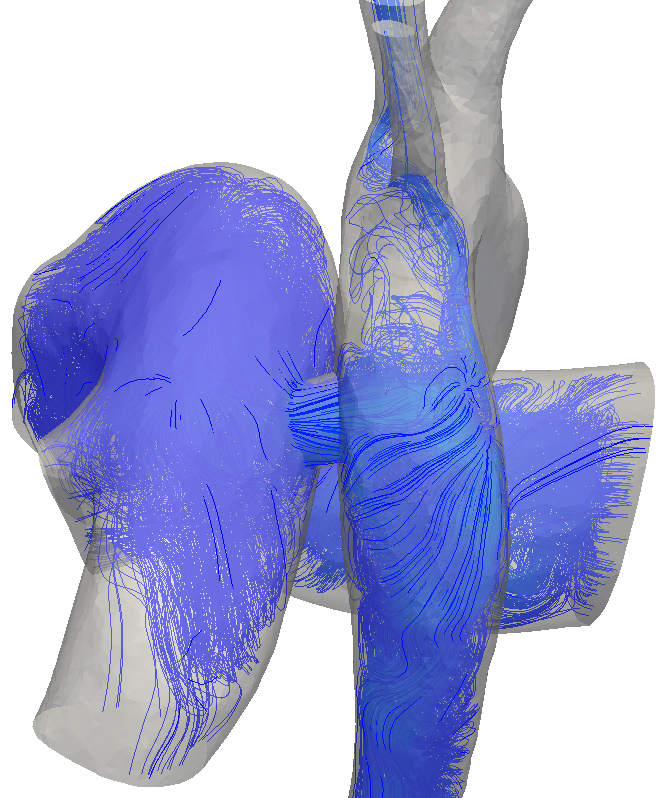}
	\hfill
    \end{subfigure}\\
       \begin{subfigure}{0.24\textwidth}
        \centering
      \caption{time = 0.46 s}
    \end{subfigure}
    \begin{subfigure}{0.24\textwidth}
         \centering
      \caption{time = 0.5 s}
    \end{subfigure}
        \begin{subfigure}{0.24\textwidth}
             \centering
      \caption{time = 0.55 s}
    \end{subfigure}
   \begin{subfigure}{0.24\textwidth}
        \centering
      \caption{time = 0.58 s}
    \end{subfigure}\\
       \begin{subfigure}[c]{1.0\textwidth}
        \centering
      \includegraphics[scale=0.50]{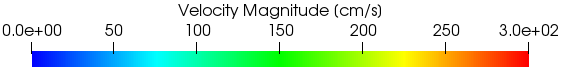}
    \end{subfigure}\\
\caption{Velocity streamlines that originate from the mPA for the 7.6 mm diameter PS at different times of the cardiac cycle.}
\label{fig:Streamlines_76mm}
\end{figure}

\begin{figure}
    \begin{subfigure}{1.0\textwidth}
      \centering
      \includegraphics[width = 0.24\textwidth]{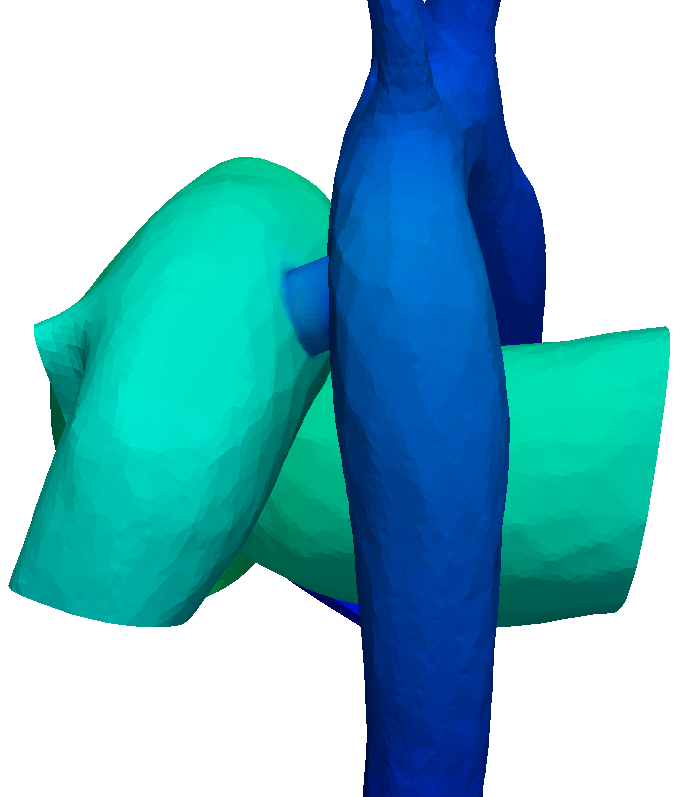}
	\hfill
      \includegraphics[width = 0.24\textwidth]{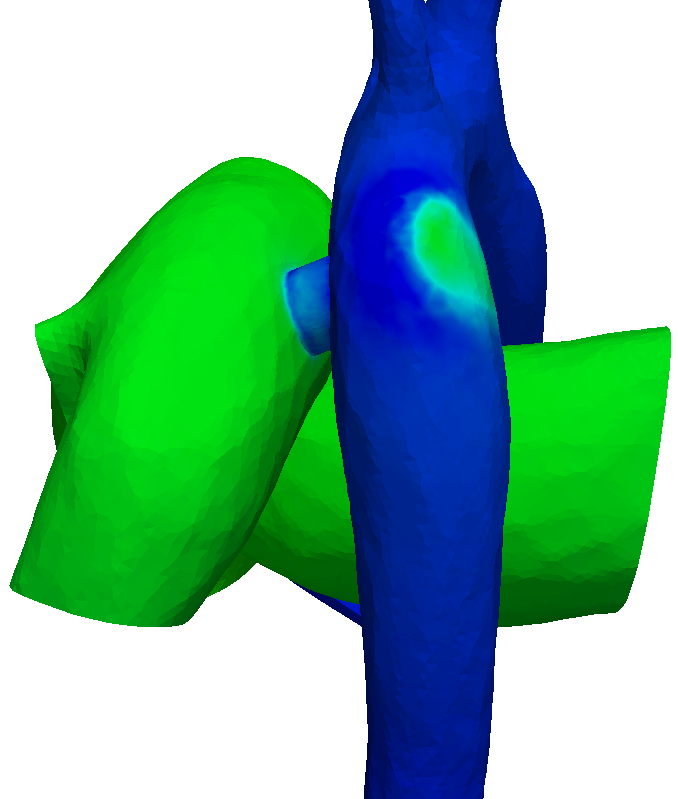}
	\hfill
      \includegraphics[width = 0.24\textwidth]{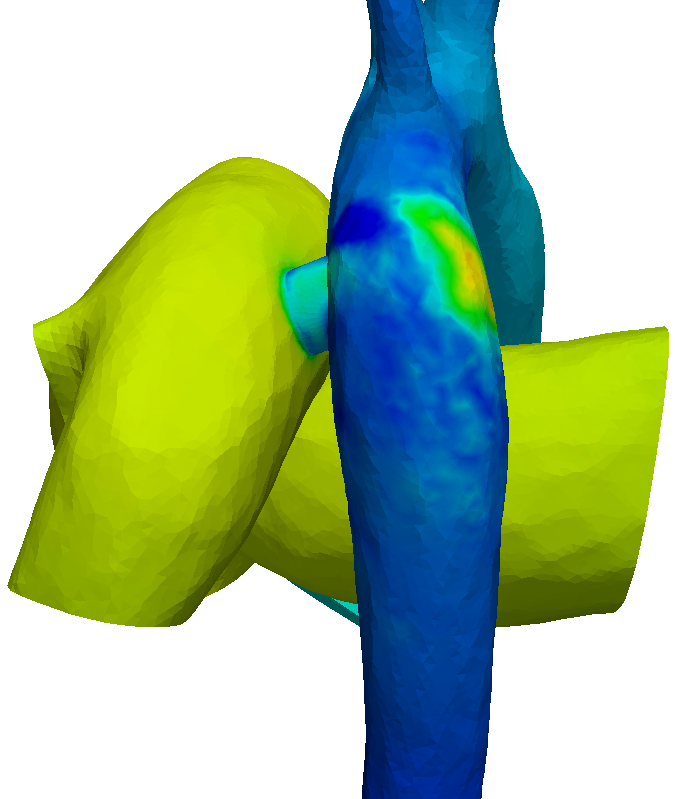}
	\hfill
      \includegraphics[width = 0.24\textwidth]{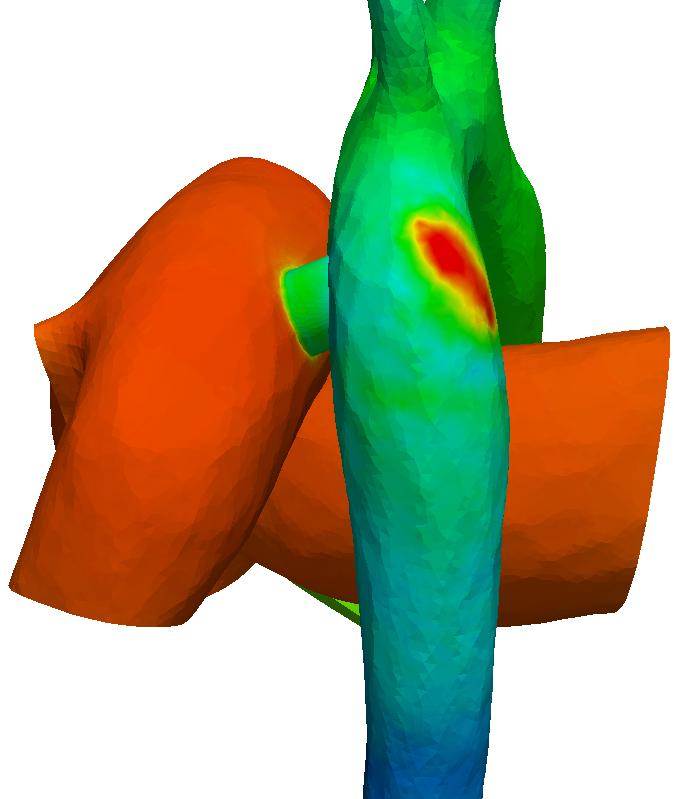}
	\hfill
    \end{subfigure}\\
    \begin{subfigure}{0.24\textwidth}
      \centering
      \caption{Pressure at 0.08 s}
    \end{subfigure}
   \begin{subfigure}{0.24\textwidth}
     \centering
      \caption{Pressure at 0.1 s}
    \end{subfigure}
    \begin{subfigure}{0.24\textwidth}
         \centering
      \caption{Pressure at 0.13 s}
    \end{subfigure}
        \begin{subfigure}{0.24\textwidth}
         \centering
      \caption{Pressure at 0.16 s}
    \end{subfigure}\\
            \begin{subfigure}{1.0\textwidth}
             \centering
      \includegraphics[scale=0.50]{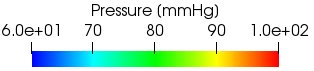}
    \end{subfigure}
    \caption{Wall pressure for the 7.6 mm diameter PS at different times within the systolic phase.}
\label{fig:Press_76mm}
    \end{figure}
    
    \begin{figure}
    \begin{subfigure}{1.0\textwidth}
      \centering
      \includegraphics[width = 0.24\textwidth]{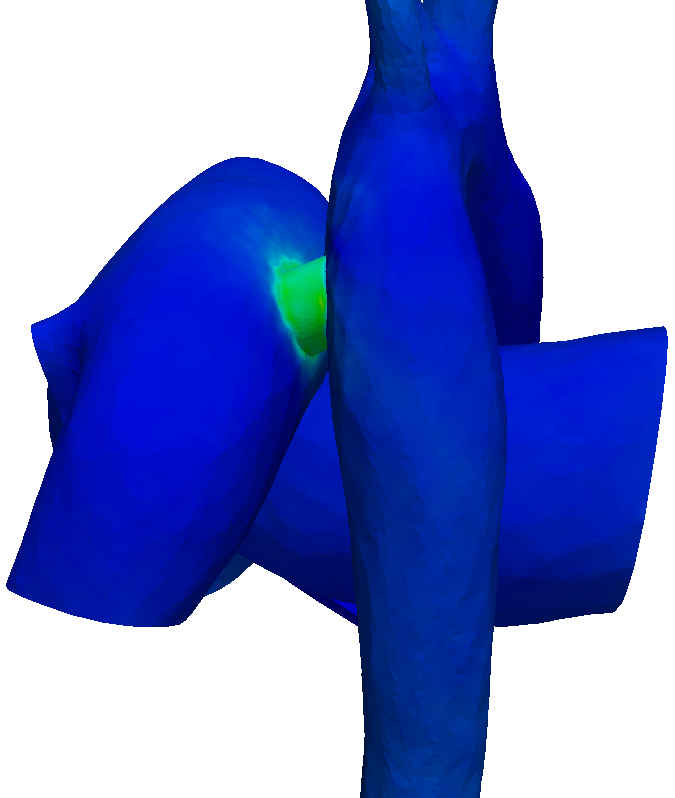}
	\hfill
      \includegraphics[width = 0.24\textwidth]{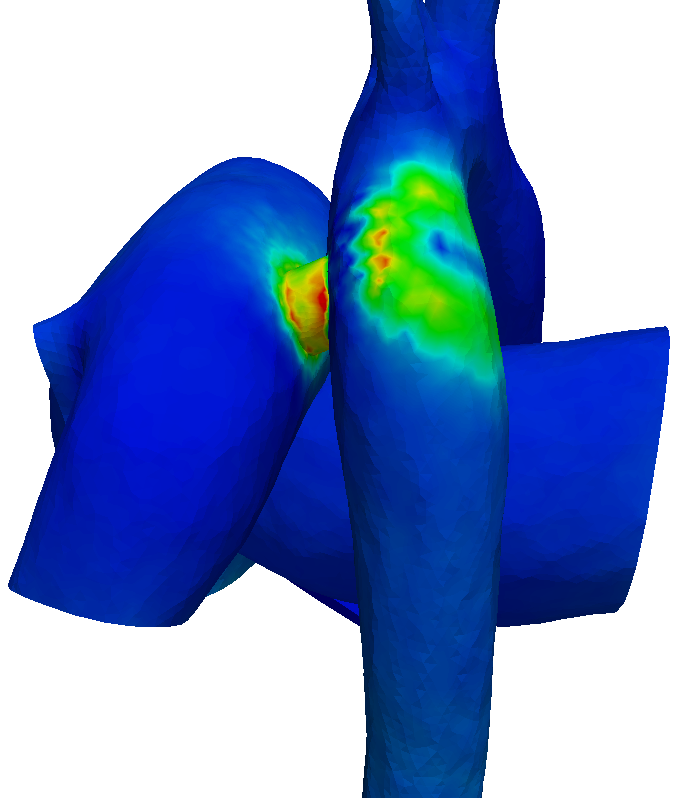}
	\hfill
      \includegraphics[width = 0.24\textwidth]{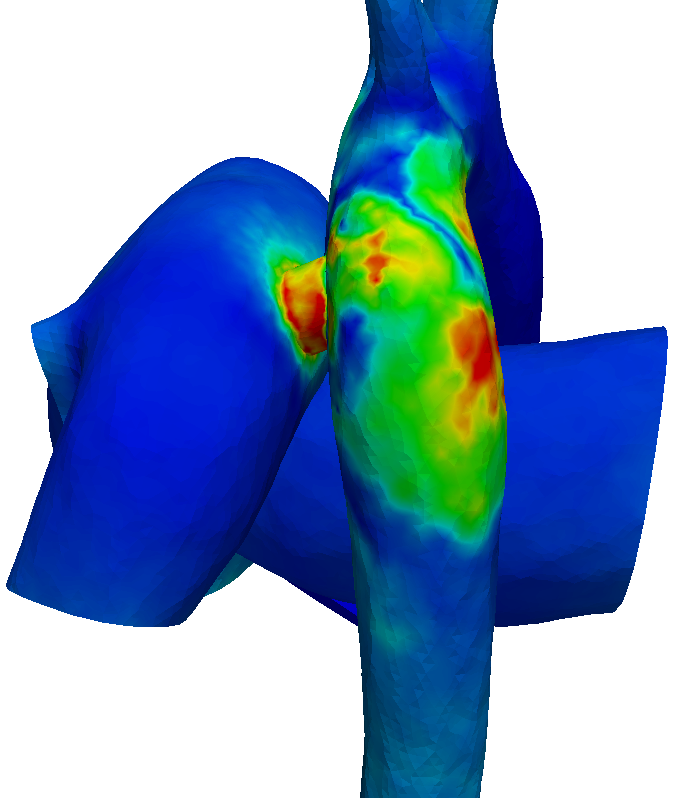}
	\hfill
      \includegraphics[width = 0.24\textwidth]{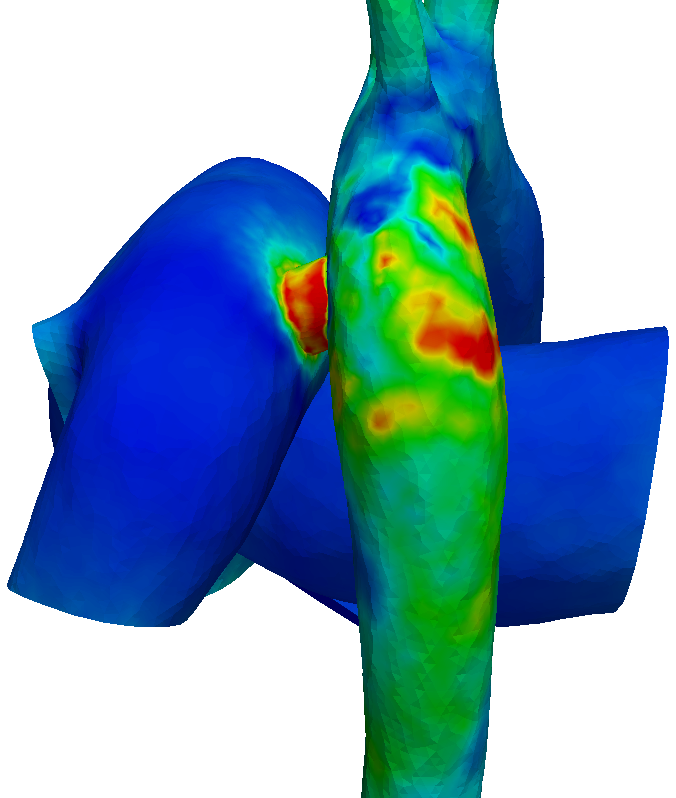}
	\hfill
    \end{subfigure}\\
        \begin{subfigure}{0.24\textwidth}
      \centering
      \caption{WSS at 0.08 s}
    \end{subfigure}
   \begin{subfigure}{0.24\textwidth}
     \centering
      \caption{WSS at 0.1 s}
    \end{subfigure}
    \begin{subfigure}{0.24\textwidth}
         \centering
      \caption{WSS at 0.13 s}
    \end{subfigure}
        \begin{subfigure}{0.24\textwidth}
         \centering
      \caption{WSS at 0.16 s}
    \end{subfigure}\\
        \begin{subfigure}{1.0\textwidth}
             \centering
      \includegraphics[scale=0.50]{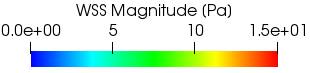}
    \end{subfigure}\\
\caption{Wall shear stress (WSS) for the 7.6 mm diameter PS at different times within the systolic phase.}
\label{fig:WSS_76mm}
\end{figure}

\subsection{Spatio-temporal haemodynamics around the PS}
\label{res_3D}
The GMM presents an advantage that local flow features in the 3D domain can be assessed. Figure \ref{fig:Streamlines_76mm} shows streamlines that originate from the mPA (thus showing only the RV contribution to the blood flow) at different points in the cardiac cycle. The streamlines are coloured by the velocity magnitude and highlight three phenomena: 
\begin{enumerate}[label=(\roman*)]
\item the flow velocities in the PS are significantly higher in comparison to those in the PAs (during peak systole the velocity magnitudes within the shunt average to approximately 225 cm/s);
\item the jet from the PS impinges with high velocity on to the opposite wall of the DAo, particularly in the systolic phase; and
\item  while most of the flow from the shunt is diverted to the distal DAo, the streamlines (with pathlines showing a similar pattern) indicate that a small amount of flow is going to the aortic arch branches.
\end{enumerate}

During the cardiac cycle, the pressure gradient across the shunt increases in systole reaching its peak at approximately peak-systole and then reduces to a slightly negative value at end-systole, quickly recovering to approximately zero pressure gradient for most of diastole (Figure \ref{fig:pre_post_pressure_ms}). The corresponding flow-rate through the shunt follows a pattern close to that of the pressure gradient (Figure \ref{fig:pre_post_flow_ms}), given the pressure gradient is driving the flow.

Figure \ref{fig:Press_76mm} shows the pressure variation in the 3D flow domain at various points in the systolic phase, and shows the high pressure created in the region of the DAo opposite to the shunt by the impinging jet. Similarly, Figure \ref{fig:WSS_76mm} shows wall shear stress (WSS) variation in the 3D region at various points in the systolic phase, and highlights high WSS regions within the DAo wall created by the shunt jet.

\begin{figure*}[!t]
    \begin{subfigure}[c] {0.33\textwidth}
      \centering
      \includegraphics[width=0.9\textwidth]{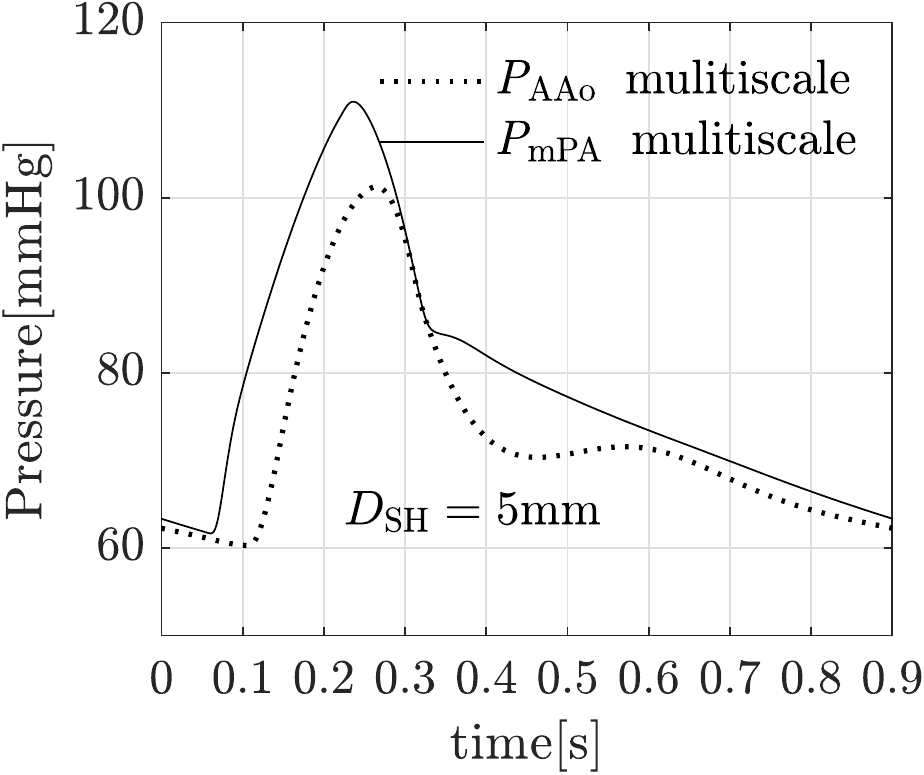}
      \caption{5mm diameter}
    \end{subfigure}
   \begin{subfigure}[c] {0.33\textwidth}
     \centering
      \includegraphics[width=0.9\textwidth]{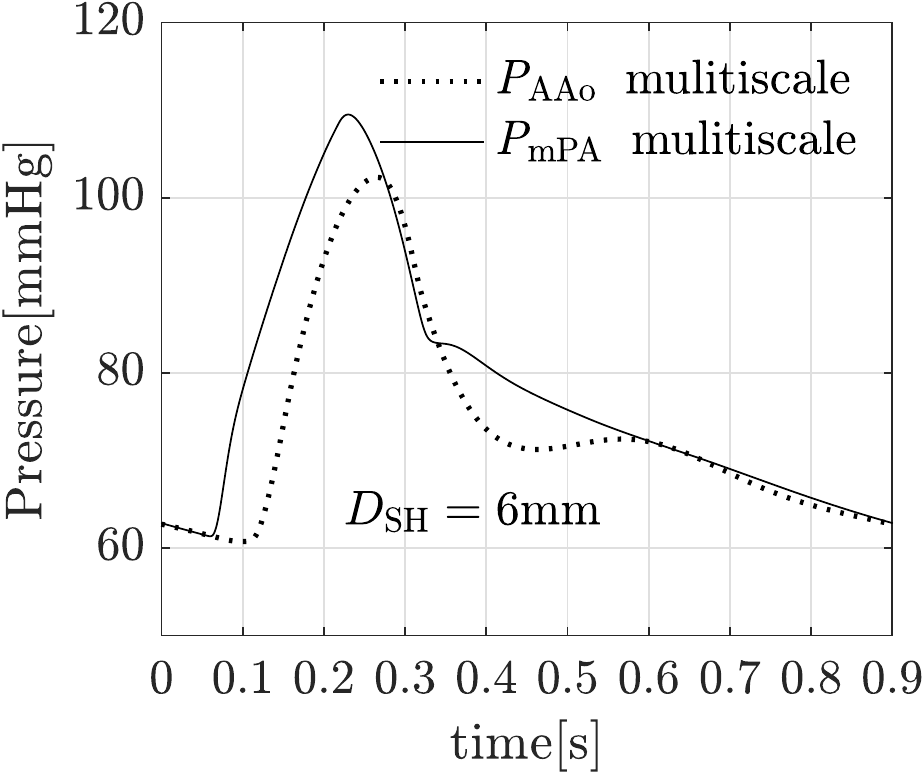}
      \caption{6mm diameter}
    \end{subfigure}
    \begin{subfigure}[c] {0.33\textwidth}
      \centering
      \includegraphics[width=0.9\textwidth]{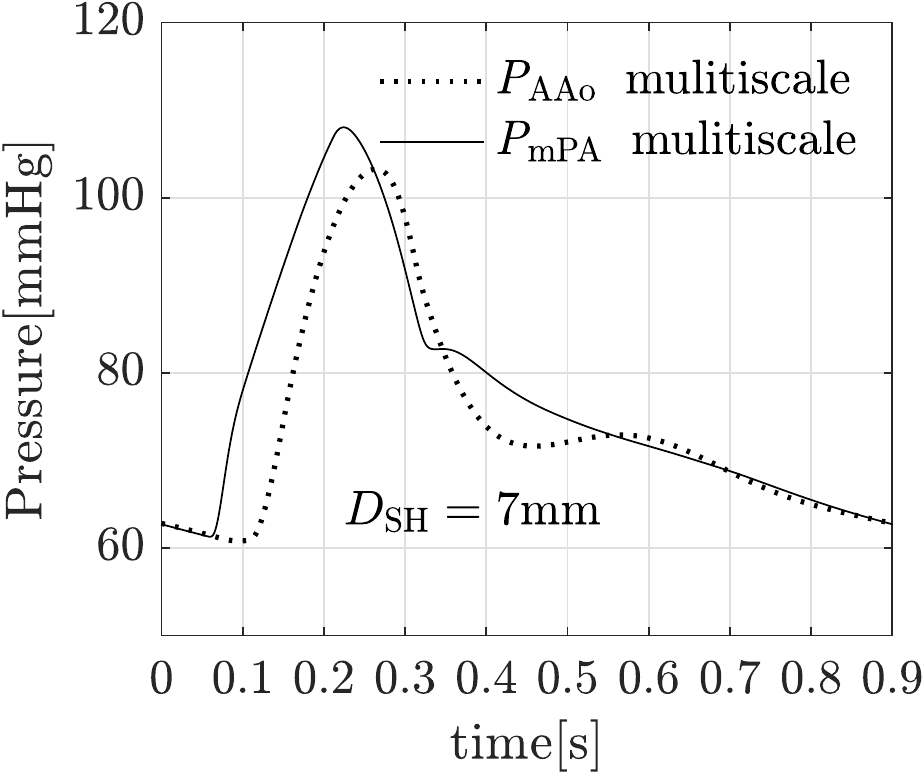}
      \caption{7mm diameter}
    \end{subfigure}\\
        \begin{subfigure}[c] {0.33\textwidth}
          \centering
      \includegraphics[width=0.9\textwidth]{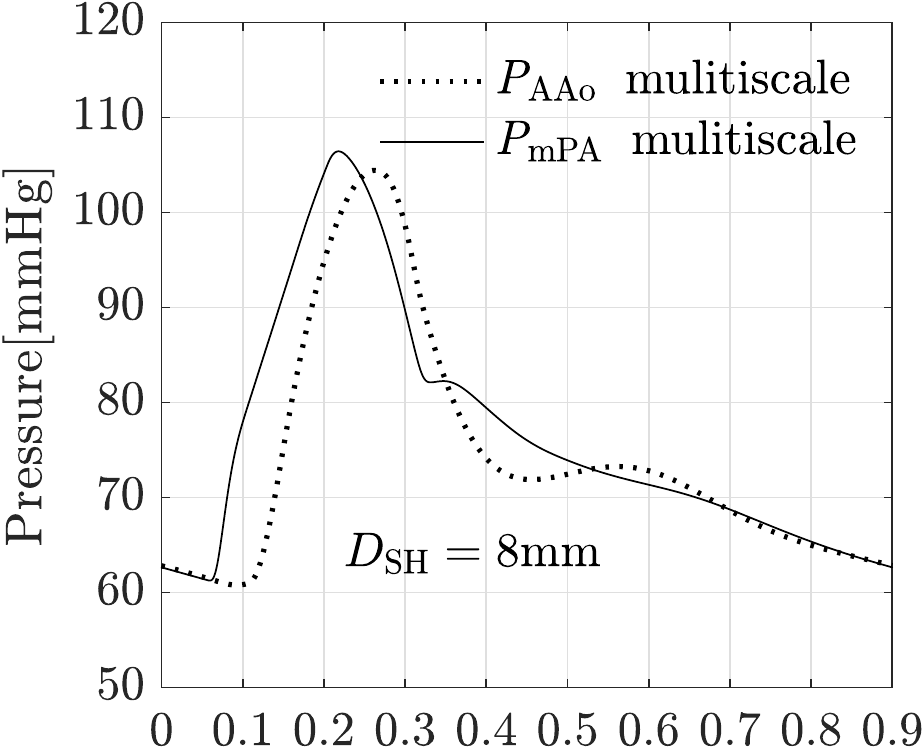}
      \caption{8mm diameter}
    \end{subfigure}
   \begin{subfigure}[c] {0.33\textwidth}
     \centering
      \includegraphics[width=0.9\textwidth]{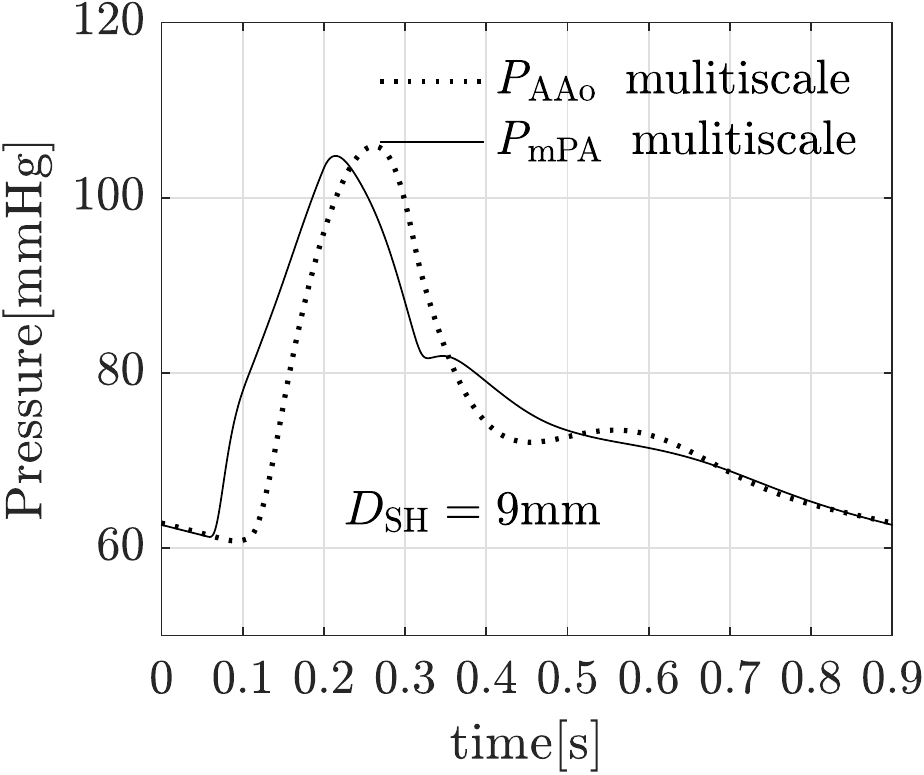}
      \caption{9mm diameter}
    \end{subfigure}
    \begin{subfigure}[c] {0.33\textwidth}
      \centering
      \includegraphics[width=0.9\textwidth]{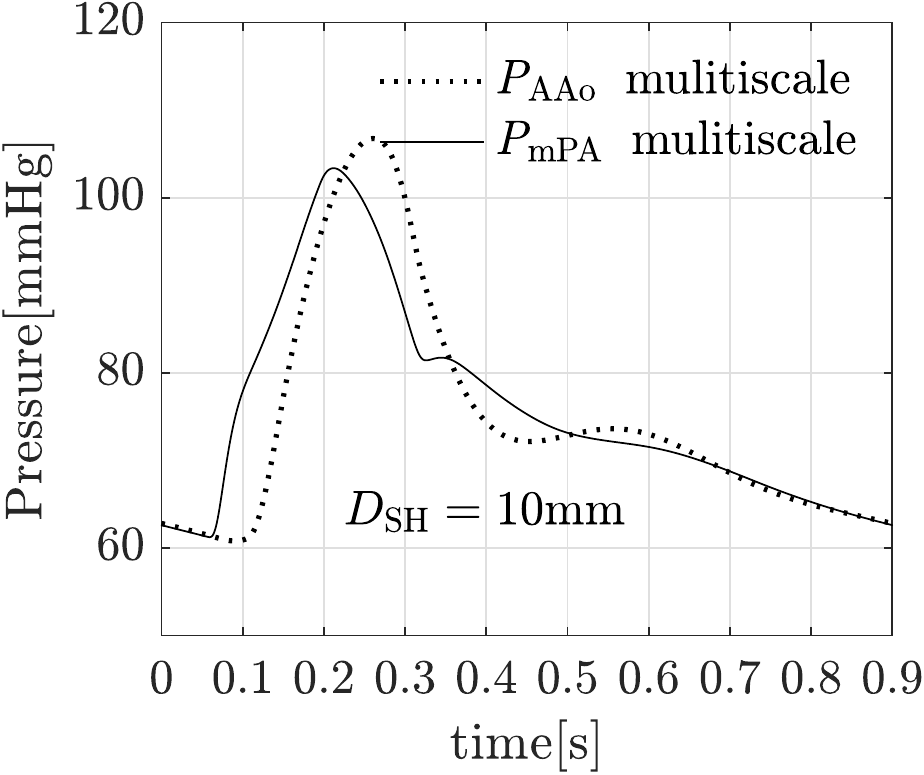}
      \caption{10mm diameter}
    \end{subfigure}\\
\caption{GMM: post-operative $P_{\rm{AAo}}$ and $P_{\rm{mPA}}$ for different shunt diameters.}
\label{fig:PaoPpa_PSdia_ms}
\end{figure*}

\begin{figure}[tb]
\centering
\begin{subfigure}{0.45\textwidth}
\centering
\includegraphics[width=1.0\textwidth]{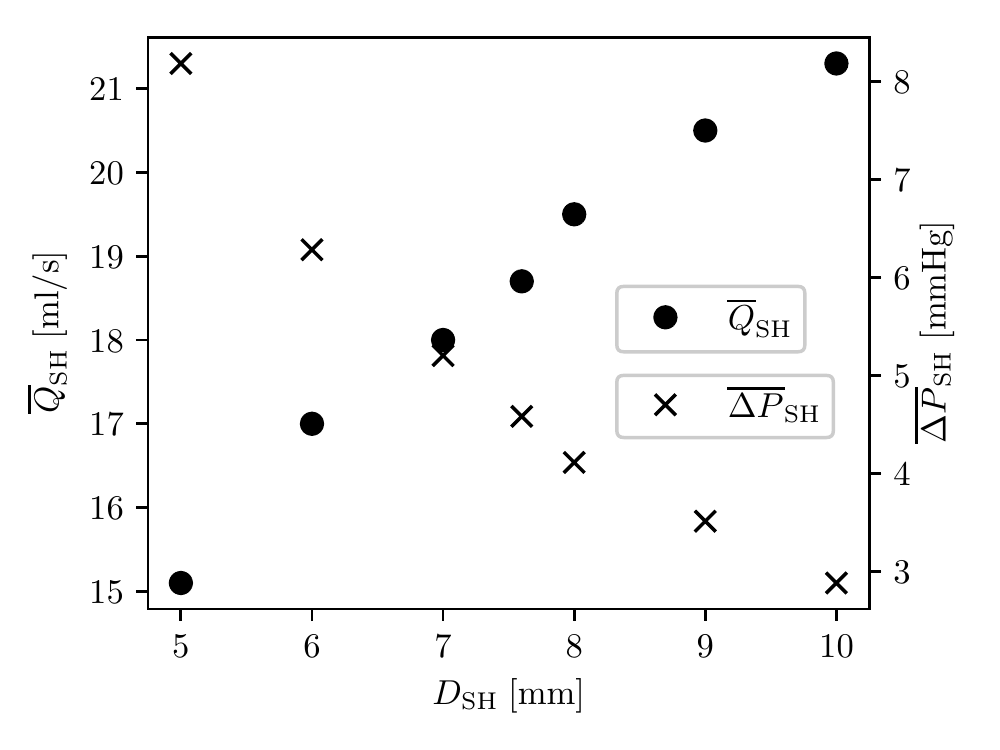}
\caption{Average flow-rate through the shunt and pressure gradient across the shunt for varying shunt diameters.}
\label{fig:linear_qsh_psh_ms}
\end{subfigure}
\begin{subfigure}{0.45\textwidth}
\centering
\includegraphics[width=0.9\textwidth]{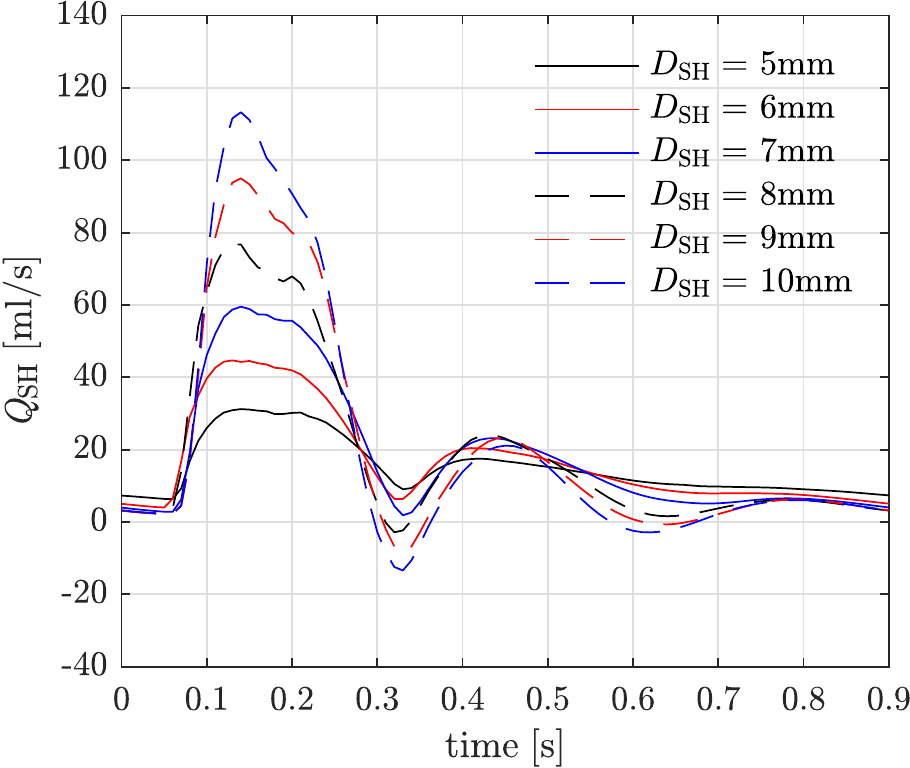}
\caption{Effect of shunt diameter on volumetric flow rate through the PS.}
\label{fig:Qshunt_dia_ms}
\end{subfigure}
\label{fig:combined_qsh_linear_ms}
\caption{GMM: effect of varying shunt diameter on flow-rate and pressure gradient across the shunt.}
\end{figure}

\begin{figure}[tb]
\centering
\begin{subfigure}{0.45\textwidth}
\includegraphics[width=1.0\textwidth]{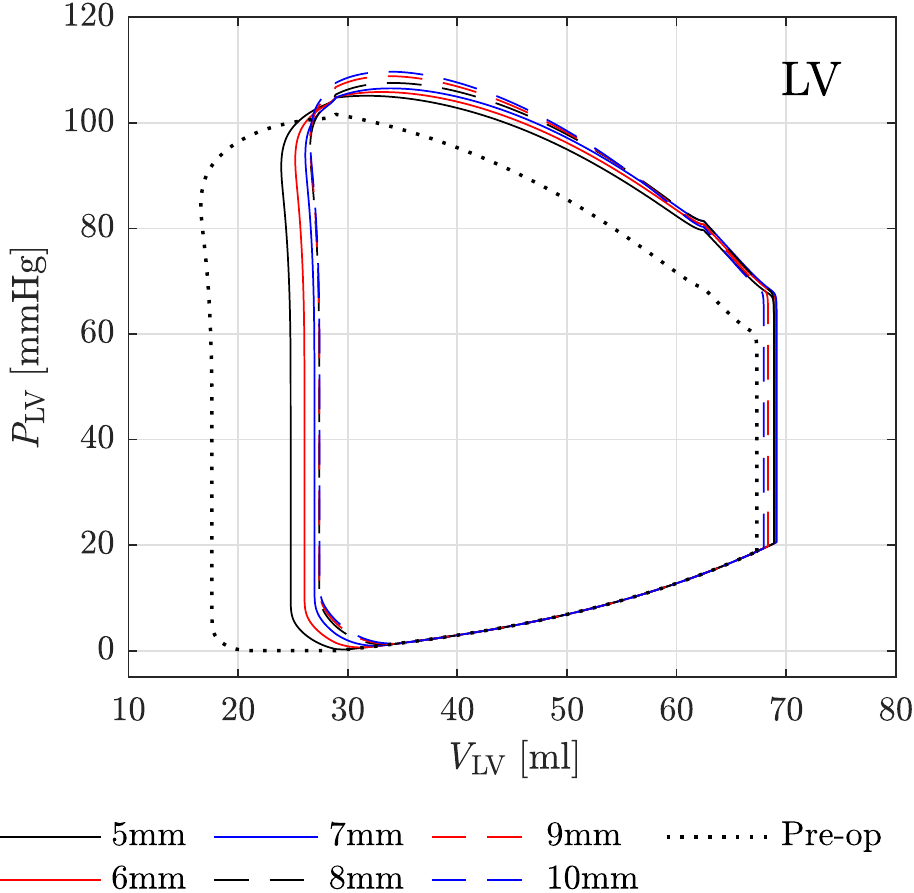}
\caption{Left ventricle}
\end{subfigure}
\hfill
\begin{subfigure}{0.45\textwidth}
\includegraphics[width=1.0\textwidth]{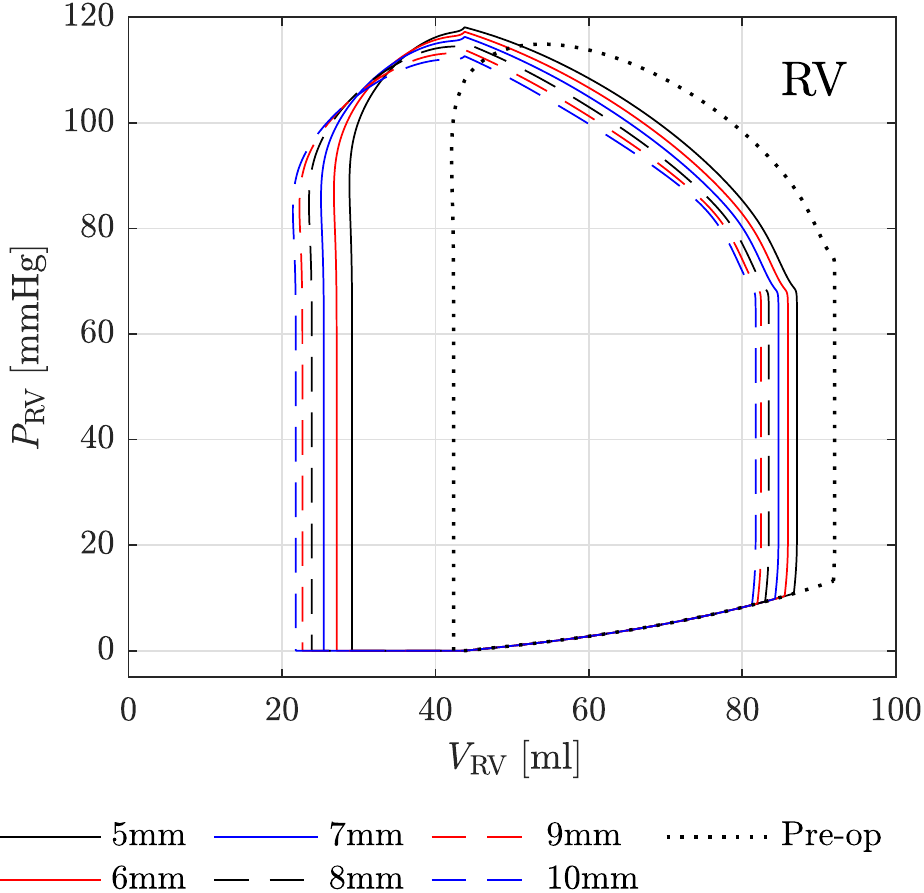}
\caption{Right ventricle}
\end{subfigure}
\caption{GMM: effect of shunt diameter on pressure-volume loops for (a) left ventricle and (b) right ventricle.}
\label{fig:PVloopLVRV_AllDia_ms}
\end{figure} 

\begin{figure}[t]
\centering
    \begin{subfigure} [c] {0.24\textwidth}
      \centering
      \caption*{$D_{\mathrm{SH}}=5$mm}
    \end{subfigure}
    \begin{subfigure} [c] {0.24\textwidth}
      \centering
      \caption*{$D_{\mathrm{SH}}=10$mm}
    \end{subfigure}\\
    \begin{subfigure} {1.0\textwidth}
      \centering
      \includegraphics[width = 0.24\textwidth]{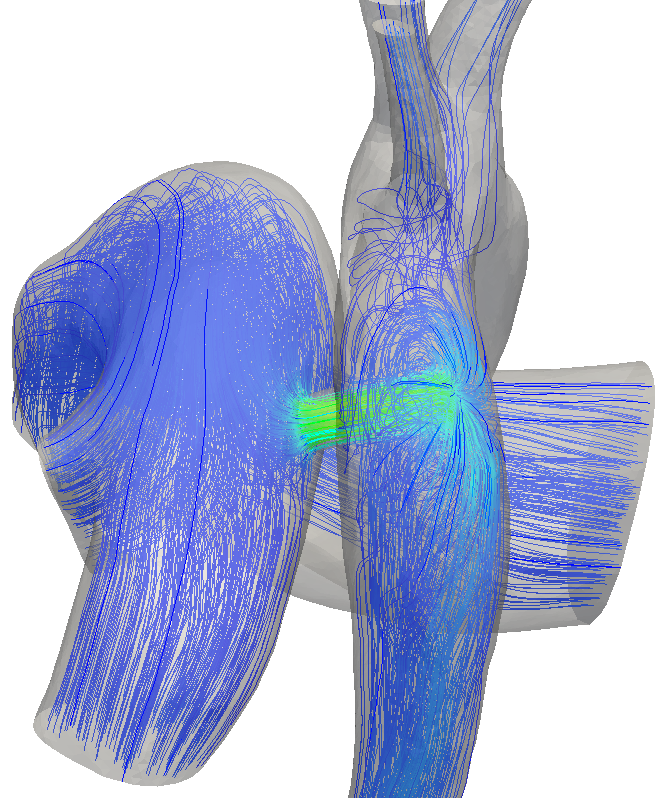}
      \includegraphics[width = 0.24\textwidth]{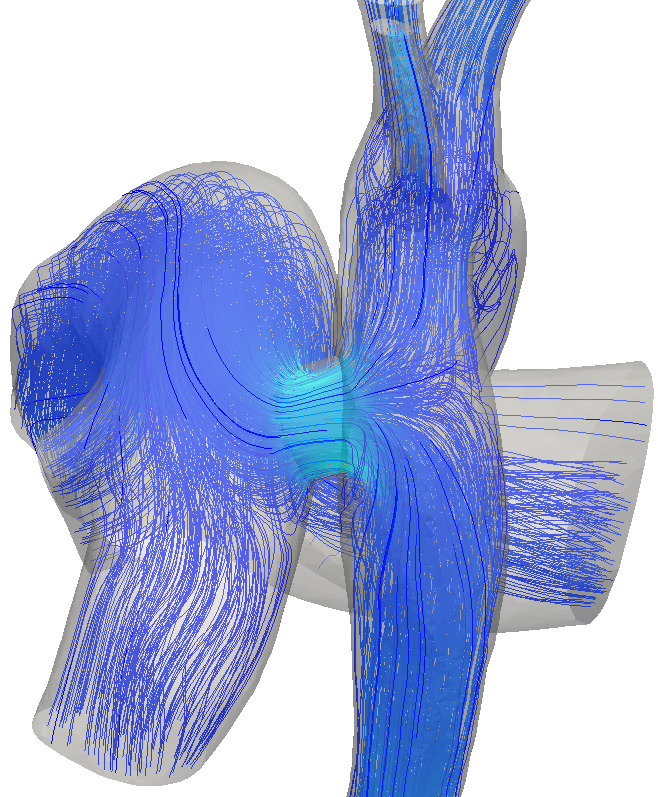}
      \caption{time = 0.09 s}
\end{subfigure}\\
    \begin{subfigure} {1.0\textwidth}
    \centering
      \includegraphics[width = 0.24\textwidth]{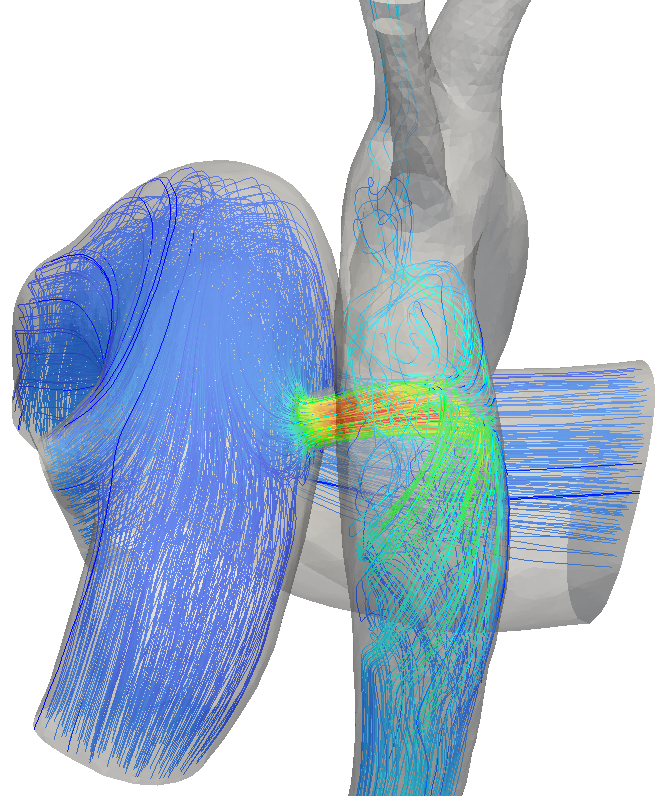}
      \includegraphics[width = 0.24\textwidth]{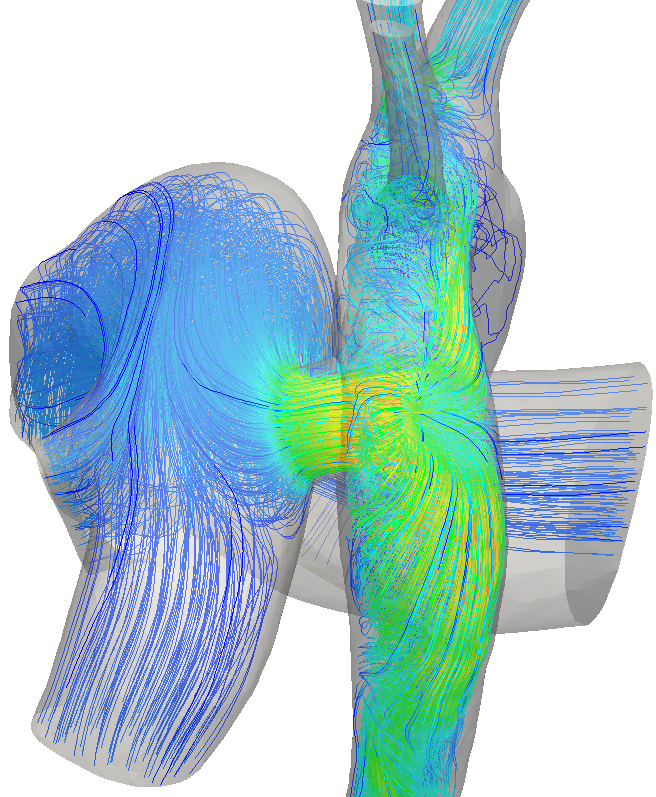}
      \caption{time = 0.15 s}
      \end{subfigure}\\[5pt]
   \begin{subfigure} [c] {1.0\textwidth}
          \centering
      \includegraphics[scale=0.45]{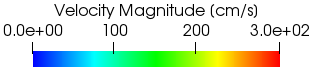}
    \end{subfigure}\\
\caption{Velocity streamlines for 5 mm and 10 mm diameter shunts.}
\label{fig:Streamlines5mm10mm}
\end{figure}

\begin{figure}
    \begin{subfigure} [c] {0.48\textwidth}
      \centering
      \caption*{Surface Pressure [mmHg]}
    \end{subfigure}
    \begin{subfigure} [c] {0.48\textwidth}
      \centering
      \caption*{Wall Shear Stress [Pa]}
    \end{subfigure}\\
    \begin{subfigure} [c] {0.24\textwidth}
      \centering
      \caption*{$D_{\mathrm{SH}}=5$mm}
    \end{subfigure}
    \begin{subfigure} [c] {0.24\textwidth}
      \centering
      \caption*{$D_{\mathrm{SH}}=10$mm}
    \end{subfigure}
    \begin{subfigure} [c] {0.24\textwidth}
      \centering
      \caption*{$D_{\mathrm{SH}}=5$mm}
    \end{subfigure}
    \begin{subfigure} [c] {0.24\textwidth}
      \centering
      \caption*{$D_{\mathrm{SH}}=10$mm}
    \end{subfigure}\\
    \begin{subfigure} {1.0\textwidth}
      \centering
      \includegraphics[width = 0.24\textwidth]{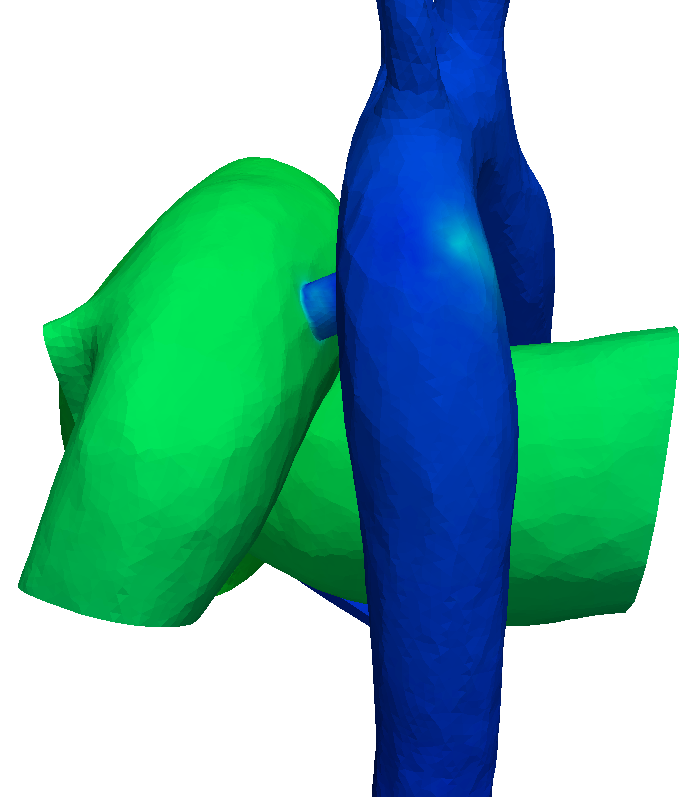}
      \hfill
      \includegraphics[width = 0.24\textwidth]{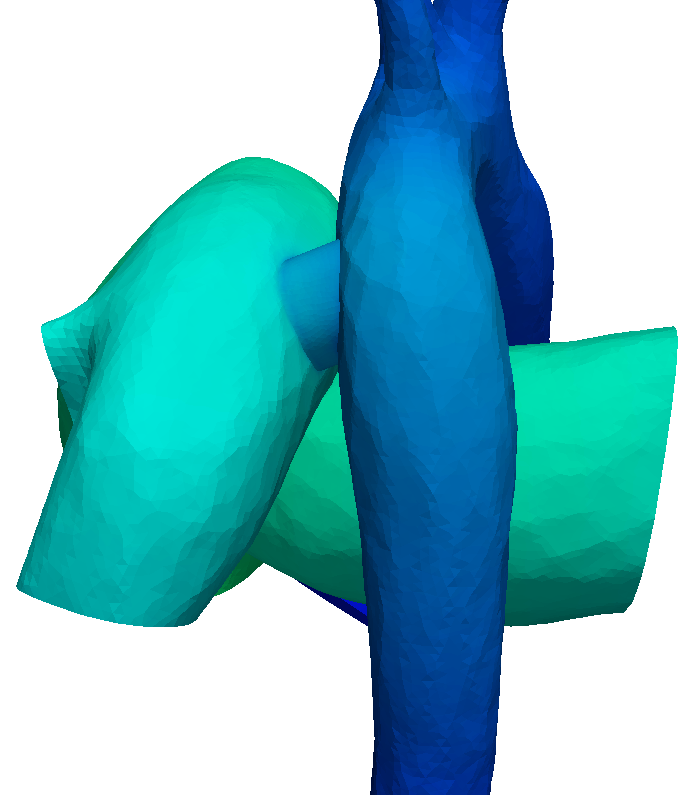}
      \hfill
      \includegraphics[width = 0.24\textwidth]{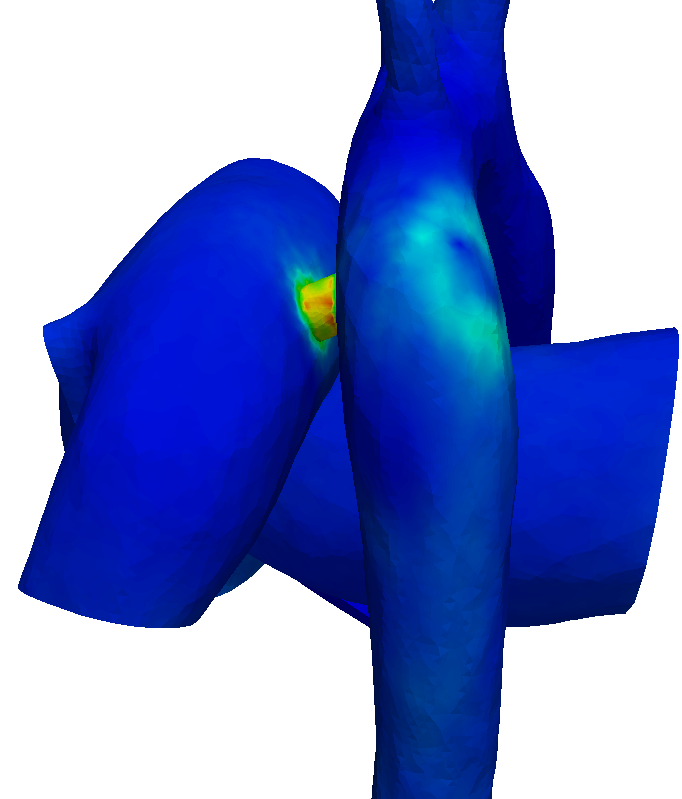}
           \hfill
      \includegraphics[width = 0.24\textwidth]{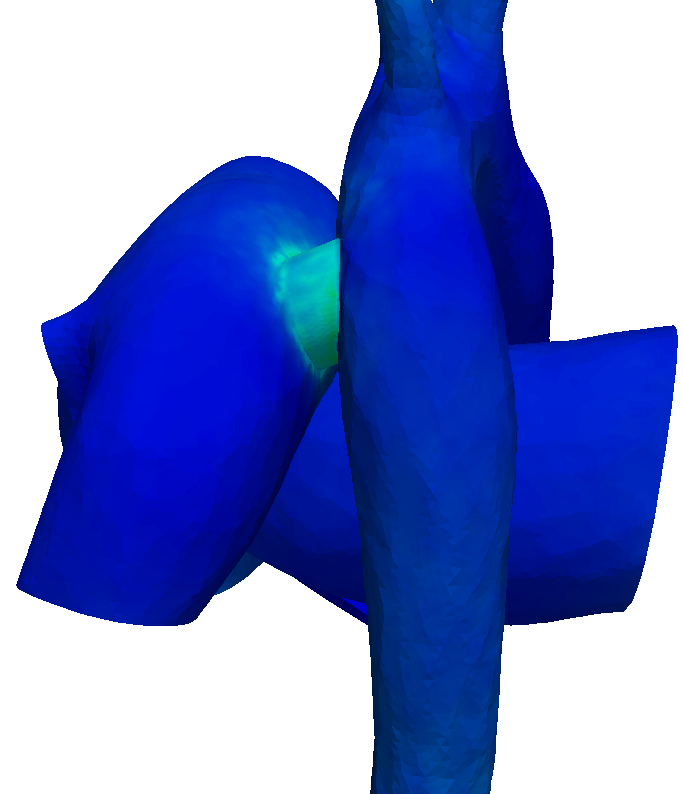}
    \caption{time = 0.09 s }    
    \end{subfigure}
    \begin{subfigure} {1.0\textwidth}
      \centering
      \includegraphics[width = 0.24\textwidth]{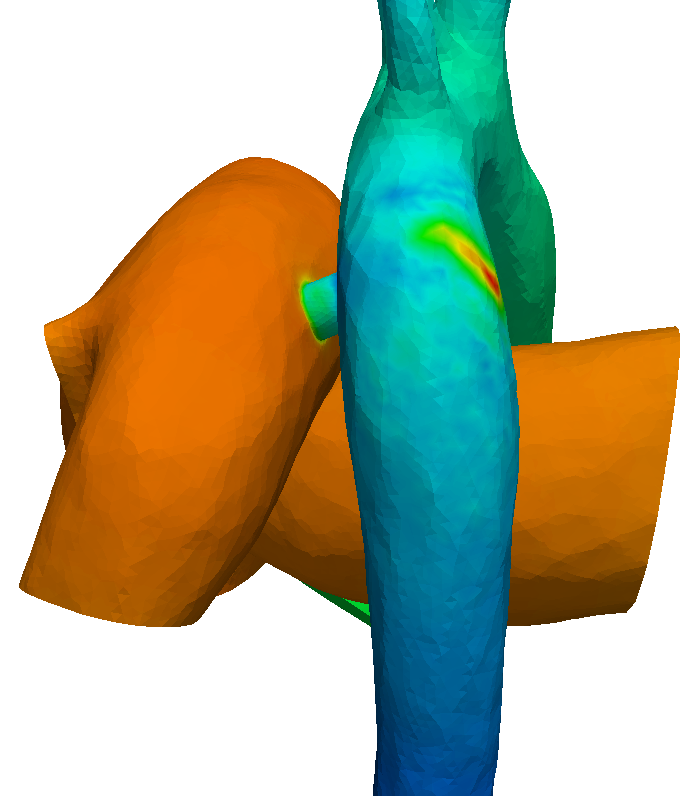}
      \hfill
      \includegraphics[width = 0.24\textwidth]{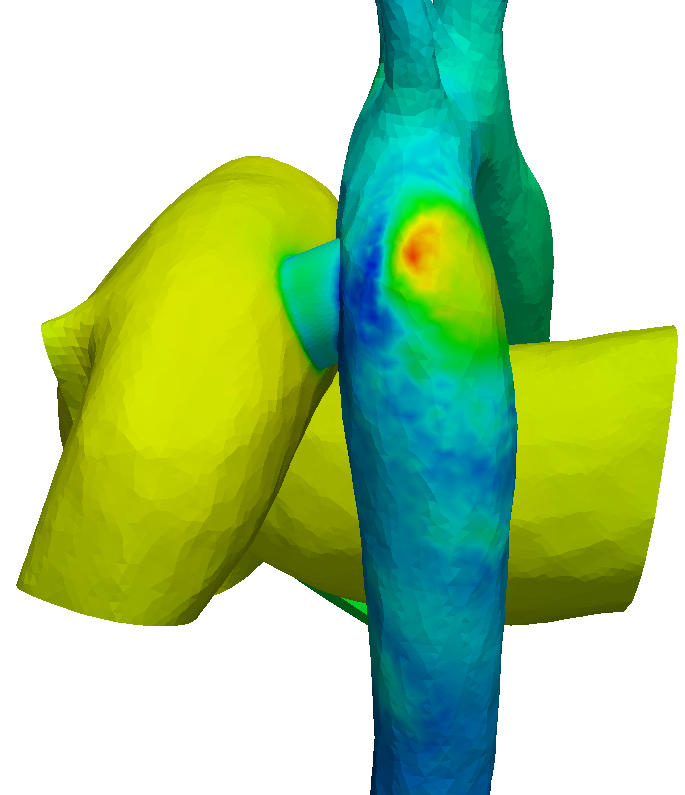}
      \hfill
      \includegraphics[width = 0.24\textwidth]{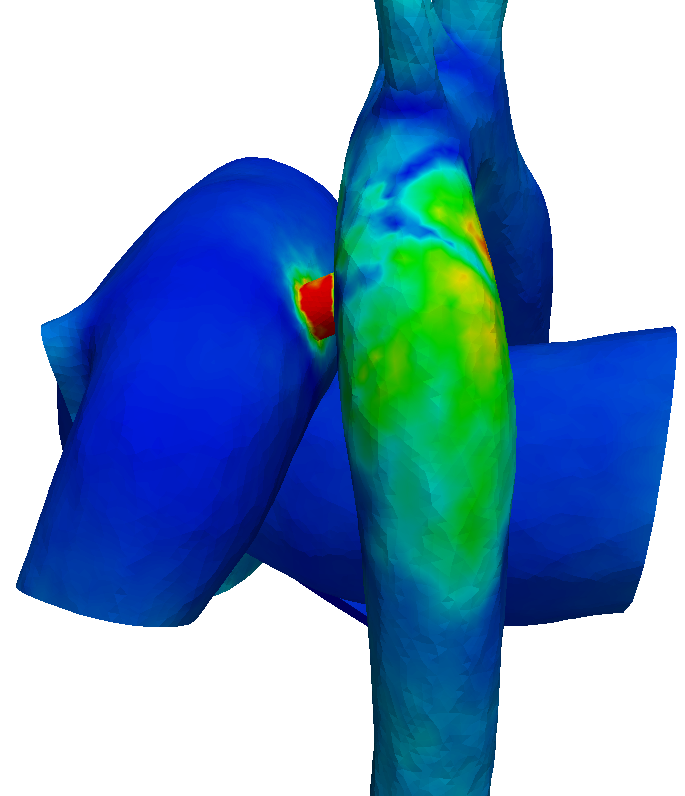}
           \hfill
      \includegraphics[width = 0.24\textwidth]{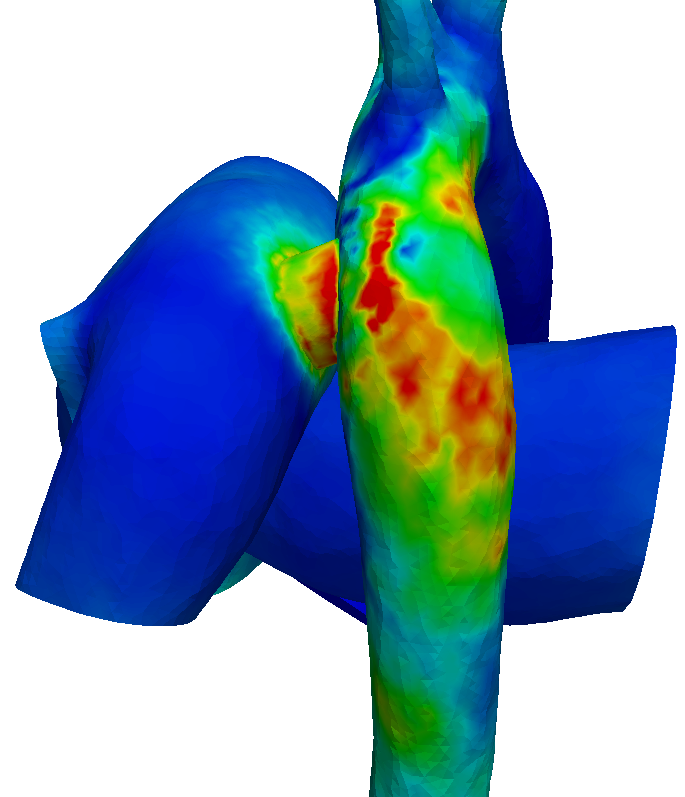}
    \caption{time = 0.15 s }    
    \end{subfigure}
      \begin{subfigure} {1.0\textwidth}
      \centering
      \includegraphics[width = 0.24\textwidth]{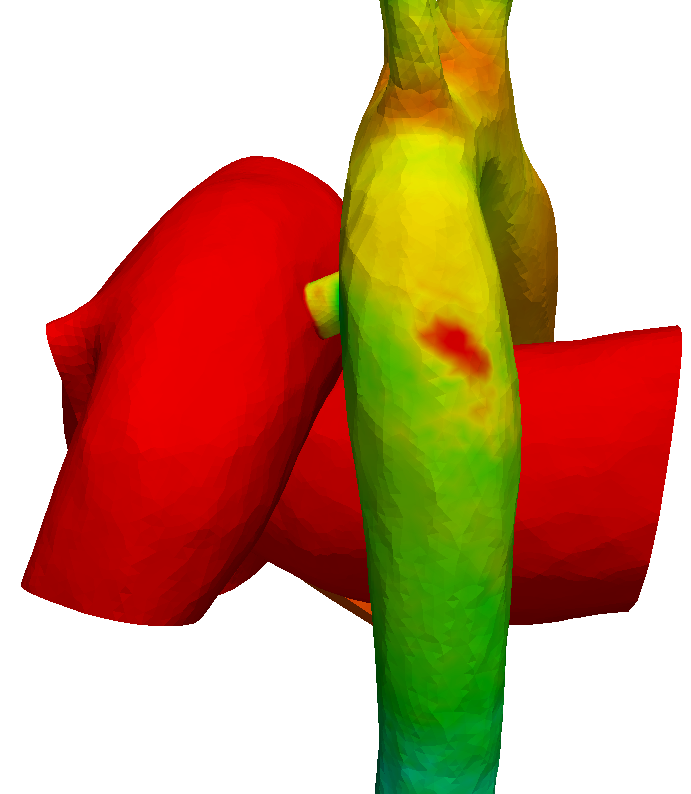}
      \hfill
      \includegraphics[width = 0.24\textwidth]{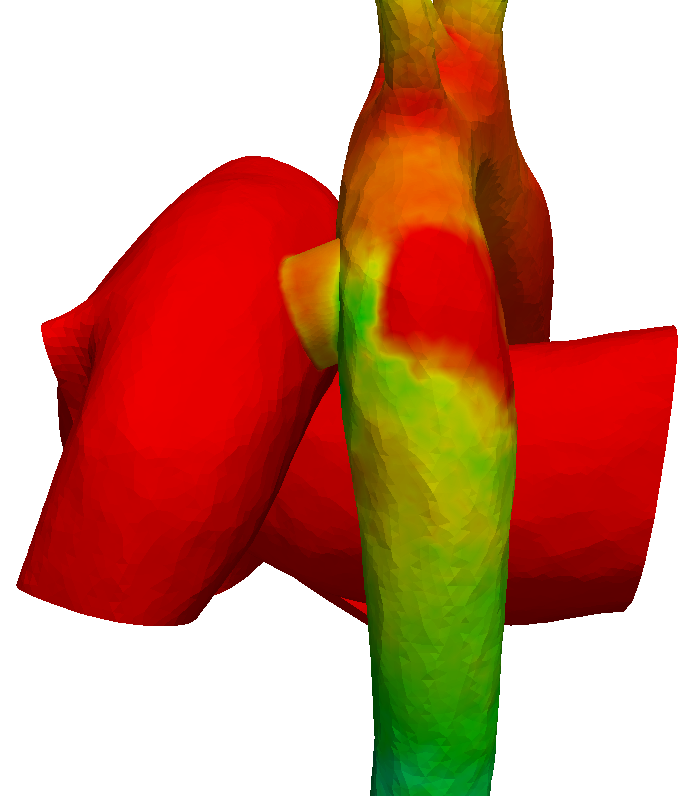}
      \hfill
      \includegraphics[width = 0.24\textwidth]{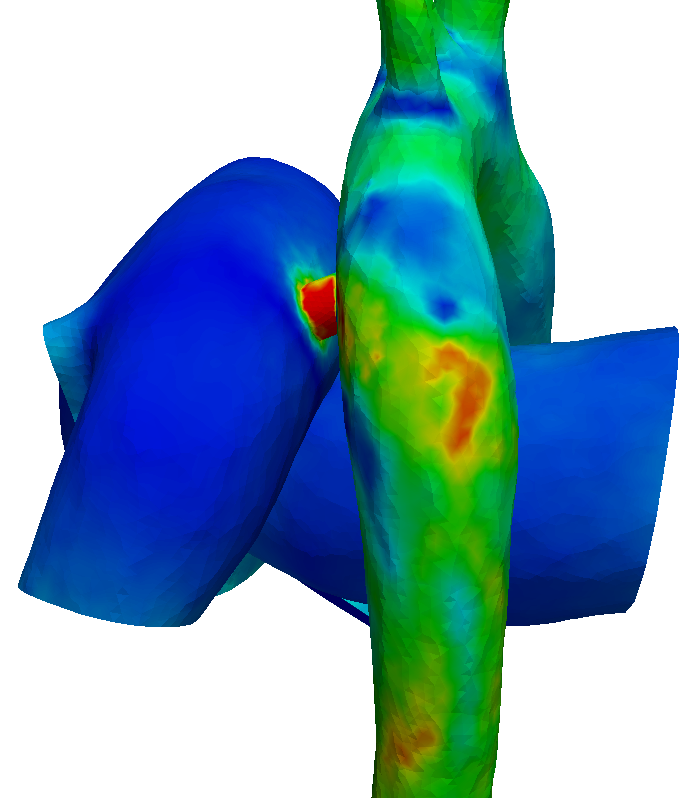}
           \hfill
      \includegraphics[width = 0.24\textwidth]{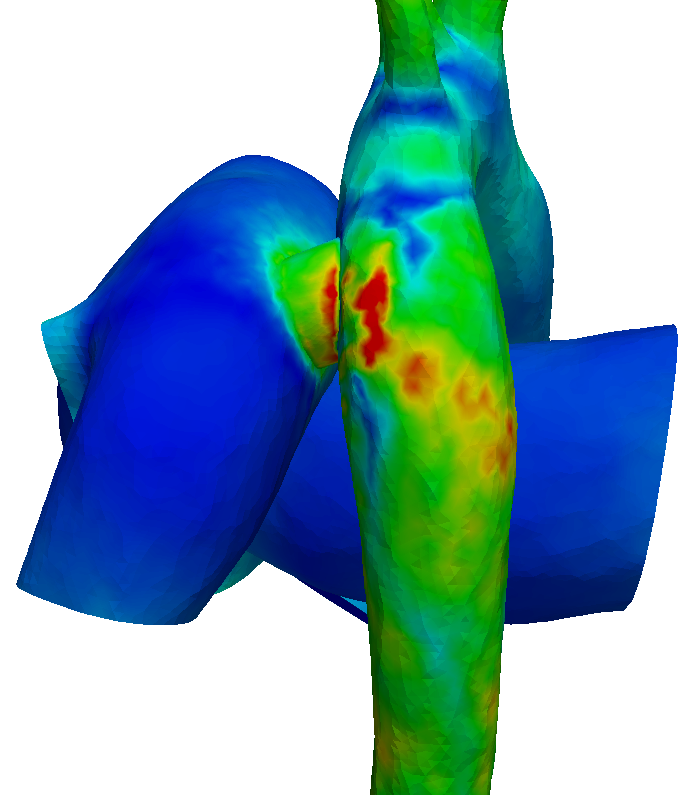}
    \caption{time = 0.22 s }    
    \end{subfigure}
      \begin{subfigure}{0.49\textwidth}
          \centering
      \includegraphics[scale=0.5]{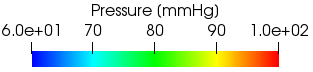}
    \end{subfigure}
          \begin{subfigure}{0.49\textwidth}
          \centering
      \includegraphics[scale=0.5]{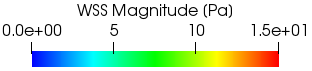}
    \end{subfigure}
\caption{Surface pressure and wall shear stress (WSS) for 5 mm and 10 mm diameter shunts at 0.09 s, 0.15 s and 0.22 s.}
\label{fig:VelPressWSS_Dia}
\end{figure}

\begin{table}
\begin{center}
\def\arraystretch{1.3}
\resizebox{1.0\textwidth}{!}{ 
\begin{tabular}{|c|ccc|ccc|cc|cc|cc|cc|}
\hline
  & \multicolumn{3}{|c|}{$P_{\rm{AAo}}$ [mmHg]} & \multicolumn{3}{|c|}{$P_{\rm{mPA}}$ [mmHg]} & \multicolumn{2}{|c|}{EDV [ml]} & \multicolumn{2}{|c|}{SV [ml]} &\multicolumn{2}{|c|}{EF} & \multicolumn{2}{|c|}{CO [L/min]}\\
\hline
 Diameter [mm] & Systolic & Diastolic & Mean & Systolic & Diastolic & Mean & LV & RV & LV & RV & LV & RV & LV & RV\\
 \hline
pre-op & 95.4 & 52.5 & 65.4 & 111.1 & 68.3 & 84.6 & 67.3 & 92.0 & 50.7 & 49.9 & 0.75 & 0.54 & 3.38 & 3.38\\ 
\hdashline
5 & 101.2 & 60.2 & 72.9 & 111.0 & 61.7 & 78.9 & 69.1 & 86.5 & 44.9& 58.2 & 0.65 & 0.67 & 2.99 & 3.88\\
6 & 102.3 & 60.7 & 73.6 & 109.5 & 61.3 & 77.9 & 69.2 & 85.2 & 43.8 & 59.0 & 0.63 & 0.69 & 2.92 & 3.93\\
7 & 103.3 & 60.8 & 74.0 & 107.9 & 61.3 & 77.2 & 69.1 & 84.0 & 43.0 & 59.4 & 0.62 & 0.71 & 2.87 & 3.96\\
\hdashline
7.6 & 104.1 &	60.8 &	74.3 &	106.9 & 61.3 & 76.8 & 68.9 & 83.4 & 42.7 & 59.6 & 0.62 & 0.71 & 2.85 & 3.97\\
\hdashline
8 & 104.6 & 60.8 & 74.4 & 106.4 & 61.3 & 76.7 & 68.7 & 83.0 & 42.4 & 59.7 & 0.62 & 0.72 & 2.83 & 3.98\\
9 & 105.9 & 60.8 & 74.7 & 104.8 & 61.3 & 76.2 & 68.2 & 82.2 & 41.9 & 60.0 & 0.61 & 0.73 & 2.79 & 4.00\\
10 & 106.9 & 60.8 & 75.0 & 103.3 & 61.2 & 75.9 & 67.8 & 81.5 & 41.4 & 60.2 & 0.61 & 0.74 & 2.76 & 4.02\\
\hline
\end{tabular}
}
\caption{GMM: $P_{\rm{AAo}}$, $P_{\rm{mPA}}$, EDV, SV, EF and CO for different PS diameters.}
\label{tab:shuntdiaoverall_ms}
\end{center}
\end{table}

\subsection{Varying shunt diameter}
\label{res_diameter}
Results for simulations of the pre-operative state and post-operative state with PS diameters varying between 5mm and 10mm are presented in Tables \ref{tab:shuntdiaoverall_ms} and  \ref{tab:ventricle_work_ms_qsh}. The length of each PS  in these simulations is constrained by the distance between the LPA and DAo such that the PS does not protrude into either vessel. 
%
The aortic pressure increases and the PA pressure decreases with increasing shunt diameter (Table  \ref{tab:shuntdiaoverall_ms}). The systolic pressures are affected significantly more than the diastolic pressures. For example, when comparing the extreme diameters of 5mm to 10mm, the systolic $P_{\rm{AAo}}$ increases by nearly 6\% and the systolic $P_{\rm{mPA}}$ decreases by 7\%.  In contrast, the diastolic  pressures are largely unaffected---the diastolic $P_{\rm{AAo}}$ and $P_{\rm{mPA}}$ both vary by less than 1\%.  

With the smaller shunt diameters of 5 mm and 6 mm, the PS is not able to reduce the PA pressures to near systemic levels and the pulmonary-to-systemic systolic arterial pressure ratio only reduces from a pre-operative value of 1.16 to post-operative values of 1.1 and 1.07 for the 5 mm and 6 mm diameter shunts, respectively (Figure \ref{fig:PaoPpa_PSdia_ms}a,b). Increasing the shunt diameter to 7 mm and 8 mm, resulted in reduction of the PA pressures to near systemic levels, with a post-operative pulmonary-to-systemic systolic arterial pressure ratio of 1.04 and 1.02, respectively (Figure \ref{fig:PaoPpa_PSdia_ms}c,d).  With further increments in shunt diameter to 9 mm and 10 mm, more blood is pumped from the pulmonary circulation to the DAo resulting in the $P_{\rm{AAo}}$ exceeding the $P_{\rm{mPA}}$ in some part of systole (Figure \ref{fig:PaoPpa_PSdia_ms}e,f).  The mean pressure gradient across the shunt decreases and the flow-rate through the shunt increases with increasing shunt diameters (Table \ref{tab:ventricle_work_ms_qsh}). Both of these show a near-linear variation in the range of diameters tested (Figure \ref{fig:linear_qsh_psh_ms}). 
The volumetric flow-rate profiles through the shunt differ for varying diameters (Figure \ref{fig:Qshunt_dia_ms}). For a diameter of 7 mm, the flow rate is positive throughout the cardiac cycle.  However, increasing the shunt diameter to 8 mm results in a small period of reversed flow, which is amplified with further increments in diameter. 
 
The ventricular PV loops show that the LV EDV remains almost constant, while its ESV increases as the shunt diameter increases (Figure \ref{fig:PVloopLVRV_AllDia_ms}a). As a result, the LV output progressively decreases with increasing shunt diameters (Table \ref{tab:shuntdiaoverall_ms}). This reduction in LV cardiac output leads to successive worsening of LV ejection fraction. For the RV, both EDV and ESV decrease with increasing shunt diameters (Figure \ref{fig:PVloopLVRV_AllDia_ms}b). Decrease in ESV is, however, larger than that in EDV resulting in larger stroke volumes, and hence higher RV outputs and improving RV ejection fractions (Table \ref{tab:shuntdiaoverall_ms}). 
An important observation from the PV loops is the change in work performed by the ventricles. The area enclosed by the PV loop can be seen as a measure of the ventricular stroke work, which is a product of the stroke volume and the mean ventricular pressure. 
The work done by the LV after PS creation decreases with increasing shunt diameters (consistent with decreasing  stroke volume pumped, however, against higher aortic pressures), while the work done by the RV increases with increasing shunt diameters up to a certain point and slightly decreases for the larger diameters (consistent with increasing stroke volume albeit against lower PA pressures) (Table \ref{tab:ventricle_work_ms_qsh}). The latter is an important observation from the point of view of RV dynamics after PS creation: together with the reduction in RV pressures due to increasing shunt diameters, the work done by the RV mostly increases due to increased volumes pumped by the ventricle \cite{delhaas2018potts}. However, in contrast to this increase in the RV stroke volume, its EDV decreases due to the PS creation, and the EDV decrease is larger for larger shunt diameters.

The 3D assessment of flow with varying PS diameters is performed for the extreme diameters of 5mm and 10mm.  Figure \ref{fig:Streamlines5mm10mm} shows velocity streamlines originating from the mPA for these two cases at two time points in systole. The corresponding surface pressures and wall shear stresses are similarly shown in Figure \ref{fig:VelPressWSS_Dia}.
One can see that flow complexity is higher with higher diameter shunts. 
The larger PS diameter results in significantly higher flow through the shunt, but this higher flow keeps the maximum velocities comparatively less affected due to the diameter increase (Figure \ref{fig:Streamlines5mm10mm}). Higher shunt diameters lead to higher pressures in the aorta, and for high diameters---9mm and 10mm shunts as seen in Figure  \ref{fig:PaoPpa_PSdia_ms}e,f--- the AAo pressure can become equal or even exceed the mPA pressure during periods of systole (also see Figure \ref{fig:VelPressWSS_Dia}c). Smaller shunt diameters lead to smaller localised regions of high pressure on the DAo wall, due to the directed and more focused shunt flow jet of smaller size (Figure \ref{fig:Streamlines5mm10mm}b).
Finally, changing the shunt diameter does not have a significant effect in reducing regions of abnormally high WSS (Figure \ref{fig:VelPressWSS_Dia}).

\begin{table}[tb]
\centering
\def\arraystretch{1.2}
\resizebox{0.57\textwidth}{!}{ 
\begin{tabular}{c|cc|cccc}
\hline
  $D_{\rm{SH}}$ & $W_{\rm{LV}}$ & ${W}_{\rm{RV}}$  & $ \overline{Q}_{\rm{SH}}$ & $\overline{V}_{\rm{SH}}$ & $\overline{\Delta P}_{\rm{SH}}$ & $Q_p/Q_s$ \\
    ${\text{[mm]}}$ & $\times 10^{3} {\text{[mmHg-ml]}}$ & $\times 10^{3} {\text{[mmHg-ml]}}$  & ${\text{[ml/s]}}$ & ${\text{[ml]}}$ & ${\text{[mmHg]}}$ & [-] \\
  \hline
    pre  & 4.13	& 4.95  & - & - & -  & 1\\
  \hdashline
  $5 $ & 3.83	& 5.80  & 15.1	& 13.6	& 8.18 & 0.78\\
  $6 $ & 3.75	& 5.86  & 17.0	& 15.3 	& 6.28 & 0.75\\
  $7 $ & 3.69	& 5.87  & 18.0	& 16.2	& 5.20 & 0.73\\
  \hdashline
  $7.6$& 3.66	& 5.87  & 18.7	& 16.9 & 4.58 & 0.72\\
  \hdashline
  $8$  & 3.65	& 5.87  & 19.5	& 17.6	& 4.11 & 0.72 \\
  $9$  & 3.64	& 5.85  & 20.5	& 18.5	& 3.51 & 0.70\\
  $10$ & 3.63	& 5.82  & 21.3	& 19.2	& 2.88 & 0.69\\
  \hline
\end{tabular}
}
\caption{GMM: work done (area under the PV-loop) by the ventricles and mean volumetric flow rate, $ \overline{Q}_{\rm{SH}}$, total volume displaced in a cardiac cycle, $\overline{V}_{\rm{SH}}$, and pressure drop, $\overline{\Delta P}_{\rm{SH}}$, across the shunt for different PS diameters.}
\label{tab:ventricle_work_ms_qsh}
\end{table}

\begin{figure}[tb]
\centering
\begin{subfigure}{0.45\textwidth}
\includegraphics[width=1.0\textwidth]{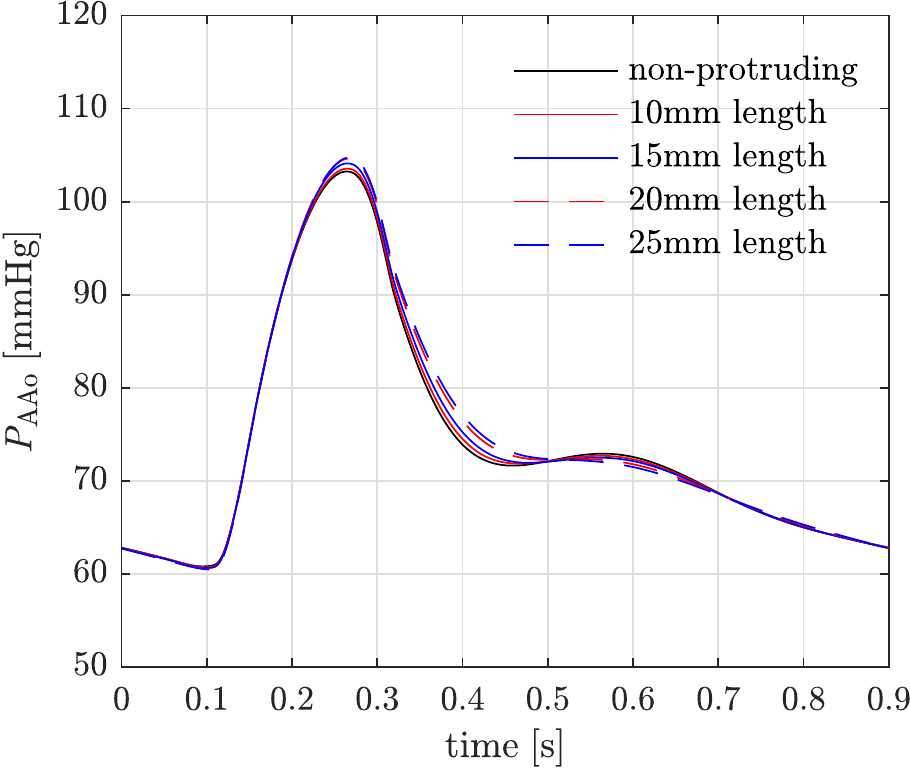}
\caption{Aorta}
\end{subfigure}
\hfill
\begin{subfigure}{0.45\textwidth}
\includegraphics[width=1.0\textwidth]{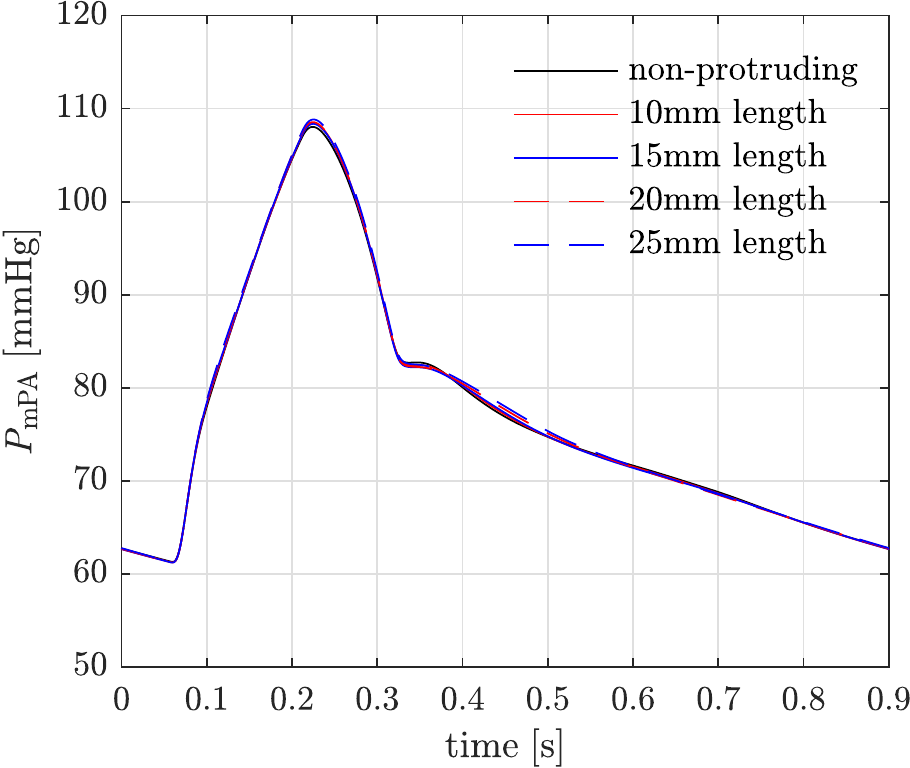}
\caption{Pulmonary artery}
\end{subfigure}
\caption{GMM: effect of shunt length on $P_{\rm{AAo}}$ and $P_{\rm{mPA}}$.}
\label{fig:PaoPpa_Profile_Lsh}
\end{figure}

\begin{figure}[tb]
\centering
\includegraphics[width=0.45\textwidth]{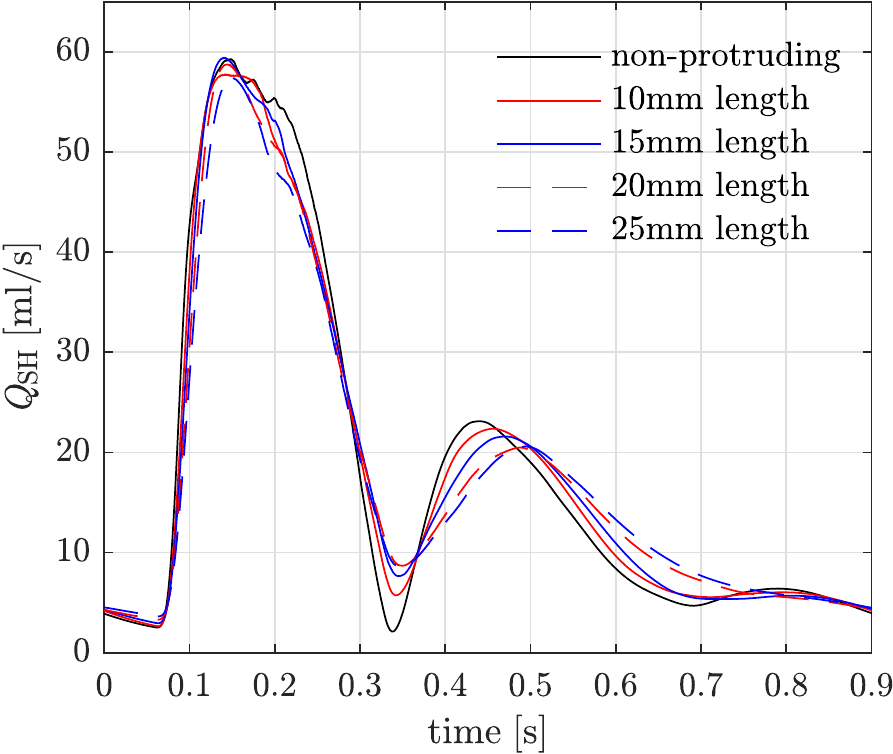}
\caption{GMM: effect of shunt length on flow across the shunt $Q_{\rm{SH}}$.}
\label{fig:Qda_Lsh}
\end{figure} 

\begin{figure*}
    \begin{subfigure} [c] {0.24\textwidth}
      \caption*{$L_{\mathrm{SH}}=10$mm}
    \end{subfigure}
   \begin{subfigure}[c] {0.24\textwidth}
     \centering
      \caption*{$L_{\mathrm{SH}}=15$mm}
    \end{subfigure}
   \begin{subfigure}[c] {0.24\textwidth}
         \centering
      \caption*{$L_{\mathrm{SH}}=20$mm}
    \end{subfigure}
   \begin{subfigure}[c] {0.24\textwidth}
             \centering
      \caption*{$L_{\mathrm{SH}}=25$mm}
    \end{subfigure}\\
       \begin{subfigure}{1.0\textwidth}
         \centering
      \includegraphics[width = 0.24\textwidth]{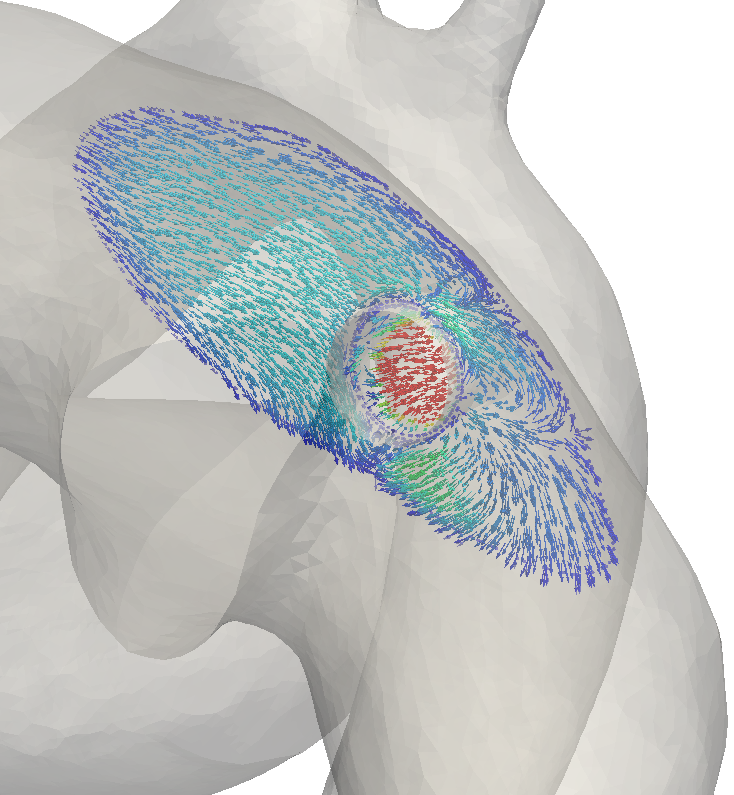}
\hfill
      \includegraphics[width = 0.24\textwidth]{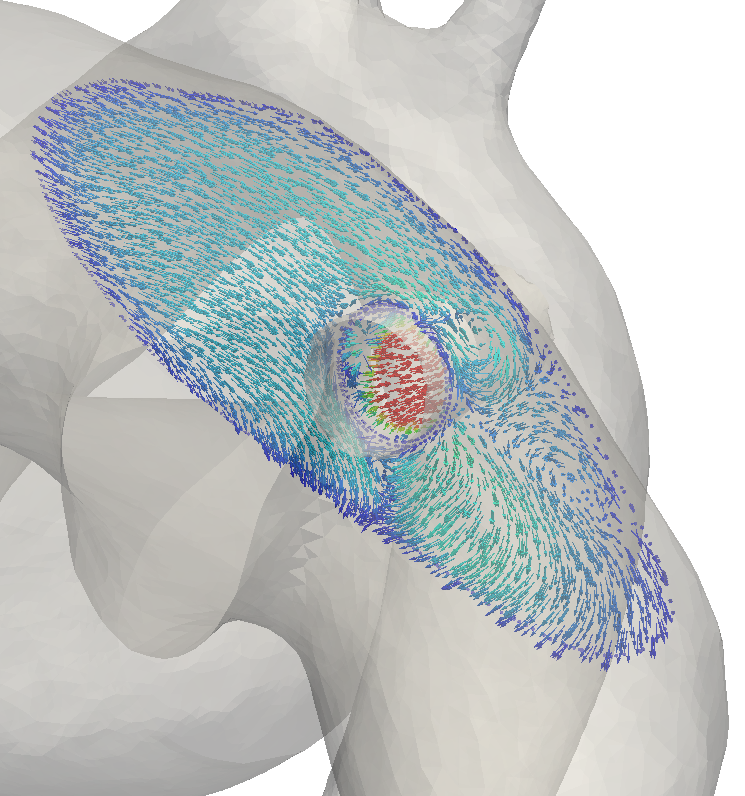}
\hfill
      \includegraphics[width = 0.24\textwidth]{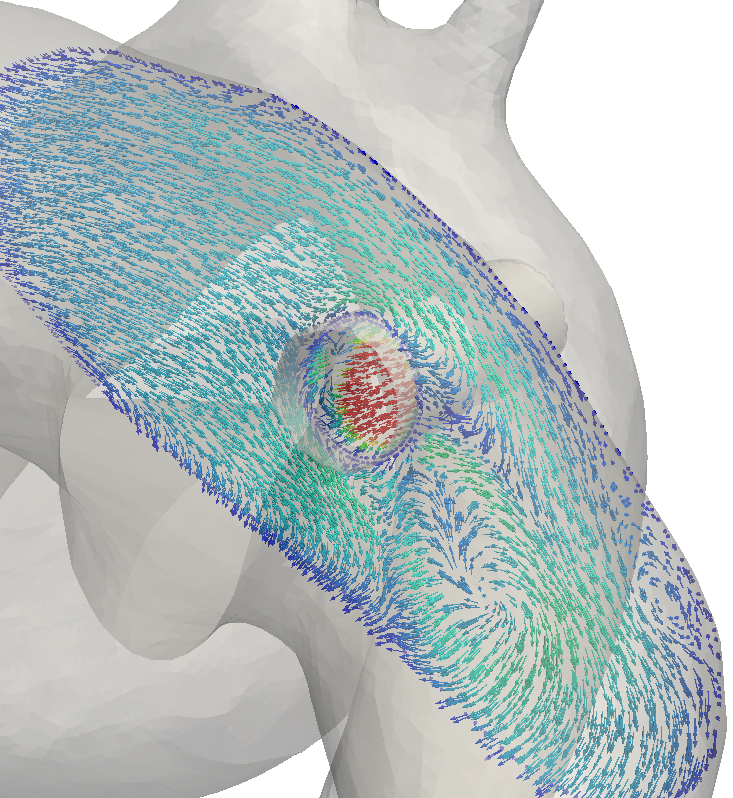}
\hfill
      \includegraphics[width = 0.24\textwidth]{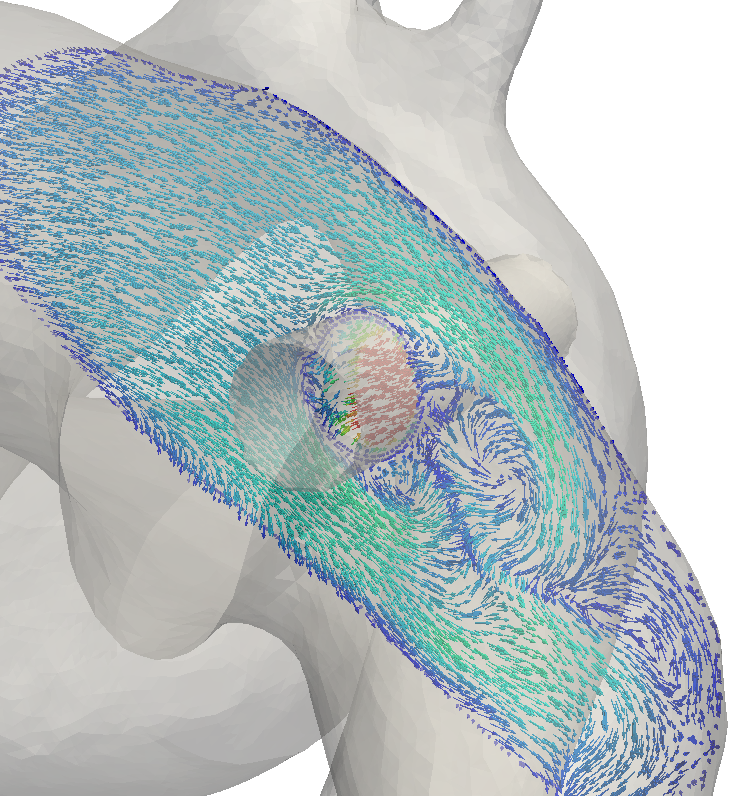}
      \caption{time = 0.25 s}
    \end{subfigure}\\
       \begin{subfigure}{1.0\textwidth}
        \centering
      \includegraphics[width = 0.24\textwidth]{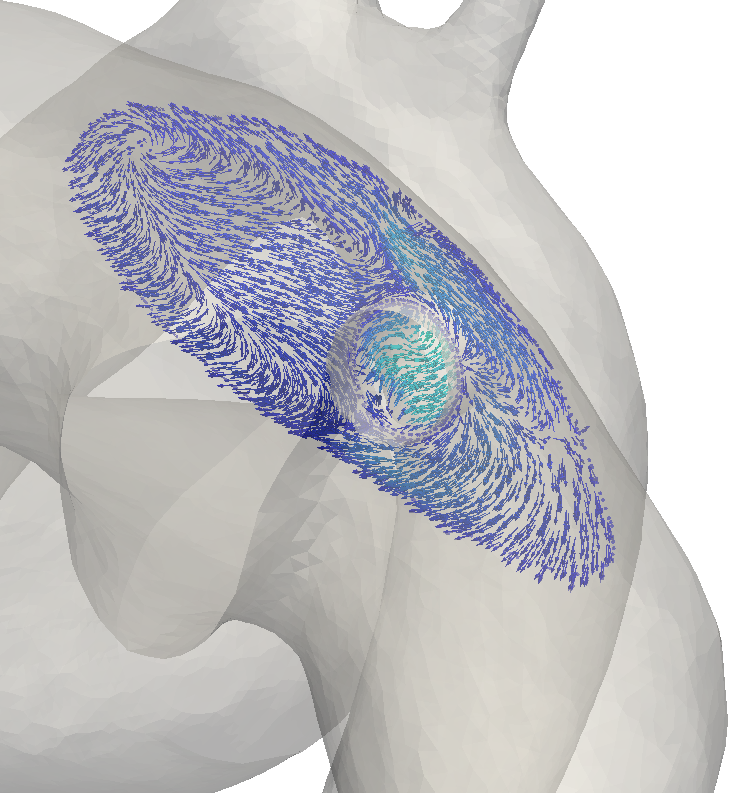}
\hfill
      \includegraphics[width = 0.24\textwidth]{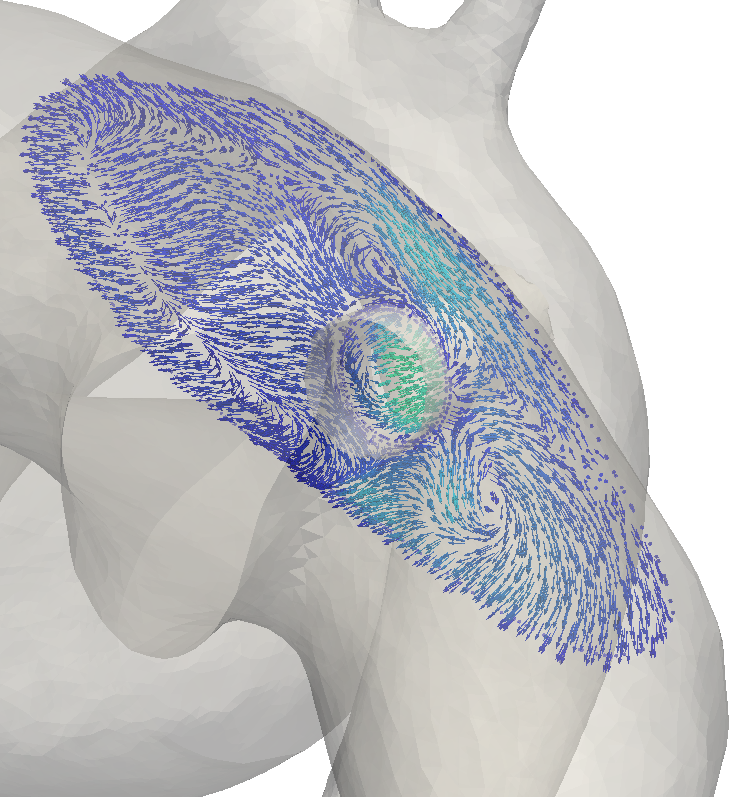}
\hfill
      \includegraphics[width = 0.24\textwidth]{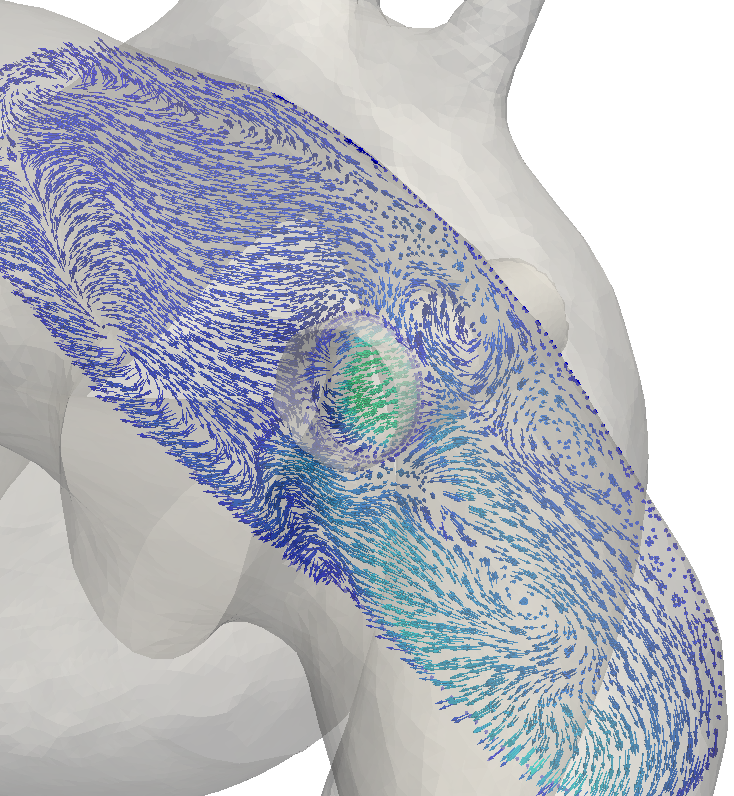}
\hfill
      \includegraphics[width = 0.24\textwidth]{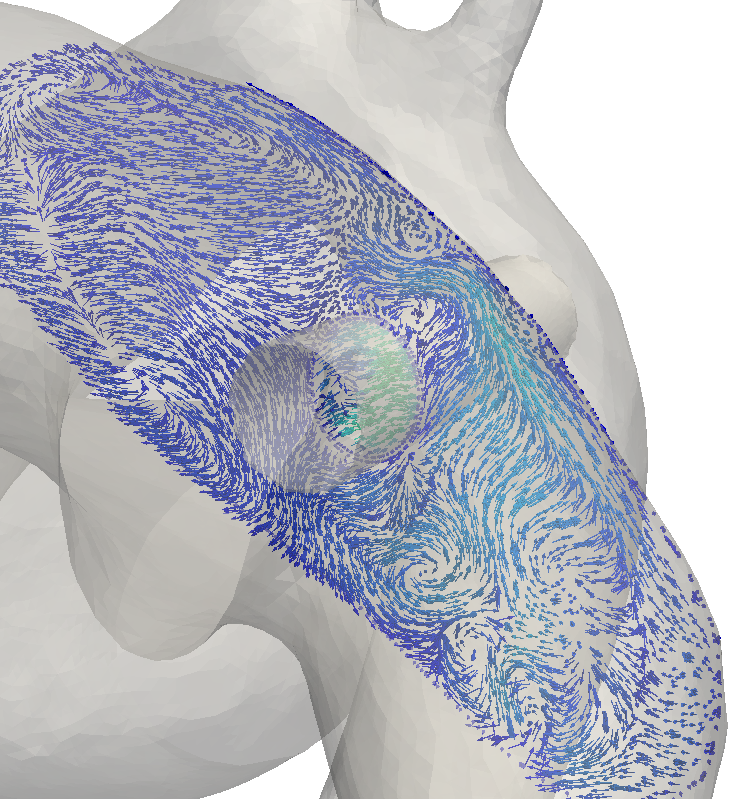}
      \caption{time = 0.35 s}
    \end{subfigure}\\[5pt]
      \begin{subfigure}{1.0\textwidth}
          \centering
      \includegraphics[scale=0.45]{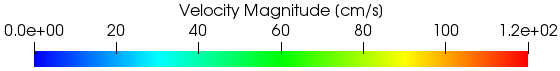}
    \end{subfigure}
    \caption{Flow field visualisation in the PA for different shunt lengths at 0.3--0.45 s.}
    \label{fig:vonkarman}
    \end{figure*}

\begin{table}[tb]
\centering
\def\arraystretch{1.2}
\resizebox{0.70\textwidth}{!}{ 
\begin{tabular}{c|ccc|ccc}
\hline
  & \multicolumn{3}{c}{$ P_{\rm{AAo}}$} & \multicolumn{3}{|c}{$P_{\rm{mPA}}$}\\
  \hline
  Length & Systolic & Diastolic & Mean & Systolic & Diastolic & Mean \\
    ${\text{[mm]}}$ & ${\text{[mmHg]}}$ & ${\text{[mmHg]}}$ & ${\text{[mmHg]}}$ & ${\text{[mmHg]}}$ & ${\text{[mmHg]}}$ & ${\text{[mmHg]}}$ \\
  \hline
$\approx$ 4.2   (no protrusion)& 103.3 & 60.8 & 74.0 & 107.9 & 61.3 & 77.2\\ 
10 & 103.6 & 60.7 & 74.1 & 108.4 & 61.2 & 77.2\\
15 & 104.1 & 60.6 & 74.3 & 108.4 & 61.2 & 77.2\\
20 & 104.7 & 60.5 & 74.5 & 108.5 & 61.2 & 77.3\\
25 & 104.8 & 60.5 & 74.6 & 108.8 & 61.2 & 77.5\\
  \hline
\end{tabular}
}
\caption{$ P_{\rm{AAo}}$ and $P_{\rm{mPA}}$ for 7 mm diameter PS with different lengths.}
\label{tab:PaoPpa_Lsh}
\end{table}

\subsection{Varying length of the communicating stent (protrusion)}
\label{res_length}
The results for changing length of the stent used to create the PS leading to protrusion into the LPA and the DAo are presented in Table \ref{tab:PaoPpa_Lsh}, which shows the PA and aortic pressures for shunts with lengths of 10, 15, 20, and 25 mm against the non-protruding PS with length of approximately 4.2 mm. The shunt diameter for all the cases of varying lengths is kept constant at 7mm. Figure \ref{fig:PaoPpa_Profile_Lsh} shows the corresponding pressure traces in the AAo and mPA. Finally, Figure \ref{fig:Qda_Lsh} illustrates the flow-rate across the shunt during the cardiac cycle with varying shunt lengths. 
Interestingly, increasing the length of the communicating stent from 10mm up to 25mm does not result in substantial increase of the pressure gradient across the PS. Similarly, volumetric flow-rate through the shunt with increasing length is also not substantially affected with the only minor changes observed in the diastole when the shunt flow-rates are small. These observations imply that at sufficiently large shunt diameter, the global haemodynamics is not significantly affected by the shunt length as it does not affect the resistance to fluid flow in the principal directions.

For the 3D assessment of flow, velocity vectors in the PA in a plane perpendicular to the PS are considered. At peak systole ($t\approx0.25$s) and peak deceleration phase ($t\approx0.35$s), these velocity vectors for all the four shunt lengths are shown in Figure \ref{fig:vonkarman}. It shows that as the protruded areas of the stent interfere with the flow in the LPA: recirculating zones of flow are formed in the  downstream region of the protrusion. Blood velocity in the eye of such von K\'{a}rm\'{a}n-type of vortices is low and these recirculating regions are larger and higher in number as the stent's protrusion length increases. Thus, while the global flow features are not affected significantly, the local flow features, as demonstrated by the 3D assessment, are substantially altered by the protrusion length.

\subsection{Stand-alone LPM output}
\label{res_lpm}
The stand-alone LPM also reliably predicts the global haemodynamics and its changes due to PS creation. While the GMM results are of higher fidelity, and are therefore used in the main text to assess physiology, the stand-alone LPM model results are of comparable, albeit of slightly lower, fidelity. In order to avoid overcrowding of the main text, the results from the stand-alone LPM model are presented separately in Appendix \ref{app:lpm}. Importantly, with the exception of the local flow-features and assessment of varying shunt lengths, the stand-alone LPM accurately predicts all other haemodynamic parameters, including those that vary due to changes in shunt diameter.

\section{Discussion}

\subsection{Model validation}
\label{dis_validation}

LPMs, known interchangeably as a zero-dimensional (0D) models, are frequently employed to analyse global haemodynamics in the circulatory system over a range of physiological and pathological conditions \citep{shi2011review}.  However, analysing complex local flow fields, typically associated with surgical/interventional procedures, often necessitates a more detailed geometric model in the region of interest.  Hence, two closed-loop computational models have been developed: a stand-alone LPM; and a GMM consisting of a reconstructed patient-specific three-dimensional (3D) flow domain, coupled to an LPM. In this study, due to a lack of the complete set of time-varying traces of pressure and flow-rate measurements in the major arteries and veins, we adopt a largely manual approach to parameter estimation. Alternatively, if such measurements are available, it would be possible to employ automated parameter estimation methods, such as those based on data-assimilation methods \cite{pant2016data,pant2017inverse,pant2014methodological}, which can also account for measurement uncertainty.

The results of Section \ref{res_comp} show that when comparing the post-operative predictions with clinical measurements of the arterial pressures, the largest relative errors are in the diastolic phase, but the absolute differences do not exceed 10 mmHg. The larger disparities in the post-operative state may be attributed to insufficiently robust parameter estimation, largely due to an incomplete set of pre-operative measurements available for the patient, the inherent uncertainty in the clinical measurements, potentially non-unique set of model parameters,  and auto-regulatory/adaptive mechanisms, which may be present in the post-operative state and not accounted by the model. Divergences between the model output and clinical measurements may also be present due to lumped representations of the heart and valve models, with inherent approximations. Notably, a previous study for PS analysis in a non patient-specific manner \cite{delhaas2018potts} utilised the three-wall segment model for the heart \cite{lumens2009three}, which explicitly accounts for ventricular interactions through the inter-ventricular septum. These interactions are ignored in this study as septal bulging has been shown to have little effect on global haemodynamics \cite{lumens2010left}, but their inclusion may play a role in achieving better agreement of model output with post-operative measurements.
Lastly, explicit modelling of 3D fluid-structure interaction may also help in reducing the discrepancies between model output and measurements. 

Although many of the aforementioned limitations can be addressed with more sophisticated models, it will also magnify the problem of estimating additional parameters for these models, while clinical measurements remain limited. Overall, given that the GMM reproduces all the pre-operative clinical measurements within 5\% differences, and that differences in post-operative predictions versus the measurements are less than 20\%, the GMM presented here as well as the stand-alone LPM depicted in Appendix \ref{app:lpm} can be considered as a reasonably good model for assessing the effects of PS creation: apart from diastolic aortic pressure which surprisingly decreases in the measurements, both models predict variations in the same direction as the measurements.


\subsection{Haemodynamic changes induced by the Potts shunt}
\label{dis_haemodynamics}
The results of Section \ref{dis_haemodynamics} show that
the connection of the high-pressure PAs with the low-pressure aorta naturally results in reduction of PA pressures, increase of aortic pressures, and a commensurate flow from the LPA to the DAo through the shunt. The changes induced in global haemodynamics, see Figure \ref{fig:volumes_ms}, are best explained by consideration of the effective resistances faced by the LV and RV in the post-operative state.  
The RV, due to the bypass created by the PS, experiences a lower effective resistance (alternatively, a lower afterload), which results in a larger stroke volume generated by the RV, explaining its higher output. Between the left and the right pulmonary circulations, the reduction in resistance occurs in the left branch, explaining the increased flow in the proximal LPA section and a reduced flow in the distal LPA and the RPA. The flow volume through the shunt is, however, larger than the increase in the RV output with the difference occurring at the cost of flow reduction in the RPA. Effectively, the PS steals all the increase in the RV output and additionally some flow from RPA, thus resulting in decreased effective pulmonary blood flow despite higher RV output when compared to the pre-operative state. 
This apparently adverse effect of the PS, in terms of reduced return of oxygenated blood from the lungs to the LV, may be beneficial in the long-term as reduced flow through the pathological pulmonary vascular bed can potentially result in regression of the arterial wall media hypertrophy similar to situation seen after correction of certain congenital heart defects \cite{hsu2016functional}.


The LV, on the other hand, due to the increased aortic pressures created by PS and because of the flow competition between the streams coming from the shunt and aortic arch, experiences a higher resistance (alternatively, a higher afterload) in the post-operative state. This results in a lower volume of blood ejected by the LV, explaining its lower output. The increase in the effective resistance for the LV occurs, however, in the DAo at the level of the PS, explaining (i) an increased flow to the aortic arch branches, and hence to the SVC, and (ii) the reduced LV contribution to flow in the DAo and to the lower body compartment. The flow in the DAo  downstream the PS is the sum of oxygenated blood flow from the LV and the deoxygenated blood flow from the RV coming through the shunt, resulting in an increased IVC flow compared to the pre-operative state.

In addition to the increased afterload for the LV after PS creation, there is an issue of reduced pulmonary venous return to the LV due to the post-ventricular right-to-left shunt. Together with reduced LV stroke volume after the PS creation, it could result in decreased perfusion of the upper body and coronary circulation with the oxygenated blood. The simulations for this particular patient demonstrate increase of the blood flow originating from the AAo to the aortic arch branches due to the higher aortic pressures. Consequently, better perfusion of the upper body with oxygenated blood is achieved after PS creation. An extrapolation of this reasoning suggests that the higher AAo pressures should also result in increase of the coronary arterial perfusion and eventually sustained (or increased) oxygen delivery to the ventricular myocardium despite the decreased LV output. There are, however, some patients who demonstrated acute deterioration of haemodynamics with cardiac arrest after PS creation, presumably due to catastrophic drop in the LV output and coronary arterial perfusion. Thus, while `compensatory' increase of coronary perfusion may be seen as a positive course of events, degree of reduction in the LV output remains an overall cause for concern as it determines the level of blood oxygenation going to the coronary circulation. While physics dictates that coronary perfusion will be higher due to aortic pressure increase after PS creation, in the scenario of extreme right-to-left shunting that induces severe reduction in LV output, the increased flow to the coronaries (and also the aortic arch branches) will occur largely retrogradely with the deoxygenated blood coming from the RV through the PS. Furthermore, in patients with long-standing severe iPAH, the myocardium of the LV has been shown to become atrophic with contractile cardiomyocyte dysfunction \cite{manders2014contractile}.  The substantial reduction in the LV output after PS creation in combination with pre-existent LV myocardial atrophy is also of potential concern, as in such iPAH patients the degree of LV stroke volume decrease will be so severe that the LV output may not compete with the deoxygenated blood flow coming through the high-pressure pulmonary circulation through the PS. This, in turn, can result in perfusion of the upper body and the coronary circulation largely by deoxygenated blood leading rapidly to myocardial ischaemia and circulatory arrest, as was seen in the patients with acute deterioration upon PS creation \cite{boudjemline2017safety}. Interestingly, simulations for the patient in this study, despite the clinical history of well tolerated creation of the PS, have demonstrated small amount of the blood flow going retrogradely from the shunt into aortic arch. Patients from the published series of percutaneously created PS, who suffered from acute circulatory deterioration during the procedure, had retrograde flow in aortic arch clearly visible on Doppler echocardiography \cite{boudjemline2017safety}, corroborating the hypothesis of circulatory arrest due to myocardial ischemia because of catastrophic decrease of LV stroke volume. Further development of the model should address the factors potentially influencing the myocardial oxygen delivery and ventricular performance in the setting of the right-to-left shunt through the PS and arterial pressure changes in the patients with severe suprasystemic iPAH and ventricular dysfunction.

As can be seen from the areas within the PV-loops, 
creation of the PS causing a post-ventricular right-to-left shunt results in approximately 11\% decrease in the work done by the LV while operating at higher pressures.
This shows that the contribution of the smaller LV stroke volume towards decreasing the workload overpowers the contribution of higher LV afterload towards increasing it. For the RV, however, even though a reduction in afterload is achieved by the PS creation, the increase in volume returning to the ventricle overpowers and results in a net 18\% increase in the RV workload, accompanied presumably by increase in myocardial oxygen consumption. These findings are consistent with the main conclusion of the previous LPM study on PS for severe PAH, where the authors also reported that the shunt does not result in unloading of the RV despite its partial decompression \cite{delhaas2018potts}. 
The increase of the work performed by the already dilated RV after PS creation is, thus, due to the increased stroke volume accompanied, however, by the significant decrease in afterload. Given the confirmed clinical improvement of the patients with severe suprasystemic iPAH who survived the PS creation on intermediate term \cite{baruteau2015palliative}, the clinical relevance of such an increased RV workload remains yet unclear: it is accompanied by an enhanced EF and a RV of lower volumes, so of less stressed fibers, which suggest enhanced function and potential for favorable remodelling. 
Nonetheless, to avoid circulatory collapse, increased workload for the RV myocardium after the PS creation should be met with increased oxygen delivery through the coronary circulation, sustained perfusion of which depends on LV performance and oxygenation level of the blood entering the coronary arteries. Oxygen delivery to the myocardium as yet, is not accounted for in the model. Inclusion of this dynamic interplay between myocardial perfusion with variable oxygen concentration, associated changes in cardiac mechanics, and the ability of RV to do more work within these constraints, should be further analysed to determine criteria for suitability of individual patients for the PS creation.


The reduction in LV output is accompanied by an increase in the operating volumes for the LA (Figure \ref{fig:pre_post_flow_ms}). Effectively, both the maximum and minimum LA volumes are higher when compared to the pre-operative state. However, their difference (maximum minus minimum) is smaller, as the volume leaving LA in each cycle is smaller due to the lower volume returning from the pulmonary veins. 
On the contrary, for the RA, both the maximum and minimum RA volumes are lower, but their difference is larger due to the higher volume returning from the systemic veins.
With an understanding of the volume changes, as described above, the changes in pressure follow an expected and predictable pattern (Figure \ref{fig:pre_post_pressure_ms}). 


\subsection{3D flow features and effects of varying shunt diameter and length}
\label{dis_3D}
The jet of flow coming from the shunt into the DAo impinges on its wall opposite to the shunt and creates high impulse forces and localised high pressure regions (Figures \ref{fig:Streamlines_76mm} and  \ref{fig:Press_76mm}), which potentially may be damaging to the tissue and over time lead to adaptive changes of the aortic wall. The relatively high velocity values in the shunt and the DAo downstream the PS lead to high wall shear stresses (Figure \ref{fig:WSS_76mm}). Such abnormally high wall shear stress may affect endothelial cell function, vascular biology, and promote development of aneurysms in the long-term \cite{dolan2013high}.
%
%
%

Finally, as discussed above, the streamlines in Figure \ref{fig:Streamlines_76mm} (with pathlines showing similar patterns) point to some deoxygenated blood originating from the mPA and traversing through the PS to the aortic arch branches. This proportion is insignificant in the current case, where post-operative flow of oxygenated blood from the LV into aortic arch branches is higher relative to pre-operative state, and, thus it is unlikely that oxygenation of the brain is compromised. 
%
%
However, as mentioned above in Section \ref{dis_haemodynamics}, excessive right-to-left shunting may result in catastrophic decrease of LV contribution to the upper body perfusion allowing deoxygenated blood from the shunt traversing retrogradely to the aortic arch branches and to the coronaries in significant amounts.
%
%
%
This is also evidenced by the streamlines in Figure \ref{fig:Streamlines5mm10mm}, where a higher right-to-left shunting achieved in a PS with larger diameters PS shows a larger proportion of streamlines traversing to the aortic arch branches. 
This finding of increasing right-to-left volume shunting with increasing PS diameters in this study (Table \ref{tab:ventricle_work_ms_qsh}) is in conflict with the previous study using the CircAdapt LPM, where authors report no change in Qp/Qs ratio for PS diameters above 7mm \cite{delhaas2018potts}. 

The results of the model reported here 
%
show a clear near-linear relationship between the shunt diameters and both cycle-averaged pressure-gradient across the shunt and flow-rate through the shunt (Figure \ref{fig:linear_qsh_psh_ms}). Thus, there is a clear potential and need to determine an optimum shunt diameter for an individual patient that balances pressure equalisation achieved by the PS and the extent of the right-to-left shunting determining the post-operative LV  output and RV workload. This balance will depend in every particular iPAH patient on different factors, which are needed to be addressed in a further study before any recommendations regarding a `universal' shunt diameter, such as a percentage of the DAo diameter, can be provided.

Figure \ref{fig:Qshunt_dia_ms} highlights the importance of shunt diameter selection to ensure that blood is pumped from the pulmonary to systemic circulations across the entire cardiac cycle, where shunt diameters larger than 8mm caused some back-flow from the DAo to the LPA during diastole. The general trend is that the amount of back-flow increases with increasing shunt diameters, although the direction of the flow across the shunt also depends on the vascular resistances, which may vary in response to different factors, such as physical or emotional exertion or inflammation. To avoid this undesirable left-to-right shunting, a valved PS may be considered \cite{baruteau2015palliative}.

The results of varying shunt lengths in Section \ref{res_length} show that while global haemodynamics is largely unaltered with changing shunt lengths, the local flow-features are significantly more disturbed due to PS protrusion. In particular, the protrusions result in von K\'{a}rm\'{a}n type of vortices, within which the blood velocities are small, and hence the risk of thrombus formation may be elevated, especially with longer protrusions.

\subsection{Role and utility of the stand-alone LPM}
\label{dis_lpm_utility}
The results from the stand-alone LPM are presented and discussed in Appendix \ref{app:lpm}. Here, it is highlighted that the LPM models display the same trends for the assessment of global haemodynamics as the GMM. This is not surprising as the global haemodynamics, also in the GMM, is determined primarily by the LPM components of the model and the shunt diameter, both of which are replicated in the stand-alone LPM. Indeed, the stand-alone LPM does not account for shunt length protrusion, but the results of section \ref{res_length} show that this has minimal effect on global haemodynamics. 

More importantly, it should be noted that since the stand-alone LPM runs in seconds, in contrast to the days needed to obtain results from the GMM, it presents an ideal tool for parameter estimation, where large series of multiple simulations are necessary to comprehensively assess and predict the haemodynamic changes. This approach has been adopted in this study, where the estimation of parameters was largely done manually with the fast run times of the stand-alone LPM presenting, thus, an immense utility. Furthermore, since the LPM captures the trends in global haemodynamics correctly, it can be used for quick assessment of how different clinically measurable pre-operative parameters would affect global response of the circulation to PS creation, and thus, determine the suitability of an individual iPAH patient for the procedure. Also, such a predictive stand-alone LPM  would provide a useful tool during clinical decision-making about optimal shunt diameter for the patient with particular set of pre-operative parameters balancing the RV partial decompression with the tolerable decrease in LV output achieved by the PS. 

The GMM was useful to create the LPM components for the shunt and for the aortic part above the shunt. The later may be established based on geometrical information, while the current shunt law might be general enough for other iPAH patients. Overall, it is only when local flow features are deemed of high importance, that the GMM presents unparalleled advantages. This however, may be of less importance in the global assessment of the suitability of an individual patient for the PS creation.


\section{Conclusions}
This work presents the development of a patient-specific geometrical multiscale computational model of blood flow without and with PS, and its stand-alone equivalent lumped-parameter model of the whole circulation. Model parametrization is based on clinical data obtained in a paediatric patient with suprasystemic idiopathic PAH who received the stent-based PS. These computational models are respectively representative and predictive of the patient-specific haemodynamics pre- and post-creation of the PS. The results show sensitivity of the local and global hemodynamics to the PS diameter, for which an optimum based on other patient-specific factors than the DAo size is probably desirable. PS protrusion is only affecting local haemodynamics, with thrombogenic flow features. Overall, this study provides insights into physiology of this complex condition and hints towards potential causes of the circulatory collapse in some patients. Complemented with oxygen delivery assessment, this model may become a useful predictive tool in clinical decision-making regarding suitability for this type of palliative treatment in an individual PAH patient with drug-resistant suprasystemic PAH.

\section{Limitations and future work}

The primary limitations of this work are that only patient has been analysed, oxygen delivery was not included in the analysis, and that regulatory/adaptive mechanisms have been ignored in modelling the haemodynamic changes. To alleviate the first limitation, a larger study is required where complete pre- and post-operative data in multiple patients are acquired, so that parameters can be automatically and uniquely estimated for each patient. Less uncertain, reliable set of measurements are key to data-driven modelling, robust calibration of the existing models, and the development of new models that account for all relevant physics. The second limitation can be addressed by incorporating models predicting oxygen saturation as proposed elsewhere \cite{delhaas2018potts}, into the GMM. These models, including a model for coronary circulation, may be immensely beneficial for getting deeper insights into the PS-induced changes in the ventricular workload and function. 
In addition, the septal shift is not included in the model, which could be addressed using the three-wall segment model describing mechanics of ventricular interaction \cite{lumens2009three}. Equalization of ventricular pressures will lead to a better shape of the LV post-operatively, but its effect on global hemodynamics against the much stronger influence of the shifts in volumes and pressures after the PS creation remains unknown. Thus, to assess the importance of LV geometry for its filling and contraction, future iterations of the model can include the three-dimensional models of the ventricles. 
The final limitation is harder to address as the effectiveness of endogenous biological compensatory mechanisms are highly individual and depends strongly on large number of cardiovascular and non-cardiac factors. In particular, a key missing component now is modelling the adaptation of the ventricular function (e.g. contractility) to changes in its output. With such an addition to the model, it will be possible to develop second order, time-dependent logic for assessing post-operative haemodynamics. The development of such a model and its relation to patient-specific characteristics, such as level of pre-operative ventricular dysfunction, PVR, SVR, etc., forms the primary area of future investigation.

\section*{Author Contributions}
AS and YB proposed the clinical assessment and SP and IVC conceptualised the mathematical and engineering models. AS processed and analysed all clinical data and edited the manuscript. GG, AN, AK, SP, and IVC contributed to the development of the numerical code and model development. GG, AK,  SP, and IVC performed preliminary multiscale analysis. AK and SP performed patient-specific parameter estimation and ran the final GMM and LPM models. AK produced the first draft. SP produced the final version of the manuscript. AK, AS, and SP produced all the Figures. AS and YB provided clinical interpretation of the results. All authors read and made improvements to the final manuscript. SP and IVC acquired funding for the study.

\section*{Acknowledgement}
This work is supported by the EPSRC grant number EP/R010811/1.

\footnotesize
\bibliographystyle{unsrt}
\bibliography{references}

\begin{thebibliography}{10}

\bibitem{ivy2013pediatric}
D~Dunbar Ivy, Steven~H Abman, Robyn~J Barst, Rolf~MF Berger, Damien Bonnet,
  Thomas~R Fleming, Sheila~G Haworth, J~Usha Raj, Erika~B Rosenzweig,
  Ingram~Schulze Neick, et~al.
\newblock Pediatric pulmonary hypertension.
\newblock {\em Journal of the American College of Cardiology}, 62(25
  Supplement):D117--D126, 2013.

\bibitem{barst2011pulmonary}
RJ~Barst, SI~Ertel, Maurice Beghetti, and DD~Ivy.
\newblock Pulmonary arterial hypertension: a comparison between children and
  adults.
\newblock {\em European Respiratory Journal}, 37(3):665--677, 2011.

\bibitem{hoeper2013definitions}
Marius~M Hoeper, Harm~Jan Bogaard, Robin Condliffe, Robert Frantz, Dinesh
  Khanna, Marcin Kurzyna, David Langleben, Alessandra Manes, Toru Satoh,
  Fernando Torres, et~al.
\newblock Definitions and diagnosis of pulmonary hypertension.
\newblock {\em Journal of the American College of Cardiology}, 62(25
  Supplement):D42--D50, 2013.

\bibitem{shu2021efficacy}
Tingting Shu, Huaqiao Chen, Lu~Wang, Wuwan Wang, Panpan Feng, Rui Xiang,
  Li~Wen, and Wei Huang.
\newblock The efficacy and safety of pulmonary vasodilators in pediatric
  pulmonary hypertension (ph): a systematic review and meta-analysis.
\newblock {\em Frontiers in pharmacology}, 12:668902, 2021.

\bibitem{baruteau2014palliative}
Alban-Elouen Baruteau, Emre Belli, Younes Boudjemline, Daniela Laux, Marilyne
  L{\'e}vy, G{\'e}rald Simonneau, Adriano Carotti, Marc Humbert, and Damien
  Bonnet.
\newblock Palliative potts shunt for the treatment of children with
  drug-refractory pulmonary arterial hypertension: updated data from the first
  24 patients.
\newblock {\em European Journal of Cardio-Thoracic Surgery}, 47(3):e105--e110,
  2014.

\bibitem{grady2016potts}
R~Mark Grady and Pirooz Eghtesady.
\newblock Potts shunt and pediatric pulmonary hypertension: what we have
  learned.
\newblock {\em The Annals of thoracic surgery}, 101(4):1539--1543, 2016.

\bibitem{sizarov2016vascular}
Aleksander Sizarov, Francesca Raimondi, Damien Bonnet, and Younes Boudjemline.
\newblock Vascular anatomy in children with pulmonary hypertension regarding
  the transcatheter potts shunt.
\newblock {\em Heart}, 102(21):1735--1741, 2016.

\bibitem{blanc2004potts}
Julie Blanc, Pascal Vouh{\'e}, and Damien Bonnet.
\newblock Potts shunt in patients with pulmonary hypertension.
\newblock {\em New England Journal of Medicine}, 350(6):623--623, 2004.

\bibitem{potts1946anastomosis}
Willis~J Potts, Sidney Smith, and Stanley Gibson.
\newblock Anastomosis of the aorta to a pulmonary artery: certain types in
  congenital heart disease.
\newblock {\em Journal of the American Medical Association}, 132(11):627--631,
  1946.

\bibitem{hansmann2017pulmonary}
Georg Hansmann.
\newblock Pulmonary hypertension in infants, children, and young adults.
\newblock {\em Journal of the American College of Cardiology},
  69(20):2551--2569, 2017.

\bibitem{boudjemline2013patent}
Younes Boudjemline, Mehul Patel, Sophie Malekzadeh-Milani, Isabelle
  Szezepanski, Marilyne L{\'e}vy, and Damien Bonnet.
\newblock Patent ductus arteriosus stenting (transcatheter potts shunt) for
  palliation of suprasystemic pulmonary arterial hypertension: a case series.
\newblock {\em Circulation: Cardiovascular Interventions}, 6(2):e18--e20, 2013.

\bibitem{boudjemline2017safety}
Younes Boudjemline, Aleksander Sizarov, Sophie Malekzadeh-Milani, Cristian
  Mirabile, Marien Lenoir, Diala Khraiche, Marilyne L{\'e}vy, and Damien
  Bonnet.
\newblock Safety and feasibility of the transcatheter approach to create a
  reverse potts shunt in children with idiopathic pulmonary arterial
  hypertension.
\newblock {\em Canadian Journal of Cardiology}, 33(9):1188--1196, 2017.

\bibitem{esch2013transcatheter}
Jesse~J Esch, Pinak~B Shah, Barbara~A Cockrill, Harrison~W Farber, Michael~J
  Landzberg, Mandeep~R Mehra, Mary~P Mullen, Alexander~R Opotowsky, Aaron~B
  Waxman, James~E Lock, et~al.
\newblock Transcatheter potts shunt creation in patients with severe pulmonary
  arterial hypertension: initial clinical experience.
\newblock {\em The Journal of Heart and Lung Transplantation}, 32(4):381--387,
  2013.

\bibitem{schafer2019close}
Michal Sch{\"a}fer, Benjamin~S Frank, D~Dunbar Ivy, Neil Wilson, Gareth~J
  Morgan, Alex~J Barker, Lorna~P Browne, Max~B Mitchell, and Uyen Truong.
\newblock Close look at the potts shunt flow hemodynamics in a patient with
  severe pulmonary hypertension: 4d-flow mri evaluation.
\newblock {\em Journal of magnetic resonance imaging: JMRI}, 49(6):1800--1802,
  2019.

\bibitem{delhaas2018potts}
Tammo Delhaas, Yvette Koeken, Heiner Latus, Christian Apitz, and Dietmar
  Schranz.
\newblock Potts shunt to be preferred above atrial septostomy in pediatric
  pulmonary arterial hypertension patients: a modeling study.
\newblock {\em Frontiers in physiology}, 9:1252, 2018.

\bibitem{Mimics}
{Mimics Innovation Suite} kernel description.
\newblock
  \url{https://www.materialise.com/en/medical/mimics-innovation-suite-22}.
\newblock Accessed: 2019-06-26.

\bibitem{chigogidze2006intervascular}
Nikolay~A Chigogidze, Jos{\'e}~I Bilbao, Michael~V Avaliani, Valery~A
  Cherkasov, Isabel Vivas, and Dmitry~I Kolesnik.
\newblock Intervascular anastomoses created by an endovascular approach:
  technical aspects and initial results in an animal study.
\newblock {\em Journal of vascular and interventional radiology},
  17(3):521--531, 2006.

\bibitem{si2015tetgen}
Hang Si.
\newblock Tetgen, a delaunay-based quality tetrahedral mesh generator.
\newblock {\em ACM Transactions on Mathematical Software (TOMS)}, 41(2):11,
  2015.

\bibitem{sahni2008adaptive}
Onkar Sahni, Kenneth~E Jansen, Mark~S Shephard, Charles~A Taylor, and Mark~W
  Beall.
\newblock Adaptive boundary layer meshing for viscous flow simulations.
\newblock {\em Engineering with Computers}, 24(3):267, 2008.

\bibitem{updegrove2017simvascular}
Adam Updegrove, Nathan~M Wilson, Jameson Merkow, Hongzhi Lan, Alison~L Marsden,
  and Shawn~C Shadden.
\newblock Simvascular: An open source pipeline for cardiovascular simulation.
\newblock {\em Annals of biomedical engineering}, 45(3):525--541, 2017.

\bibitem{lan2018re}
Hongzhi Lan, Adam Updegrove, Nathan~M Wilson, Gabriel~D Maher, Shawn~C Shadden,
  and Alison~L Marsden.
\newblock A re-engineered software interface and workflow for the open-source
  simvascular cardiovascular modeling package.
\newblock {\em Journal of biomechanical engineering}, 140(2):024501, 2018.

\bibitem{shi2011review}
Yubing Shi, Patricia Lawford, and Rodney Hose.
\newblock Review of zero-d and 1-d models of blood flow in the cardiovascular
  system.
\newblock {\em Biomedical engineering online}, 10(1):33, 2011.

\bibitem{pant2013multiscale}
Sanjay Pant, Benoit Fabr{\`e}ges, Jean-Fr{\'e}d{\'e}ric Gerbeau, and Irene~E
  Vignon-Clementel.
\newblock A multiscale filtering-based parameter estimation method for
  patient-specific coarctation simulations in rest and exercise.
\newblock In {\em International Workshop on Statistical Atlases and
  Computational Models of the Heart}, pages 102--109. Springer, 2013.

\bibitem{pant2014methodological}
Sanjay Pant, Benoit Fabr{\`e}ges, J-F Gerbeau, and IE~Vignon-Clementel.
\newblock A methodological paradigm for patient-specific multi-scale cfd
  simulations: from clinical measurements to parameter estimates for individual
  analysis.
\newblock {\em International journal for numerical methods in biomedical
  engineering}, 30(12):1614--1648, 2014.

\bibitem{arts1991relation}
Theo Arts, PH~Bovendeerd, Frits~W Prinzen, and Robert~S Reneman.
\newblock Relation between left ventricular cavity pressure and volume and
  systolic fiber stress and strain in the wall.
\newblock {\em Biophysical journal}, 59(1):93--102, 1991.

\bibitem{bovendeerd2006dependence}
Peter~HM Bovendeerd, Petra Borsje, Theo Arts, and Frans~N van De~Vosse.
\newblock Dependence of intramyocardial pressure and coronary flow on
  ventricular loading and contractility: a model study.
\newblock {\em Annals of biomedical engineering}, 34(12):1833--1845, 2006.

\bibitem{pant2016data}
Sanjay Pant, Chiara Corsini, Catriona Baker, Tain-Yen Hsia, Giancarlo Pennati,
  Irene~E Vignon-Clementel, Modeling of~Congenital Hearts Alliance
  (MOCHA)~Investigators, et~al.
\newblock Data assimilation and modelling of patient-specific single-ventricle
  physiology with and without valve regurgitation.
\newblock {\em Journal of biomechanics}, 49(11):2162--2173, 2016.

\bibitem{pant2017inverse}
Sanjay Pant, Chiara Corsini, Catriona Baker, Tain-Yen Hsia, Giancarlo Pennati,
  and Irene~E Vignon-Clementel.
\newblock Inverse problems in reduced order models of cardiovascular
  haemodynamics: aspects of data assimilation and heart rate variability.
\newblock {\em Journal of The Royal Society Interface}, 14(126):20160513, 2017.

\bibitem{mynard2012simple}
JP~Mynard, MR~Davidson, DJ~Penny, and JJ~Smolich.
\newblock A simple, versatile valve model for use in lumped parameter and
  one-dimensional cardiovascular models.
\newblock {\em International Journal for Numerical Methods in Biomedical
  Engineering}, 28(6-7):626--641, 2012.

\bibitem{migliavacca2000computational}
Francesco Migliavacca, Gabriele Dubini, Giancarlo Pennati, Riccardo
  Pietrabissa, Roberto Fumero, Tain-Yen Hsia, and Marc~R de~Leval.
\newblock Computational model of the fluid dynamics in systemic-to-pulmonary
  shunts.
\newblock {\em Journal of biomechanics}, 33(5):549--557, 2000.

\bibitem{migliavacca2001modeling}
Francesco Migliavacca, Giancarlo Pennati, Gabriele Dubini, Roberto Fumero,
  Riccardo Pietrabissa, Gonzalo Urcelay, Edward~L Bove, Tain-Yen Hsia, and
  Marc~R de~Leval.
\newblock Modeling of the norwood circulation: effects of shunt size, vascular
  resistances, and heart rate.
\newblock {\em American Journal of Physiology-Heart and Circulatory
  Physiology}, 280(5):H2076--H2086, 2001.

\bibitem{moghadam2011comparison}
Mahdi~Esmaily Moghadam, Yuri Bazilevs, Tain-Yen Hsia, Irene~E Vignon-Clementel,
  Alison~L Marsden, et~al.
\newblock A comparison of outlet boundary treatments for prevention of backflow
  divergence with relevance to blood flow simulations.
\newblock {\em Computational Mechanics}, 48(3):277--291, 2011.

\bibitem{moghadam2013modular}
Mahdi~Esmaily Moghadam, Irene~E Vignon-Clementel, Richard Figliola, Alison~L
  Marsden, Modeling of~Congenital Hearts Alliance (MOCHA)~Investigators, et~al.
\newblock A modular numerical method for implicit 0d/3d coupling in
  cardiovascular finite element simulations.
\newblock {\em Journal of Computational Physics}, 244:63--79, 2013.

\bibitem{ladisa2011computational}
John~F LaDisa, C~Alberto~Figueroa, Irene~E Vignon-Clementel, Hyun Jin~Kim, Nan
  Xiao, Laura~M Ellwein, Frandics~P Chan, Jeffrey~A Feinstein, and Charles~A
  Taylor.
\newblock Computational simulations for aortic coarctation: representative
  results from a sampling of patients.
\newblock {\em Journal of biomechanical engineering}, 133(9), 2011.

\bibitem{spilker2010tuning}
Ryan~L Spilker and Charles~A Taylor.
\newblock Tuning multidomain hemodynamic simulations to match physiological
  measurements.
\newblock {\em Annals of biomedical engineering}, 38(8):2635--2648, 2010.

\bibitem{mackenzie2013decreased}
Robert~V Mackenzie~Ross, Mark~R Toshner, Elaine Soon, Robert Naeije, and Joanna
  Pepke-Zaba.
\newblock Decreased time constant of the pulmonary circulation in chronic
  thromboembolic pulmonary hypertension.
\newblock {\em American Journal of Physiology-Heart and Circulatory
  Physiology}, 305(2):H259--H264, 2013.

\bibitem{chemla2016time}
D~Chemla, K~Zhu, V~Castelain, P~Attal, M~Humbert, P~Herv{\'e}, and E~Lau.
\newblock Time-constant of the pulmonary arterial windkessel in precapillary
  pulmonary hypertension: A reappraisal.
\newblock {\em Am J Respir Crit Care Med}, 193:A7334, 2016.

\bibitem{baretta2014patient}
Alessia Baretta.
\newblock {\em Patient-specific modeling of the cardiovascular system for
  surgical planning of single-ventricle defects}.
\newblock PhD thesis, Politecnico di Milano, 2014.

\bibitem{kilner2009pulmonary}
Philip~J Kilner, Rossella Balossino, Gabriele Dubini, Sonya~V Babu-Narayan,
  Andrew~M Taylor, Giancarlo Pennati, and Francesco Migliavacca.
\newblock Pulmonary regurgitation: the effects of varying pulmonary artery
  compliance, and of increased resistance proximal or distal to the compliance.
\newblock {\em International journal of cardiology}, 133(2):157--166, 2009.

\bibitem{presson1998anatomic}
Robert~G Presson~Jr, Said~H Audi, Christopher~C Hanger, Gerald~M Zenk,
  Richard~A Sidner, John~H Linehan, Wiltz~W Wagner~Jr, and Christopher~A
  Dawson.
\newblock Anatomic distribution of pulmonary vascular compliance.
\newblock {\em Journal of Applied Physiology}, 84(1):303--310, 1998.

\bibitem{spilker2007morphometry}
Ryan~L Spilker, Jeffrey~A Feinstein, David~W Parker, V~Mohan Reddy, and
  Charles~A Taylor.
\newblock Morphometry-based impedance boundary conditions for patient-specific
  modeling of blood flow in pulmonary arteries.
\newblock {\em Annals of biomedical engineering}, 35(4):546--559, 2007.

\bibitem{lumens2009three}
Joost Lumens, Tammo Delhaas, Borut Kirn, and Theo Arts.
\newblock Three-wall segment (triseg) model describing mechanics and
  hemodynamics of ventricular interaction.
\newblock {\em Annals of biomedical engineering}, 37(11):2234--2255, 2009.

\bibitem{lumens2010left}
Joost Lumens, Daniel~G Blanchard, Theo Arts, Ehtisham Mahmud, and Tammo
  Delhaas.
\newblock Left ventricular underfilling and not septal bulging dominates
  abnormal left ventricular filling hemodynamics in chronic thromboembolic
  pulmonary hypertension.
\newblock {\em American Journal of Physiology-Heart and Circulatory
  Physiology}, 299(4):H1083--H1091, 2010.

\bibitem{hsu2016functional}
Chih-Hsin Hsu, Jun-Neng Roan, Jyh-Hong Chen, and Chen-Fuh Lam.
\newblock Functional improvement and regression of medial hypertrophy in the
  remodeled pulmonary artery after correction of systemic left-to-right shunt.
\newblock {\em Scientific reports}, 6(1):1--10, 2016.

\bibitem{manders2014contractile}
Emmy Manders, Harm-Jan Bogaard, M~Louis Handoko, Marielle~C van~de Veerdonk,
  Anne Keogh, Nico Westerhof, Ger~JM Stienen, Cristobal~G Dos~Remedios, Marc
  Humbert, Peter Dorfm{\"u}ller, et~al.
\newblock Contractile dysfunction of left ventricular cardiomyocytes in
  patients with pulmonary arterial hypertension.
\newblock {\em Journal of the American College of Cardiology}, 64(1):28--37,
  2014.

\bibitem{baruteau2015palliative}
Alban-Elouen Baruteau, Emre Belli, Younes Boudjemline, Daniela Laux, Marilyne
  L{\'e}vy, G{\'e}rald Simonneau, Adriano Carotti, Marc Humbert, and Damien
  Bonnet.
\newblock Palliative potts shunt for the treatment of children with
  drug-refractory pulmonary arterial hypertension: updated data from the first
  24 patients.
\newblock {\em European Journal of Cardio-Thoracic Surgery}, 47(3):e105--e110,
  2015.

\bibitem{dolan2013high}
Jennifer~M Dolan, John Kolega, and Hui Meng.
\newblock High wall shear stress and spatial gradients in vascular pathology: a
  review.
\newblock {\em Annals of biomedical engineering}, 41(7):1411--1427, 2013.

\end{thebibliography}

\clearpage

\appendix

\section*{Appendices}

\section{Results from the stand-alone LPM and their comparison against the GMM}
\label{app:lpm}

\subsection{Shunt model for post-operative stand-alone LPM}

\begin{figure}[tb]
\centering
\includegraphics[width=0.5\linewidth]{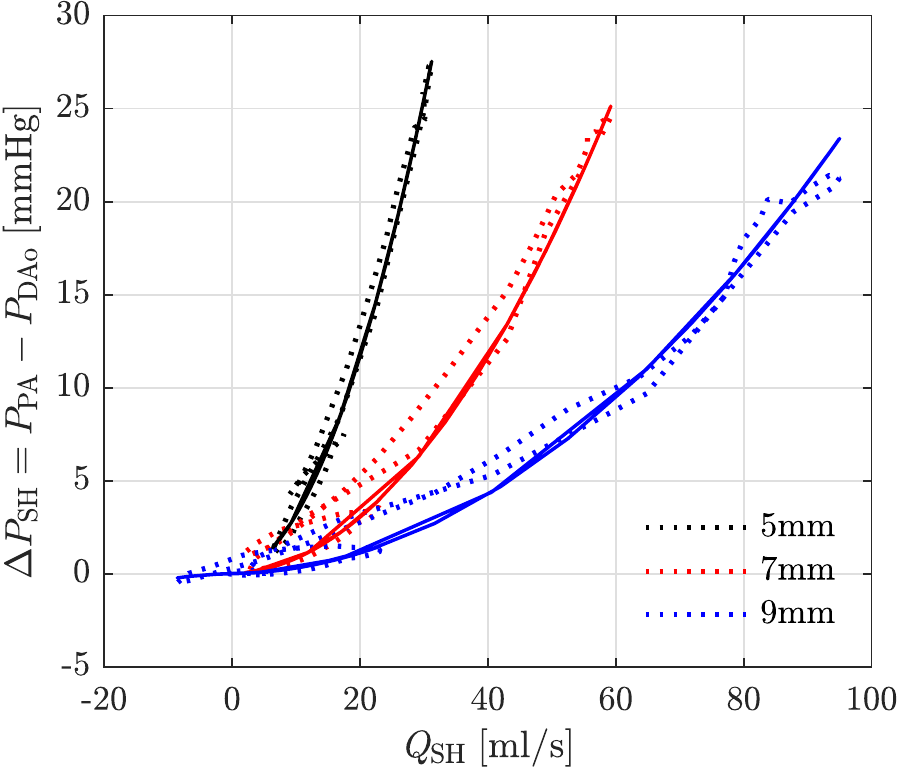}
\caption{Shunt pressure drop against shunt flow-rate over a cardiac cycle: regression (solid line) of GMM (dotted line) solutions for 5, 7 and 9 mm shunt diameters.}
\label{fig:RegressionAllDia}
\end{figure}

For the post-operative stand-alone LPM, the proportionality constants in Equation \eqref{shunt_law_reduced} are derived from regression analysis on the GMM solutions for all non-protruding shunt geometries with diameters varying from 5, 7, and 9 mm, see Figure \ref{fig:RegressionAllDia}.  This results in values of $k_1=4.106$ g/s and a $k_2=2.224$ g/cm\textsuperscript{3}.

\subsection{Stand-alone LPM results for 7.6mm diameter shunt}
Tables \ref{tab:prepost_press_data_lpm} shows the pre- and post-operative stand-alone LPM results against the clinical measurements (to be compared with Tables \ref{tab:prepost_press_data_ms}, which shows the GMM results). 
Figure \ref{fig:PaoAndPpa_lpm} shows the pre-operative pressures in the pulmonary artery and the ascending aorta from the stand-alone LPM against the measurements (to be compared against Figure \ref{fig:PaoAndPpa_ms}, which shows the GMM results).
Figures \ref{fig:pre_post_pressure_lpm}, \ref{fig:pre_post_flow_lpm}, and \ref{fig:volumes_lpm} show pressure traces including PV-loops, flow-rate traces, and volume displaced in one cardiac cycle through the circulatory system in the LPM solution (to be compared against Figures \ref{fig:pre_post_pressure_ms}, \ref{fig:pre_post_flow_ms}, and \ref{fig:volumes_ms}, respectively, which show the GMM solution)

\subsection{Stand-alone LPM results for varying shunt diameters}
\begin{figure}[tb]
\centering
\begin{subfigure}{0.45\textwidth}
\centering
\includegraphics[width=1.0\textwidth]{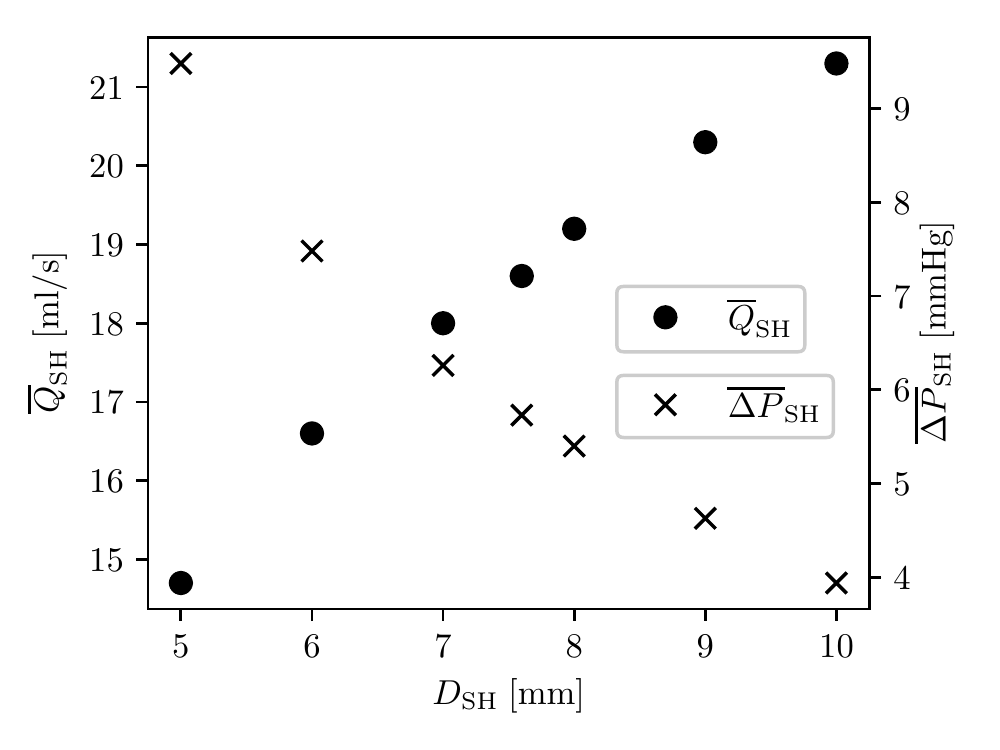}
\caption{Average flow-rate through the shunt and pressure gradient across the shunt for varying shunt diameters}
\label{fig:linear_qsh_psh_lpm}
\end{subfigure}
\begin{subfigure}{0.45\textwidth}
\centering
\includegraphics[width=0.9\textwidth]{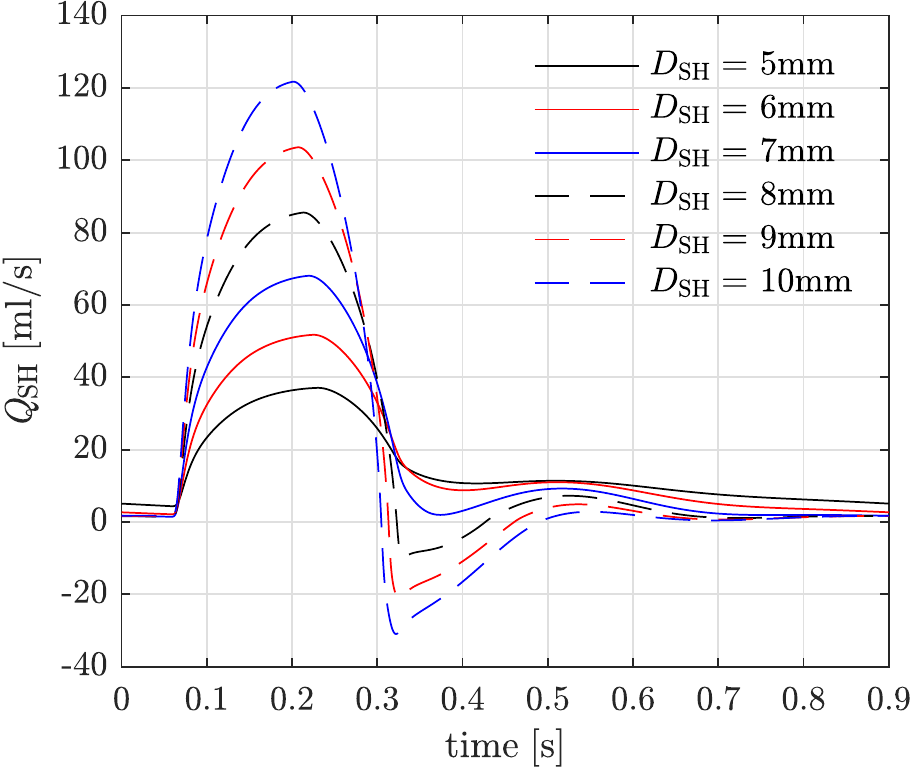}
\caption{Effect of shunt diameter on volumetric flow rate through the PS}
\label{fig:Qshunt_dia_lpm}
\end{subfigure}
\label{fig:combined_qsh_linear_lpm}
\caption{Stand-alone LPM: effect of varying shunt diameter on flow-rate and pressure gradient across the shunt.}
\end{figure}

Table \ref{tab:shuntdiaoverall_lpm} shows the key haemodynamic metrics with varying shunt diameters from the stand-alone LPM solution (to be compared against Table \ref{tab:shuntdiaoverall_ms}, which shows the GMM results). Figure \ref{fig:PaoPpa_PSdia_lpm} shows the pressure traces in the aorta and pulmonary artery for varying shunt diameters in the LPM solution (to be compared with Figure \ref{fig:PaoPpa_PSdia_ms}, which shows the GMM solution). Figure \ref{fig:linear_qsh_psh_lpm} shows how the mean pressure gradient and mean flow-rate across the shunt vary with shunt diameters (to be compared with Figure \ref{fig:linear_qsh_psh_ms}, which shows the GMM solution). Figure \ref{fig:Qshunt_dia_lpm} shows the the flow-rate profiles in the shunt and in the descending aorta for the LPM solution against varying shunt diameters (to be compared against Figure \ref{fig:Qshunt_dia_ms}, which shows the GMM solution). Figure \ref{fig:PVloopLVRV_AllDia_lpm} shows the PV loop variation in the LPM solution for the left and right ventricles with varying shunt diameters (to be compared against Figure \ref{fig:PVloopLVRV_AllDia_ms}, which shows the GMM solution). Table \ref{tab:ventricle_work_lpm_qsh} shows the corresponding areas under the ventricular PV-loops,  mean flow-rate through the shunt, volume displaced in one cardiac cycle through the shunt, mean pressure-drop across the shunt, and $Q_p/Q_s$ (to be compared against Table \ref{tab:ventricle_work_ms_qsh}, which shows the GMM solution). 

\subsection{Stand-alone LPM compared to the measurements for 7.6mm diameter shunt}

Tables\ref{tab:prepost_press_data_lpm} shows that the stand-alone LPM reproduces the pre-operative patient-specific haemodynamics with an accuracy comparable to that of the GMM.  Indeed, with the exception of the diastolic $P_{\rm{mPA}}$, which differs from the clinical measurement by 2.6\%, the stand-alone LPM displays improved correlation with the measured pre-operative pressures in the pulmonary artery and the aorta.  Similarly, the stand-alone LPM is in good agreement with the measured pre-operative blood flow volumes ejected from the ventricles, reproducing the EDV, SV and CO within 0.5\% of the measurements.  Additionally, the pre-operative volumetric flow rate profiles derived computationally from the stand-alone LPM for the MiV and TrV in Figure \ref{fig:volumes_lpm} accurately capture the normal LV filling and the impaired RV filling indicated in the echocardiographic velocity tracings in Figure \ref{fig:MVVandTCV}(c) and (d).

Post-operatively, the stand-alone LPM with a 7.6 mm diameter PS displays similar trends to the GMM solution when compared to the measured pressures in the pulmonary artery and the aorta. The systolic pressures generated computationally by the LPM show improved correlation with the measurements, differing by approximately 3.5\%, with the pulmonary-to-systemic systolic arterial pressure ratio comparable to the clinical data in both pre-operative and post-operative states.  With regard to the mean and diastolic pressures, the stand-alone LPM reproduces the mean $P_{\rm{mPA}}$ and $P_{\rm{AAo}}$ within 2\% and 6\% of the measurements, respectively, and the diastolic $P_{\rm{mPA}}$ and $P_{\rm{AAo}}$ within 15\% and 21\% of the measured data, respectively.  Although the stand-alone LPM displays slightly greater variation to the measured mean and diastolic pressures, the trends are comparable to the GMM solution, especially given the uncertainties and variability within the clinical measurements.

\subsection{Stand-alone LPM compared to the GMM}
\subsubsection{7.6 mm diameter shunt}
\begin{figure}[htbp]
\centering
\includegraphics[width=0.5\linewidth]{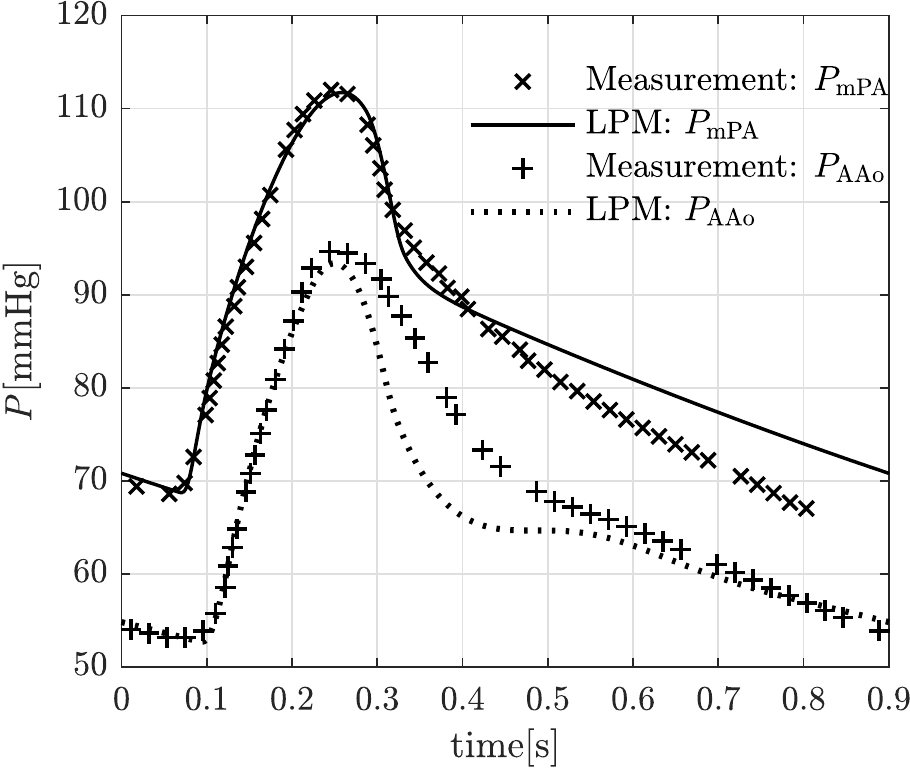}
\caption{Comparison of pre-operative pressure waveforms generated by the stand-alone LPM model agains the measurements over one cardiac cycle.}
\label{fig:PaoAndPpa_lpm}
\end{figure}

As demonstrated in Figure \ref{fig:PaoAndPpa_lpm}, the computationally derived pre-operative pressures and cardiac chamber volumes show strong congruence with a mean average percentage error (MAPE) over one cardiac cycle of less than 2.5\% when comparing the GMM and LPM solutions.  The post-operative aortic and pulmonary pressure waveforms in Figure \ref{fig:pre_post_pressure_lpm} (with key features reported in Table \ref{tab:prepost_press_data_lpm}) show that, despite differences in the diastolic portion of the pressure waveform, the stand-alone LPM is in good agreement with the GMM solution, with a MAPE of less than 3\% over one cardiac cycle.

Comparing the post-operative computational solutions in Figures \ref{fig:pre_post_pressure_lpm}, \ref{fig:pre_post_flow_lpm}, and \ref{fig:volumes_lpm} to those in \ref{fig:pre_post_pressure_ms}, \ref{fig:pre_post_flow_ms}, and \ref{fig:volumes_ms}, respectively, the two models show the greatest congruence for the pulmonary branch arteries and the supra-aortic branch arteries, with a MAPE of less than 2\% over one cardiac cycle.  The MAPE between the two computational models increases to 4.5\% for the DAo, which is potentially due to the simplified PS model in the stand-alone LPM.  Variation between the stand-alone LPM and GMM is heightened for the venous pressures, with a MAPE of 6-8\% for the SVC and IVC and 15\% for the LPV and RPV.  The computational models display reasonable congruence for the ventricular volumes, with the greatest deviation occurring during diastole and a MAPE of approximately 7\% over one cardiac cycle.  However, greater variation between the computational models is evident for the atrial volumes, with a MAPE of 10-12\%, which is attributed to the disparities between the computational solutions for the venous pressures. 

The LPM shows a mean shunt flow of 18.6 ml/s and a mean pressure gradient the shunt of 5.73 mmHg. These values for the GMM solution are 18.7 ml/s and 4.58 mmHg, respectively. The LPM flow-rate profile in the stand-alone LPM, Figure \ref{fig:pre_post_flow_lpm},  shows a similar trend to that of the GMM solution, Figure \ref{fig:pre_post_flow_ms}. However, the LPM solution differs from GMM solution in the following two respects, even though the mean flow-rate is nearly identical: (i) in the systole phase, the LPM flow-rate through the shunt is more symmetric and smoother; and (ii) the oscillatory solution is less pronounced in the diastolic phase in the LPM solution.

\subsubsection{Varying shunt diameters}

\begin{figure*}[!t]
    \begin{subfigure}[c] {0.33\textwidth}
      \centering
      \includegraphics[width=0.9\textwidth]{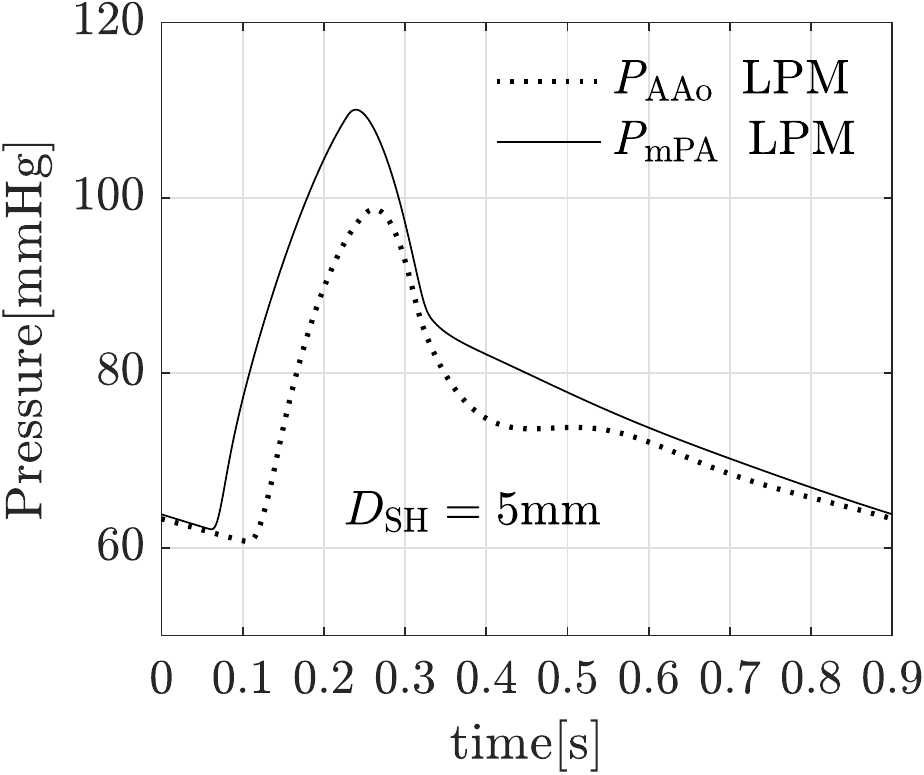}
      \caption{5mm diameter}
    \end{subfigure}
   \begin{subfigure}[c] {0.33\textwidth}
     \centering
      \includegraphics[width=0.9\textwidth]{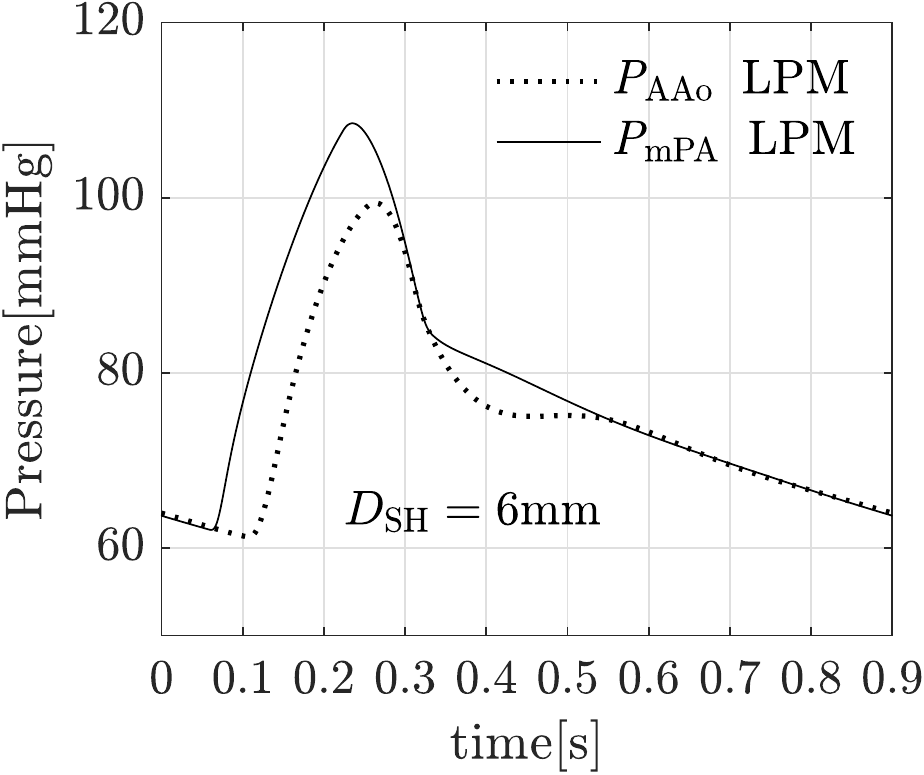}
      \caption{6mm diameter}
    \end{subfigure}
    \begin{subfigure}[c] {0.33\textwidth}
      \centering
      \includegraphics[width=0.9\textwidth]{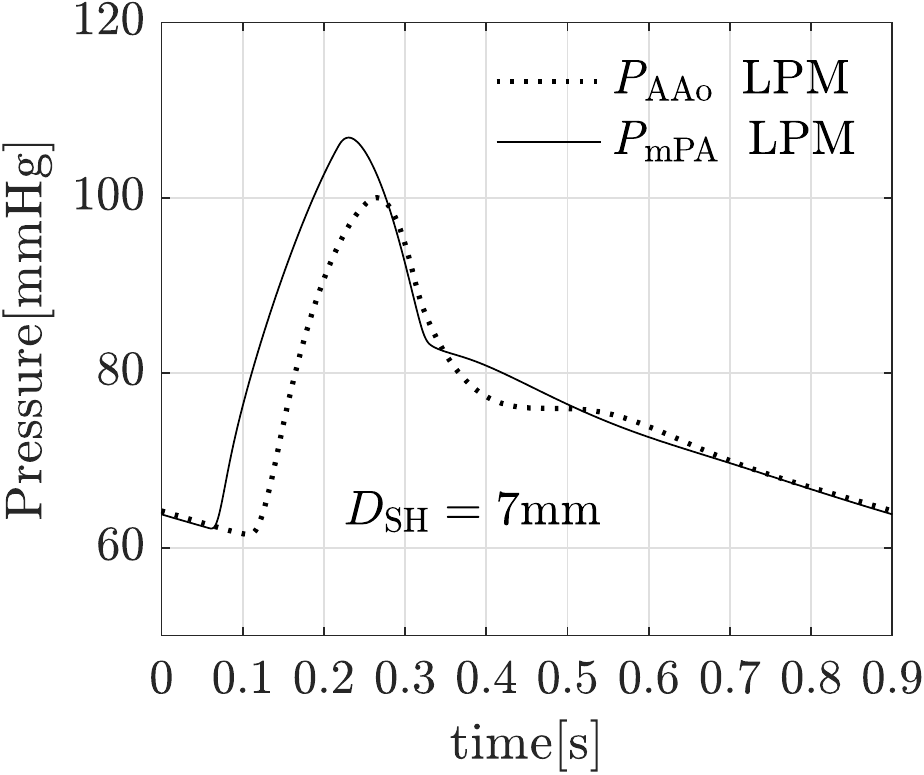}
      \caption{7mm diameter}
    \end{subfigure}\\
        \begin{subfigure}[c] {0.33\textwidth}
          \centering
      \includegraphics[width=0.9\textwidth]{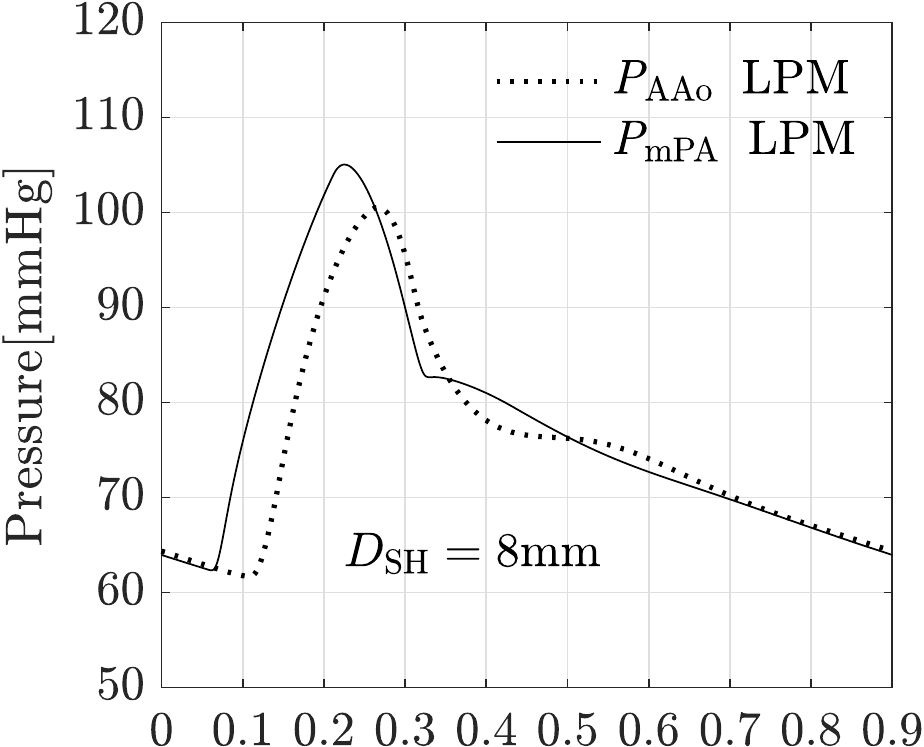}
      \caption{8mm diameter}
    \end{subfigure}
   \begin{subfigure}[c] {0.33\textwidth}
     \centering
      \includegraphics[width=0.9\textwidth]{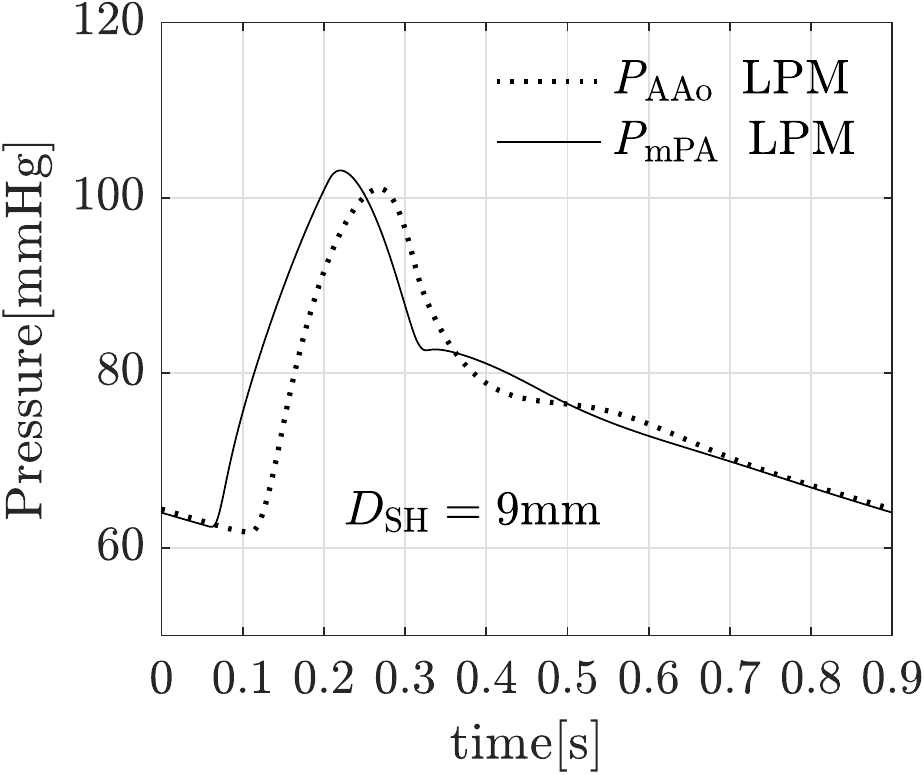}
      \caption{9mm diameter}
    \end{subfigure}
    \begin{subfigure}[c] {0.33\textwidth}
      \centering
      \includegraphics[width=0.9\textwidth]{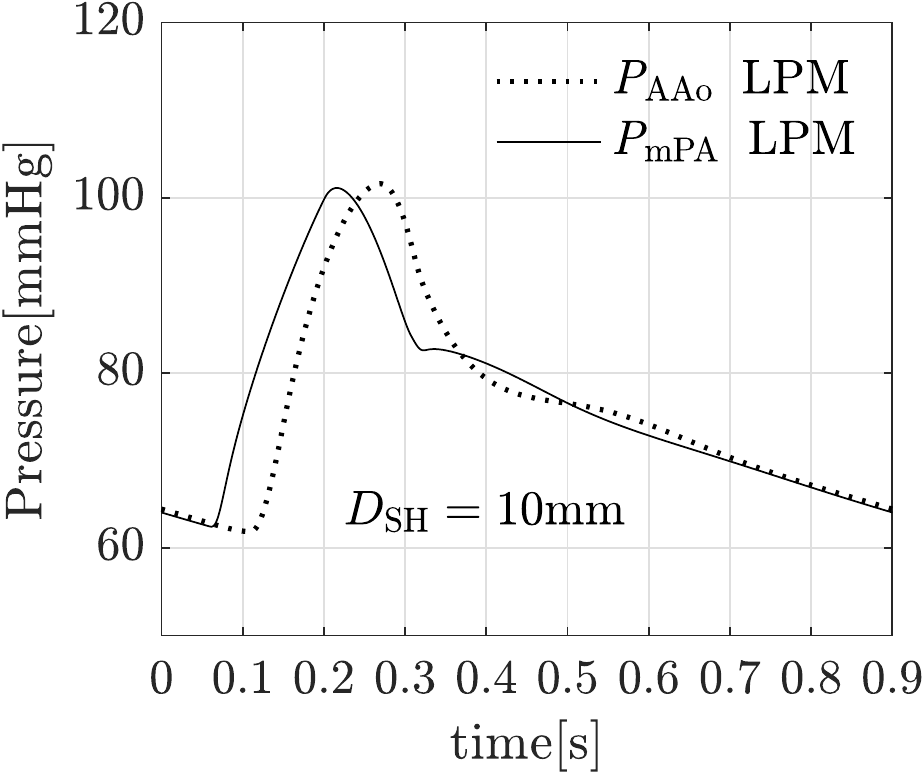}
      \caption{10mm diameter}
    \end{subfigure}\\
\caption{Stand-alone LPM: post-operative $P_{\rm{AAo}}$ and $P_{\rm{mPA}}$ for different shunt diameters.}
\label{fig:PaoPpa_PSdia_lpm}
\end{figure*}

\begin{figure}[tb]
\centering
\begin{subfigure}{0.45\textwidth}
\includegraphics[width=1.0\textwidth]{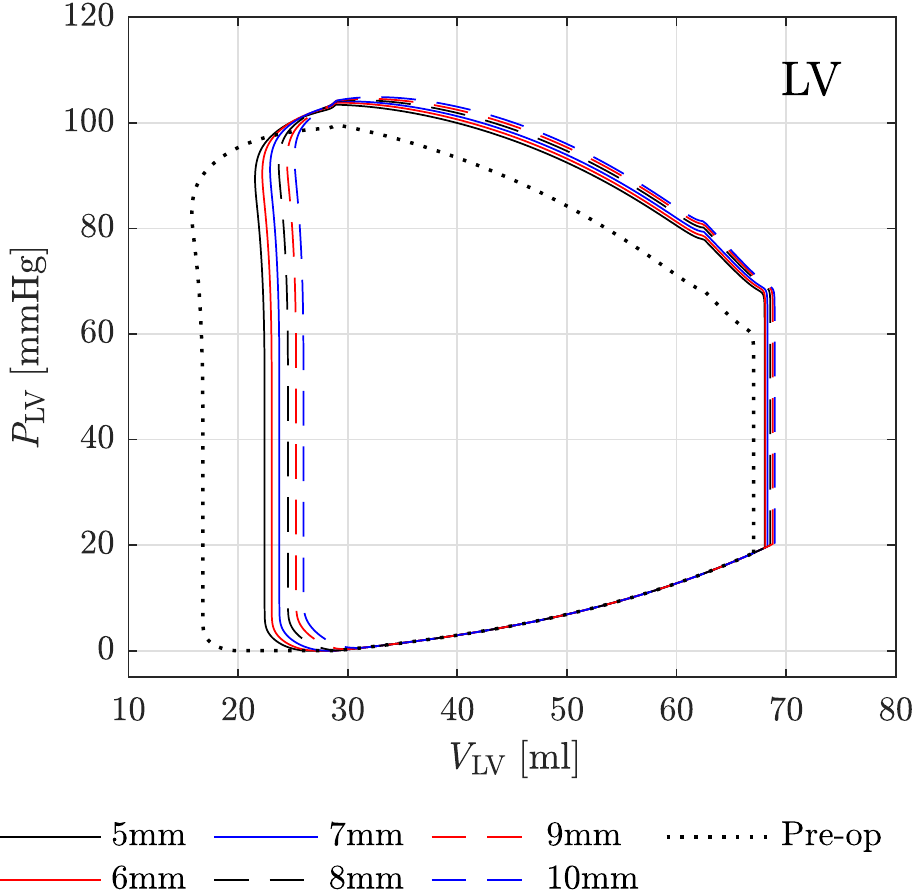}
\caption{Left ventricle}
\end{subfigure}
\hfill
\begin{subfigure}{0.45\textwidth}
\includegraphics[width=1.0\textwidth]{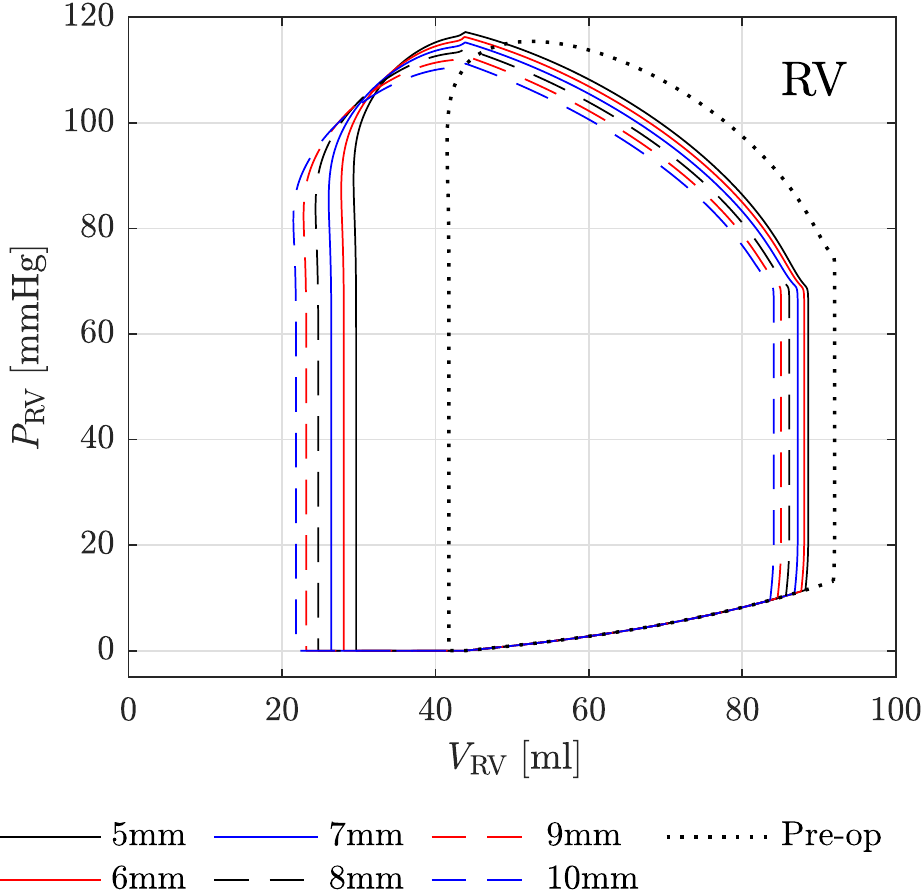}
\caption{Right ventricle}
\end{subfigure}
\caption{Stand-alone LPM: effect of shunt diameter on pressure-volume loops for (a) left ventricle and (b) right ventricle.}
\label{fig:PVloopLVRV_AllDia_lpm}
\end{figure}

Comparing Table \ref{tab:shuntdiaoverall_ms} and Table \ref{tab:shuntdiaoverall_lpm}, it is evident that the stand-alone LPM displays similar trends to the GMM solution.  The pressures in the PA and aorta are relatively consistent between the two computational models, although the LPM under-predicts the systolic pressures in comparison to the GMM solution, particularly for the aorta.  Also, contrary to the GMM solution, the LPM indicates a marginal and negligible increase in diastolic $P_{\rm{mPA}}$ with increasing shunt diameter.  The variation of pressures over the cardiac cycles show an identical trend in between the two solutions, see Figures \ref{fig:PaoPpa_PSdia_lpm} and  \ref{fig:PaoPpa_PSdia_ms}. Tables \ref{tab:ventricle_work_lpm_qsh} and \ref{tab:ventricle_work_ms_qsh} show that the LPM solution closely mimics the GMM solution for both the mean flow-rate through the shunt and the pressure gradient across the shunt. The flow-rate profiles through the shunt, as shown in Figures \ref{fig:Qshunt_dia_lpm} and \ref{fig:Qshunt_dia_ms}, show similar trends, with discrepancies identical to those observed for the 7.6mm diameter case in the above section. Notably, the LPM solution is smoother, more symmetric in the systole phase, over-damped in the diastole phase, and over-predicts the negative flow rate through the shunt at the end of systole relative to the GMM solution. The peak flow-rate and the mean flow-rate through the shunt follow each other closely between the two solutions. Comparing the PV loops obtained through the two solutions, Figures \ref{fig:PVloopLVRV_AllDia_lpm} and \ref{fig:PVloopLVRV_AllDia_ms}, the observed trends are similar with one notable exception. The LV EDV shows an increasing trend in the LPM as the shunt diameter is increased, contrary to the GMM solution. However, the changes in LV EDV in both cases, although in different directions, are minimal. Finally, Tables \ref{tab:ventricle_work_lpm_qsh} and 
\ref{tab:ventricle_work_ms_qsh} confirm that the trends shown by the area within the PV-loops and $Q_p/Q_s$ are similar between the stand-alone LPM and the GMM.

\subsection{Overall assessment}
The primary observation when comparing the stand-alone LPM and GMM solutions is that the they both display similar trends for changes induced by the introduction of the Potts shunt. Given that the stand-alone LPM run-times are in seconds compared to those in days for the GMM solution, the stand-alone LPM provides a great tool for quickly assessing the effect of changing parameters on global haemodynamics. Such parametric sweep studies can be used to uncover parameter combinations and patient characteristics that lead to excessive reduction in LV cardiac output, thus helping in defining risk indices for the Potts shunt procedure. Furthermore, since the stand-alone LPM also captures the trends in variation of post-operative output with varying shunt diameters, it can also be used a quick tool to assess the range of diameters suitable for a given combination of patient parameters. Lastly, it is important to note that the global haemodynamics in the GMM are driven primarily by the LPM components (see section \ref{dis_lpm_utility}), and hence post-operative discrepancies between the GMM and the stand-alone LPM can be further reduced by improving both the 0D shunt model and the 0D components that represent the 3D geometries.

\begin{table} [tb]
\begin{center}
\def\arraystretch{1.2}
\resizebox{1.0\textwidth}{!}{ 
\begin{tabular}{|c|ccc|ccc|cc|cc|cc|cc|}
\hline
\multicolumn{15}{|c|}{\textbf{Pre-operative measurements and model output (parameter estimation)}}\\
\hline
 & \multicolumn{3}{c}{$P_{\rm{AAo}}$ [mmHg]} & \multicolumn{3}{|c|}{$P_{\rm{mPA}}$ [mmHg]} & \multicolumn{2}{c}{EDV [ml]} & \multicolumn{2}{|c}{SV [ml]} & \multicolumn{2}{|c}{EF} & \multicolumn{2}{|c|}{CO [L/min]}\\
\hline
 & Systolic & Diastolic & Mean & Systolic & Diastolic & Mean & LV & RV & LV & RV & LV & RV & LV & RV\\
 \hline
Measurement & 94.0 & 53.0 & 69.0 & 112.0 & 67.0 & 85.0 & 66.9 & 91.8 & 51.0 & 51.0 & 0.76 & 0.55 & 3.4 & 3.4\\
LPM & 93.4 & 52.7 & 65.7 & 111.7 & 68.7 & 85.1 & 67.0 & 92.0 & 51.2 & 51.2 & 0.76 & 0.56 & 3.41 & 3.41\\
\hline
\multicolumn{15}{|c|}{\textbf{Post-operative measurements and model output (validation)}}\\
\hline
Measurement & 97.0 & 51.0 & 71.0 & 102.0 & 54.0 & 76.0 & - & - & - & - & - & - & - & - \\
LPM & 100.3 &	61.6 & 75.2	& 105.8 &	62.3 &	77.6 & 68.5 & 86.7 & 45.1 & 61.8 & 0.66 & 0.71 & 3.00 & 4.12 \\
\hline
\end{tabular}
}
\caption{Pre- and post-operative $P_{\rm{AAo}}$, $P_{\rm{mPA}}$, EDV, SV, EF and CO from the stand-alone LPM against clinical measurements.}
\vspace{0.5cm}
\label{tab:prepost_press_data_lpm}
\end{center}
\end{table}

\begin{table}
\begin{center}
\def\arraystretch{1.3}
\resizebox{1.0\textwidth}{!}{ 
\begin{tabular}{|c|ccc|ccc|cc|cc|cc|cc|}
\hline
  & \multicolumn{3}{|c|}{$P_{\rm{AAo}}$ [mmHg]} & \multicolumn{3}{|c|}{$P_{\rm{mPA}}$ [mmHg]} & \multicolumn{2}{|c|}{EDV [ml]} & \multicolumn{2}{|c|}{SV [ml]} &\multicolumn{2}{|c|}{EF} &  \multicolumn{2}{|c|}{CO [L/min]}\\
\hline
 Diameter [mm] & Systolic & Diastolic & Mean & Systolic & Diastolic & Mean & LV & RV & LV & RV & LV & RV & LV & RV\\
 \hline
pre-op & 93.4 & 52.7 & 65.7 & 111.7 & 68.7 & 85.1 & 67.0 & 92.0 & 51.2 & 50.5 & 0.76 & 0.55 & 3.42 & 3.42\\ 
\hdashline
5 & 98.7 & 60.7 & 73.5 & 110.1 & 62.1 & 79.0 & 68.0 & 88.6 & 46.5& 59.3 & 0.68 & 0.67 & 3.10 & 3.95\\
6 & 99.4 & 61.3 & 74.4 & 108.5 & 62.1 & 78.2 & 68.1 & 88.1 & 45.9 & 60.3 & 0.67 & 0.69 & 3.06 & 4.02\\
7 & 100.0 & 61.5 & 75.0 & 106.9 & 62.2 & 77.7 & 68.3 & 87.2 & 45.4 & 61.1 & 0.66 & 0.70 & 3.02 & 4.07\\
\hdashline
7.6 & 100.3 &	61.6 & 75.2	& 105.8 &	62.3 &	77.6 & 68.5 & 86.7 & 45.1 & 61.8 & 0.66 & 0.71 & 3.00 & 4.12 \\
\hdashline
8 & 100.5 & 61.7 & 75.3 & 105.1 & 62.3 & 77.4 & 68.6 & 86.1 & 44.8 & 61.7 & 0.65 & 0.72 & 2.99 & 4.12\\
9 & 101.1 & 61.8 & 75.6 & 103.1 & 62.4 & 77.1 & 68.8 & 85.0 & 44.3 & 62.2 & 0.64 & 0.73 & 2.95 & 4.15\\
10 & 101.6 & 61.9 & 75.8 & 101.1 & 62.4 & 76.8 & 69.0 & 84.1 & 43.8 & 62.6 & 0.64 & 0.74 & 2.92 & 4.17\\
\hline
\end{tabular}
}
\caption{Stand-alone LPM: $P_{\rm{AAo}}$, $P_{\rm{mPA}}$, EDV, SV, EF and CO for different PS diameters.}
\label{tab:shuntdiaoverall_lpm}
\end{center}
\end{table}

\begin{table}[tb]
\centering
\def\arraystretch{1.2}
\resizebox{0.57\textwidth}{!}{ 
\begin{tabular}{c|cc|cccc}
\hline
  $D_{\rm{SH}}$ & $W_{\rm{LV}}$ & ${W}_{\rm{RV}}$ & $ \overline{Q}_{\rm{SH}}$ & $\overline{V}_{\rm{SH}}$ & $\overline{\Delta P}_{\rm{SH}}$ & $Q_p/Q_s$\\
    ${\text{[mm]}}$ & $\times 10^{3} {\text{[mmHg-ml]}}$ & $\times 10^{3} {\text{[mmHg-ml]}}$  & ${\text{[ml/s]}}$ & ${\text{[ml]}}$ & ${\text{[mmHg]}}$ & [-]\\
  \hline
    pre  & 4.13	& 5.01 & - & - & - & 1\\
  \hdashline
  $5 $ & 3.94	& 5.84  & 14.7	& 13.2	& 9.48 & 0.78\\
  $6 $ & 3.91	& 5.91  & 16.6	& 15.0	& 7.48 & 0.76\\
  $7 $ & 3.87	& 5.95  & 18.0	& 16.2	& 6.26& 0.74\\
  \hdashline
  $7.6$& 3.85	& 5.97  & 18.6	& 16.7	& 5.73& 0.73\\
  \hdashline
  $8$  & 3.83	& 5.97  & 19.2	& 17.3	& 5.40& 0.72\\
  $9$  & 3.82	& 5.97  & 20.3	& 18.3	& 4.63& 0.71\\
  $10$ & 3.78	& 5.95  & 21.3	& 19.2	& 3.94& 0.70\\
  \hline
\end{tabular}
}
\caption{Stand-alone LPM: work done (area under the PV-loop) by the left ventricle ${W}_{\rm{LV}}$ and the right ventricle ${W}_{\rm{RV}}$ for varying shunt diameters.}
\label{tab:ventricle_work_lpm_qsh}
\end{table}

\begin{figure}
\centering
\includegraphics[width=0.87\linewidth]{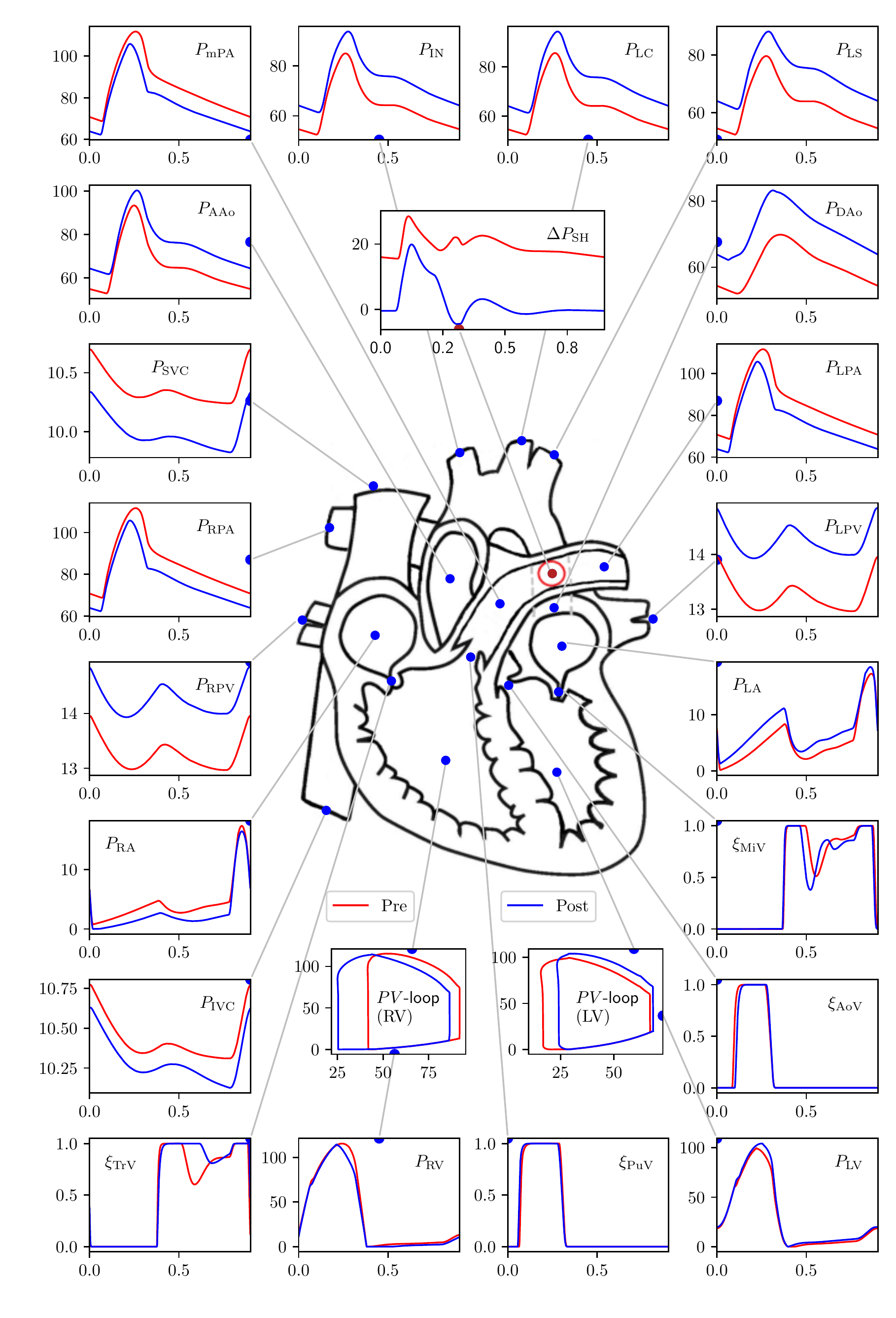}
\caption{Stand-alone LPM results for the 7.6 mm diameter PS showing pre- to post-operative changes in pressure at key locations in the arterial network. PV loops are additionally included. In the PV loop plots, the x-axis represents volume [ml] and y-axis represents pressure [mmHg]. In all other plots the x-axis represents time [s] and the y-axis represents pressure [mmHg]. The valve parameters $\xi$ are dimensionless. For a key to symbol nomenclature, please see Figure \ref{fig:multiscalemodel}. $\Delta P_{\mathrm{SH}}$ represents the pressure gradient across the PS.}
\label{fig:pre_post_pressure_lpm}
\end{figure}

\begin{figure}
\centering
\includegraphics[width=0.87\linewidth]{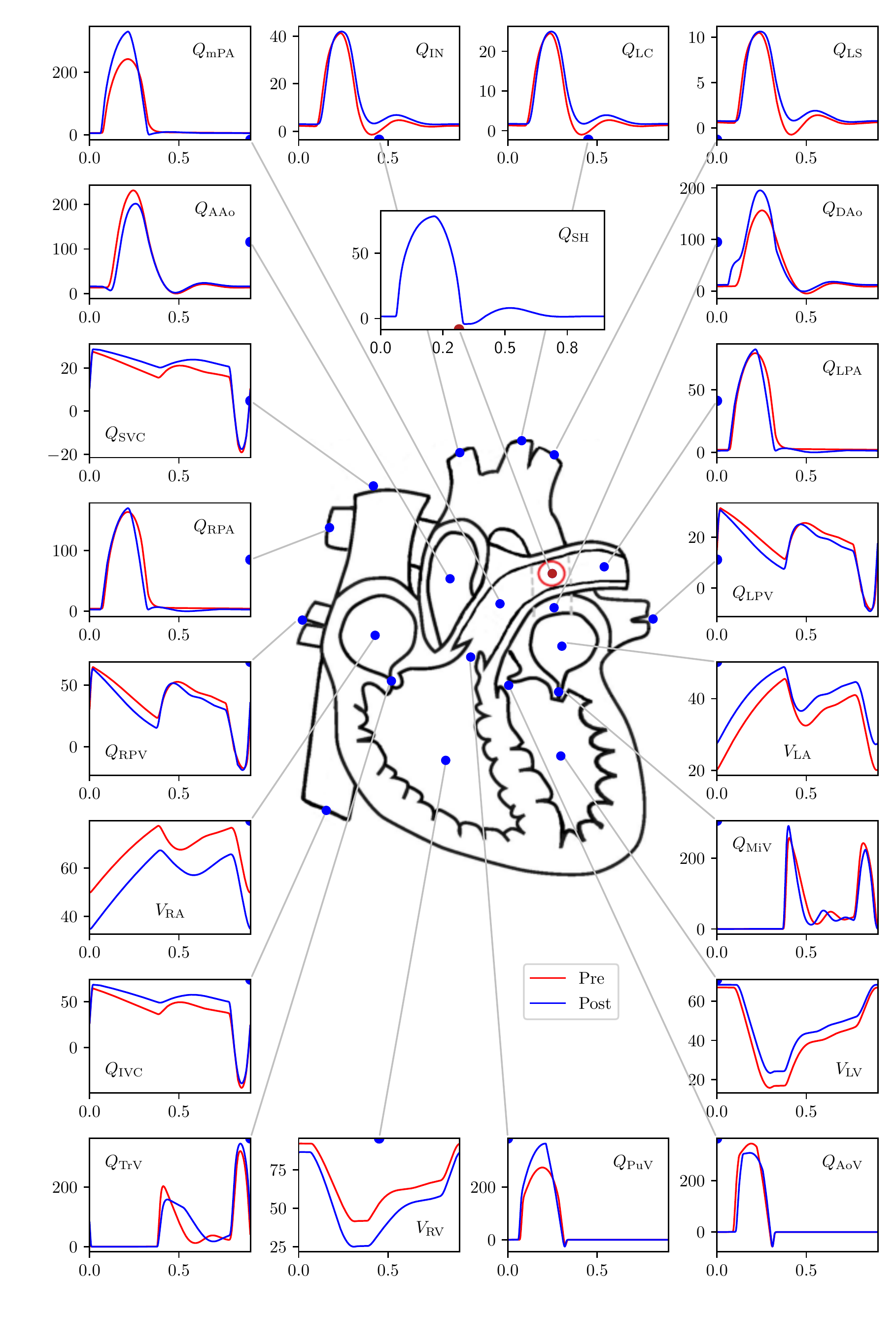}
\caption{Stand-alone LPM results for the 7.6 mm diameter PS showing pre- to post-operative changes in flow-rate and volumes at key locations in the arterial network. In all the plots the x-axis represents time, and y-axis for volumes, $V_{(\cdot)}$, is in [ml], while for the flow-rates, $Q_{(\cdot)}$, is in [ml/s]. For a key to symbol nomenclature, please see Figure \ref{fig:multiscalemodel}.}
\label{fig:pre_post_flow_lpm}
\end{figure}

\begin{figure}
\hspace{-1cm}
\includegraphics[width=1.1\textwidth]{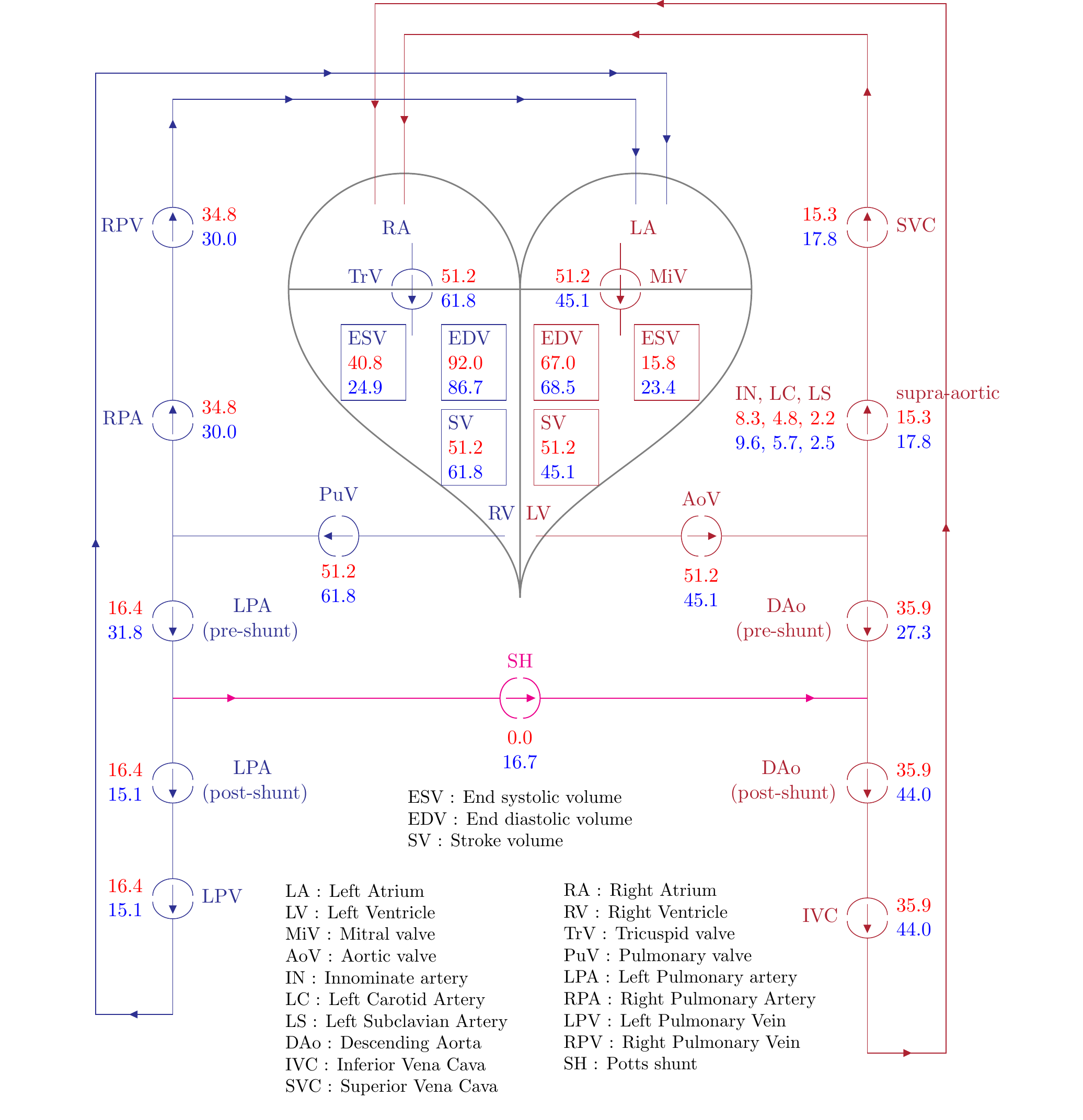}
\caption{Stand-alone LPM: Pre- to post-operative changes in volume of blood flowing in one cardiac cycle through the circulatory system. All numerical values are for volumes in ml, and values in red represent pre-operative state while those in blue represent post-operative state.}
\label{fig:volumes_lpm}
\end{figure}

\clearpage
\section{Modelling hypertension by changing contractility instead of activation curve}
\label{app:contractility}
\begin{figure}
\centering
\includegraphics[width=0.87\linewidth]{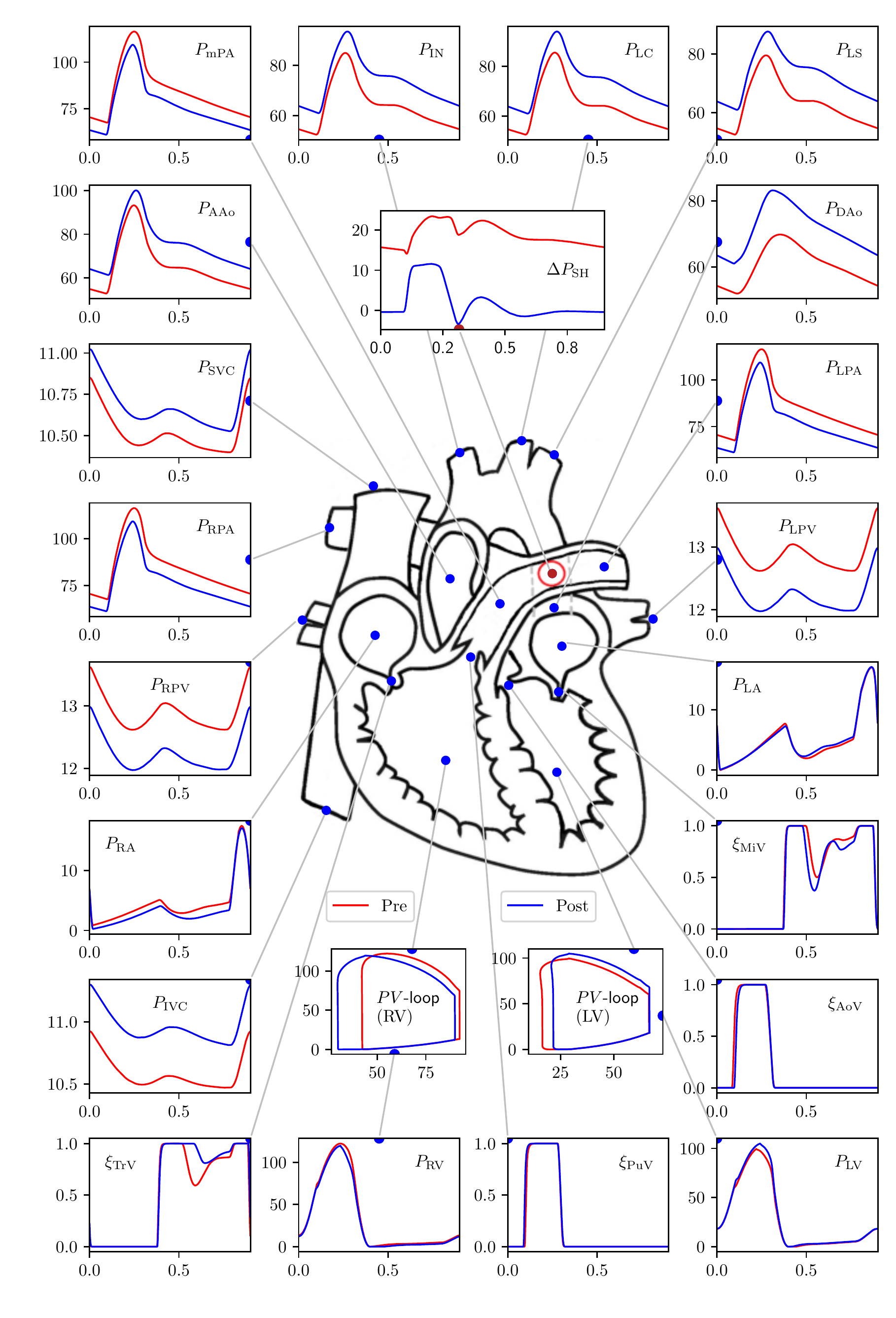}
\caption{Stand-alone LPM results (with parameter modification of Appendix \ref{app:contractility}) for the 7.6 mm diameter PS: pre- to post-operative changes in pressure at key locations in the arterial network. PV loops are additionally included. In the PV loop plots, the x-axis represents volume [ml] and y-axis represents pressure [mmHg]. In all other plots the x-axis represents time [s] and the y-axis represents pressure [mmHg]. The valve parameters $\xi$ are dimensionless. For a key to symbol nomenclature, please see Figure \ref{fig:multiscalemodel}. $\Delta P_{\mathrm{SH}}$ represents the pressure gradient across the PS.}
\label{fig:pre_post_pressure_lpmAo}
\end{figure}

\begin{figure}
\centering
\includegraphics[width=0.87\linewidth]{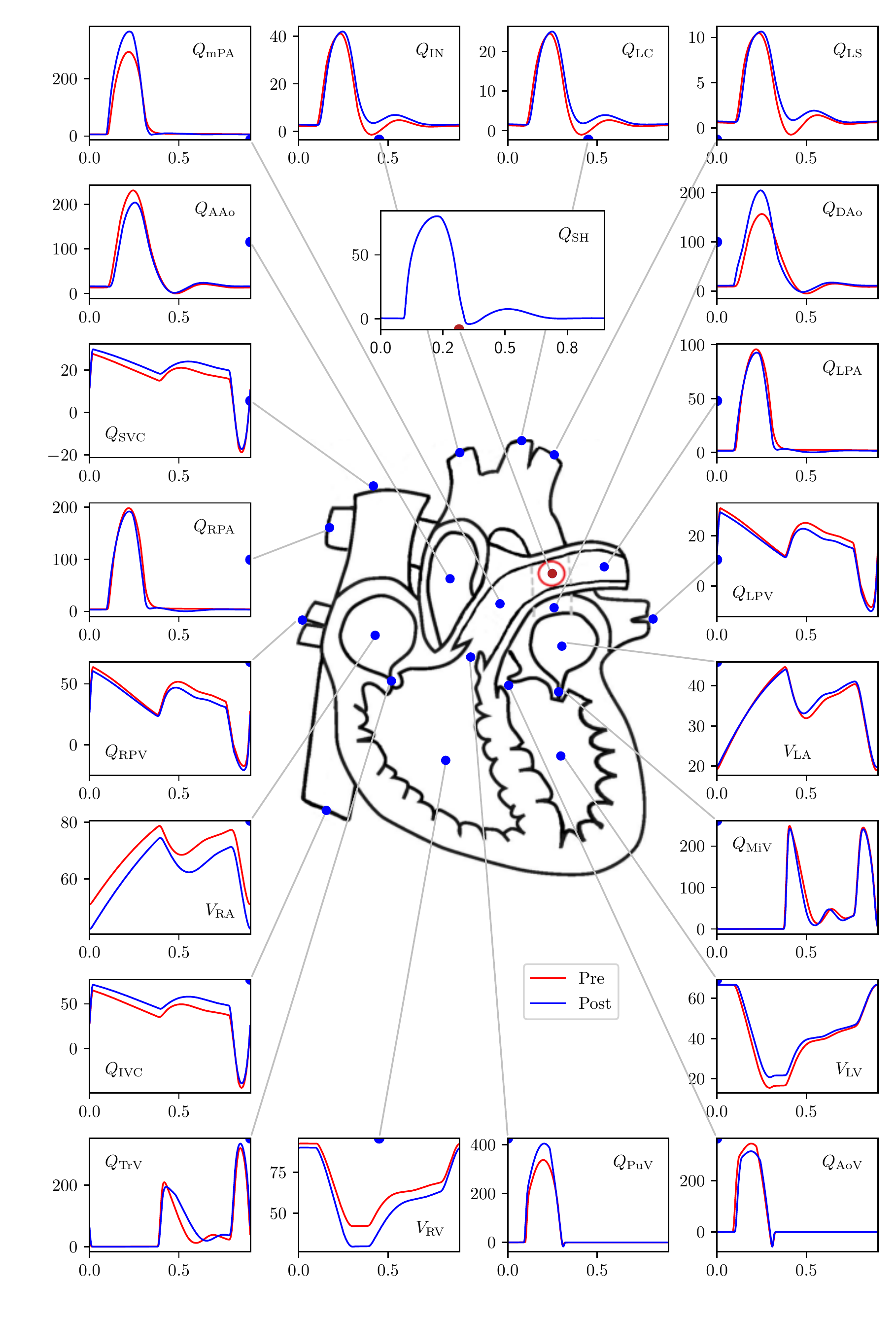}
\caption{Stand-alone LPM results (with parameter modification of Appendix \ref{app:contractility}) for the 7.6 mm diameter PS: pre- to post-operative changes in flow-rate and volumes at key locations in the arterial network. In all the plots the x-axis represents time, and y-axis for volumes, $V_{(\cdot)}$, is in [ml], while for the flow-rates, $Q_{(\cdot)}$, is in [ml/s]. For a key to symbol nomenclature, please see Figure \ref{fig:multiscalemodel}.}
\label{fig:pre_post_flow_lpmAo}
\end{figure}

\begin{figure}
\hspace{-1cm}
\includegraphics[width=1.1\textwidth]{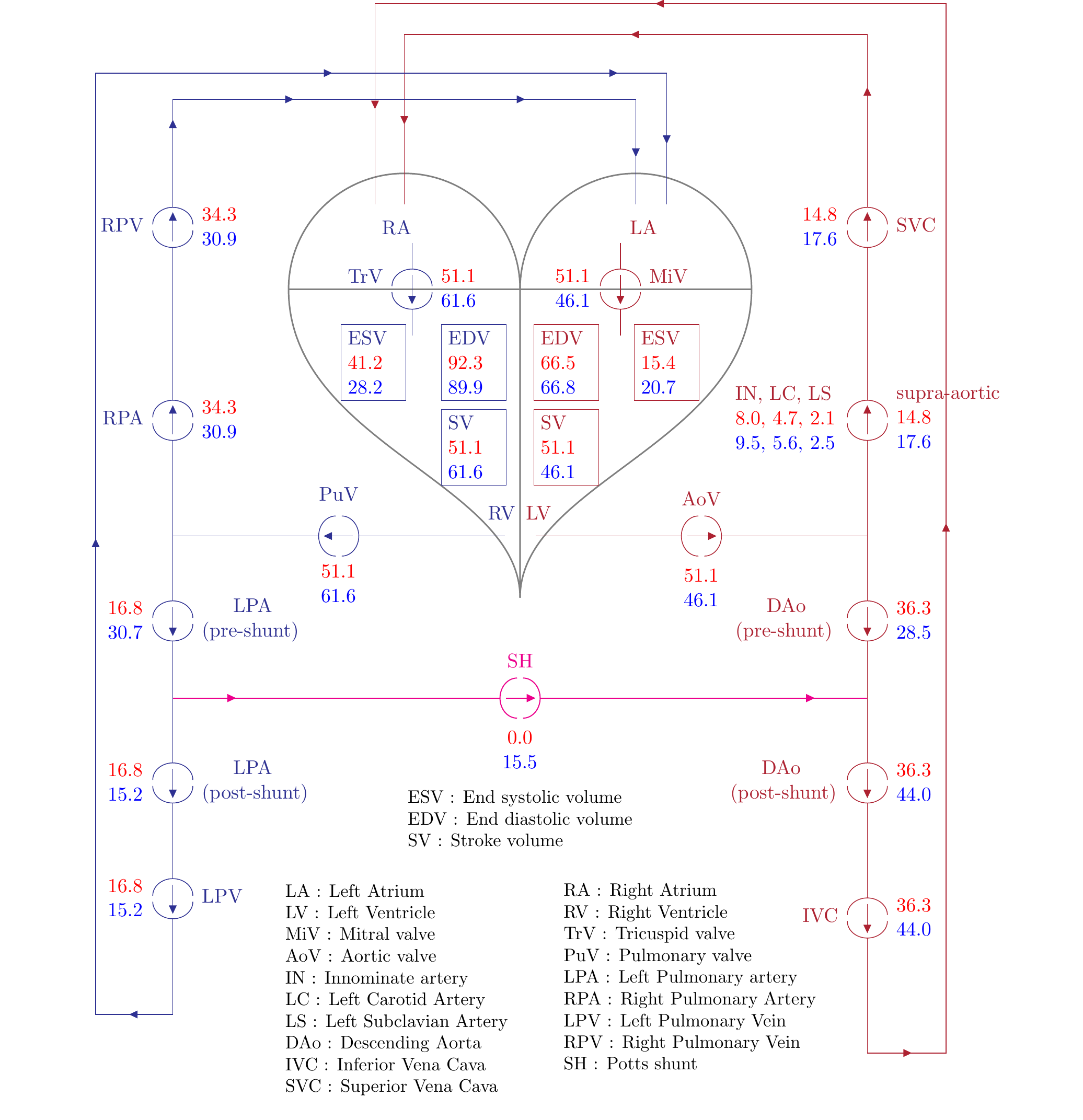}
\caption{Stand-alone LPM (with parameter modification of Appendix \ref{app:contractility}): pre- to post-operative changes in volume of blood flowing in one cardiac cycle through the circulatory system. All numerical values are for volumes in ml, and values in red represent pre-operative state while those in blue represent post-operative state.}
\label{fig:volumes_lpmAo}
\end{figure}

In Section \ref{sec:two_ways}, two methods to model pulmonary artery hypertension were outlined. The active stresses in the RV can produce higher pressures through equation \eqref{eqn:active_stress} either by the different shape of the time-dependent activation  $g(t_a)$ or through a change in contractility of the RV $c$ (alternatively $T_{a0}$). While the former approach is followed in the main text, here the results for the latter approach are presented. Thus, the parameter $E_a^{\mathrm{RV}}$, which was tuned to a value of 0.48 for a stronger activation curve, is changed to 1.05, and the contractility $c$ is increased by 15\% to 1.15. All other parameters are kept identical to the ones in the main text. The stand-alone LPM results for this configuration are shown in Figures \ref{fig:pre_post_pressure_lpmAo}, \ref{fig:pre_post_flow_lpmAo}, and \ref{fig:volumes_lpmAo}, which show the pressure traces including PV-loops, flow-rate traces, and volume displaced in one cardiac cycle through the circulatory system, respectively. Comparing these to the configuration previous configuration---Figures \ref{fig:pre_post_pressure_lpm}, \ref{fig:pre_post_flow_lpm}, and \ref{fig:volumes_lpm}, respectively, for the stand-alone LPM;  and Figures \ref{fig:pre_post_pressure_ms}, \ref{fig:pre_post_flow_ms}, and \ref{fig:volumes_ms}, respectively, for the GMM solution---it is concluded that the trends of change from pre- to post-operative state are similar, and that the choice of high pressure modelling in the RV has not influenced the major conclusions presented in the main text.

\end{document}